# Mechanistic Insights into Chemical Exchange during the Signal Amplification by Reversible Exchange Sensitization of Pyruvate

Charbel D. Assaf†[a], Vladimir V. Zhivonitko*†[b], Amaia Vicario[c], Alexander A. Auer[d], Simon B. Duckett[e], Jan-Bernd Hövener[a] and Andrey N. Pravdivtsev*[a]

[a]Section Biomedical Imaging (SBMI), Molecular Imaging North Competence Center (MOINCC), Department of Radiology and Neuroradiology, University Hospital Schleswig-Holstein, Kiel University, *Am Botanischen Garten 14/18*, 24118 Kiel, Germany. E-mail: andrey.pravdivtsev@rad.uni-kiel.de

[b]NMR Research Unit, Faculty of Science, University of Oulu, P.O. Box 3000, Oulu, Finland. E-mail: vladimir.zhivonitko@oulu.fi

[c]University of Bath, Faculty of Science, Claverton Down, Bath, BA2 7AY, UK

[d]Max-Planck-Institut für Kohlenforschung, Kaiser-Wilhelm-Platz 1, 45470 Mülheim an der Ruhr.

[e]Centre for Hyperpolarization in Magnetic Resonance (CHyM), Department of Chemistry, University of York, Heslington YO10 5NY, UK

†equal contribution

*corresponding authors

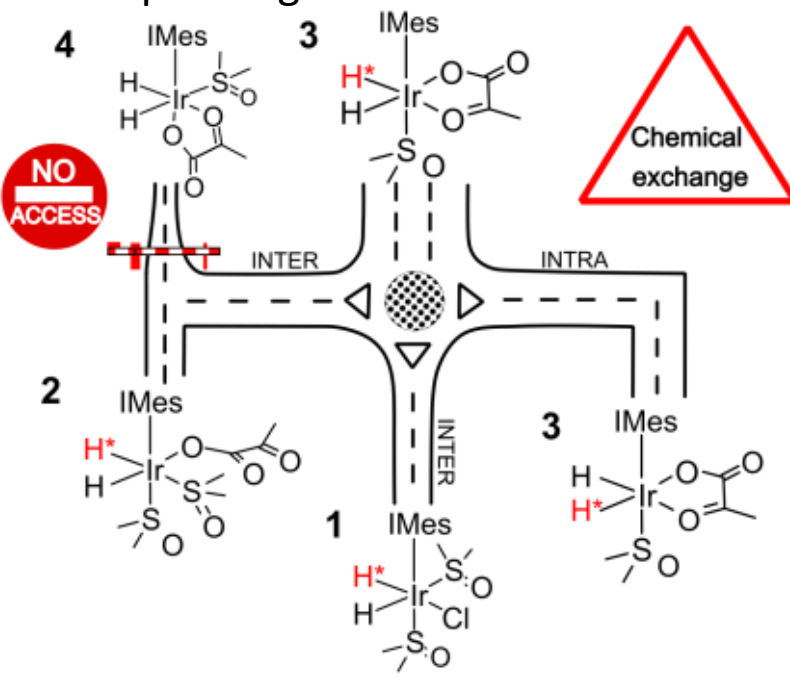

**ABSTRACT:** Signal amplification by reversible exchange (SABRE) is a nuclear spin hyperpolarization technique in which the transient interaction of parahydrogen ($pH_2$) and a target substrate with an iridium complex leads to polarization transfer to the substrate. Here, we use a parahydrogen-enhanced, spin-selective NMR method to investigate pyruvate binding, which is combined with exchange-model fitting and DFT calculations. Our study reveals several key findings that reshape the current understanding of SABRE: (a) intramolecular hydrogen exchange of the hydrides, occurring faster than pyruvate or $H_2$ loss; (b) the discovery of a novel stable $[Ir(H)_2(IMes)(\kappa^1\text{-pyr})(DMSO)_2]$ complex; and (c) the potential role of counterions (here $Na^+$) in Ir-pyruvate binding. Previously unknown insights into complex kinetics and distributions as a function of temperature, [DMSO], [pyruvate], and hydrogen pressure are presented. The methods demonstrated here, exemplified by SABRE, provide a framework that is expected to guide future research in the field.

## INTRODUCTION

For more than a decade, [1-$^{13}$C]pyruvate has been the most prominent tracer for hyperpolarized metabolic magnetic resonance imaging[1–3]. The reason for this is a sufficiently long polarization lifetime, a sufficiently high applicable dose, rapid in vivo metabolic transformation into lactate, alanine, bicarbonate, and $CO_2$, and their association with inflammation[3] and tumor malignancy[4,5] as a result of the Warburg effect[6]. To achieve the necessary increase in MRI detectability of pyruvate and its metabolic products, the dissolution dynamic nuclear polarization (dDNP) has emerged as the most successful translational method.[7] This method can produce up to 60% of $^{13}$C polarization and deliver millimolar concentrations of pyruvate for in vivo use following a 30 to 60 minute polarization and dissolution process, but it requires costly equipment. However, the hyperpolarization community widely aspires to develop swifter and more cost-effective alternatives for hyperpolarizing biomolecules,[8] with one prominent approach being parahydrogen-based methods[9–11].

Parahydrogen ($pH_2$)-based hyperpolarization methods offer faster and more cost-efficient alternatives. Hydrogenative parahydrogen-induced polarization (PHIP), more specifically, its side-arm hydrogenation variant (PHIP-SAH)[12–14] has demonstrated great potential as a viable alternative to dDNP, which is underpinned by the high-yield synthesis of essential precursors[12,13,15–18]. Additionally, it has achieved high $^{13}$C polarization levels of pyruvate around 20%[19]. The non-hydrogenative variant of PHIP, signal amplifica-tion by reversible exchange (SABRE)[20], offers another route to pyruvate hyperpolarization[19,21,22]. In SABRE, a ligand (e.g., pyruvate) and $pH_2$ are temporarily associated with an Ir complex, whose spin-spin interactions allow converting the spin alignment of $pH_2$ into the polarization of a ligand. It enables direct hyperpolarization without altering the substrate, allowing rapid polarization buildup within seconds under mild conditions. However, current implementations typically achieve lower $^{13}C$ polarization levels of a few percent for pyruvate and require careful control of exchange dynamics to optimize the performance. The strength of the spin-spin interaction and the lifetime of the complex play critical roles in maximizing polarization transfer efficiency[23–25]; the latter is controlled by tuning the temperature[26] or ligand identity[27].

In the most common SABRE process, the exchanging ligand is exemplified by a two-electron donor, such as pyridine[28]. For a better understanding of the system, it is desirable to measure ligand loss rates for the dissociative exchange pathway following the SABRE catalysts, such as $[Ir(H)_2(IMes)(pyridine)_3]Cl$[29–31]. Some recent studies in this area have deployed in-field PHIP to improve sensitivity, selective pulses to enhance selectivity and magnetization transfer efficiency, and used long-lived $^{15}N$ signals to measure lower exchange rates[26,32–38]. Obtaining these ligand-exchange rates is valuable, as they also provide benchmarking data for the subsequent *in silico* optimization of hyperpolarization conditions[39,40]. However, only a few reported studies focus on determining the $H_2$ exchange rate[32], as this chemical exchange process typically occurs after substrate loss and is complicated by the role that a dihydrogen-dihydride complex plays in the exchange mechanism[29,32]. In the $[Ir(H)_2(IMes)(pyridine)_3]Cl$-type systems, it is broadly accepted that the exchange of pyridine-type ligands is dissociative (like in $S_N1$), while the exchange of $H_2$ that happens after pyridine loss is associative (like in $S_N2$)[32,41].

Pyruvate (pyr), however, differs significantly from a pyridine-like ligand, as it can act as a mono- or bidentate ligand. In the case of SABRE, though, a co-ligand, typically dimethyl sulfoxide (DMSO), is required to be present within the metal's coordination sphere in addition to the N-heterocyclic carbenes (NHC). The need for a co-ligand and the diverse binding of pyruvate addition significantly complicates the speciation of the resulting metal complex, and several important complexes have been previously proposed to exist in solution.[22] These included *bis*-DMSO complex $[Ir(Cl)(H)_2(DMSO)_2(IMes)]$ (**1**),[42] and $[Ir(H)_2(\kappa^1\text{-}\kappa^2\text{-}pyr)(DMSO)(IMes)]$, which was initially proposed to exist as two isomers[22] where pyruvate binds in the equatorial plane (**3**) or bridges equatorial-axial sites (**4**) (**Fig. 1**). Previous density functional theory (DFT) predictions[43] indicated that isomer **4** has a Gibbs free energy 7.6 kJ/mol higher than isomer **3**. Such a significant energy difference implies much lower equilibrium concentrations of **4** than of **3**.

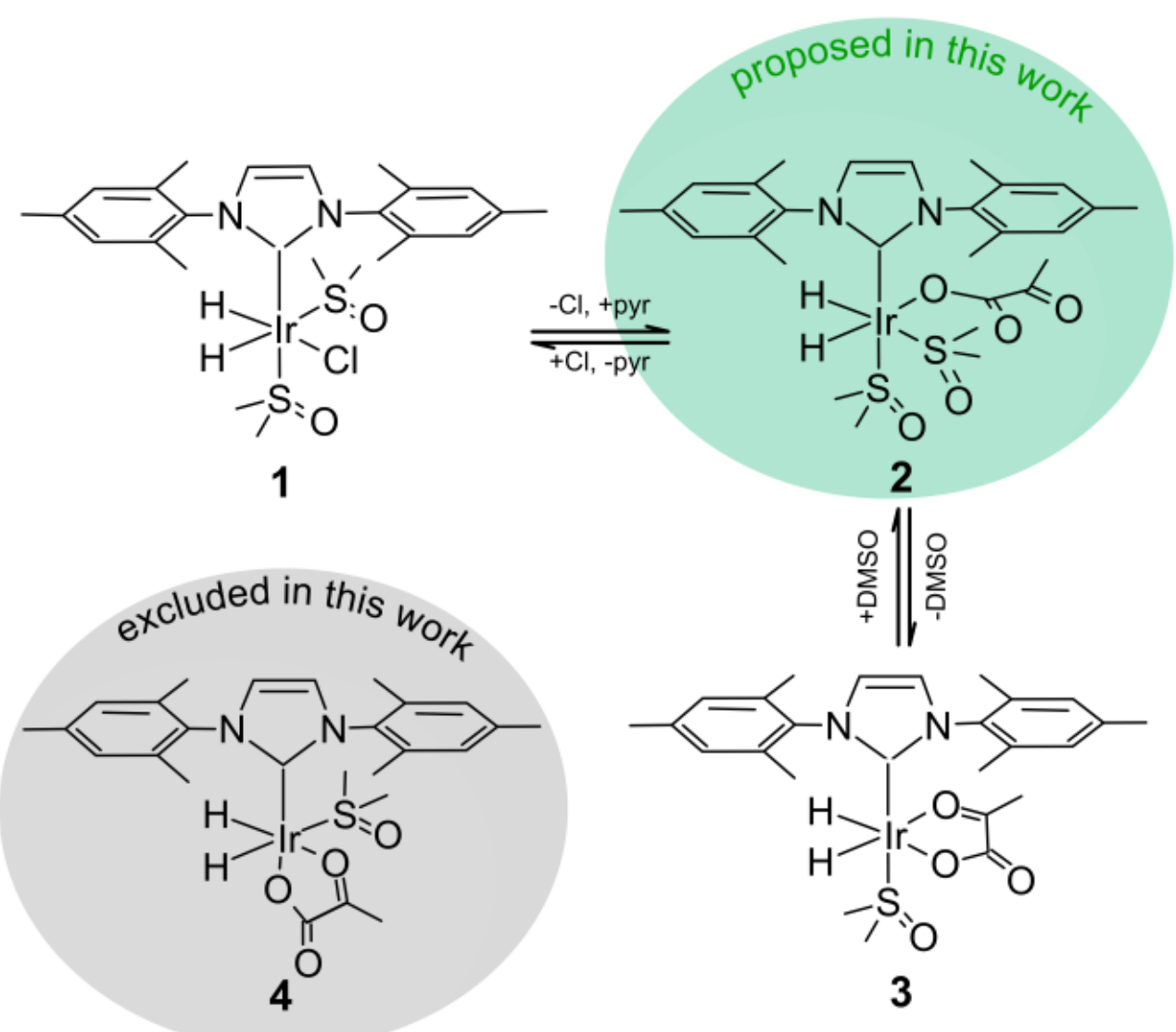

**Fig. 1. Revised speciation of Ir complexes in pyruvate SABRE.** Structures of complexes **1**-**4**. Complexes **1-3** are formed when IrIMes reacts with DMSO, pyruvate, and $H_2$ in methanol. Previously, **4** was assumed to be populated to a high level and in exchange with **1** and **3**; however, we show here that it is actually absent and energetically unfavorable compared to its isomer **3**, while **2** is present instead. Therefore, **4** should be excluded from consideration for the discussed pyruvate SABRE composition.

Herein, we show that our experimental results clearly contradict the previous assumptions, revealing significant shortcomings in the existing mechanistic model. In particular, we found that the $^{1}H$ and $^{13}C$ NMR signals, previously associated with complex **4** in the literature,[22,27] can grow to levels exceeding those of **3** as the DMSO concentration increases or as the temperature decreases, highlighting the significant flaws in the existing mechanistic model.

Given this contradiction and the importance of hyperpolarizing pyruvate using a cost-efficient SABRE method, we herein report a comprehensive reexami-nation of the mechanistic transformations of the pyruvate SABRE system. Importantly, by collecting 2D NMR and chemical-exchange data, we show that the previously proposed complex **4** (**Fig. 1**) is not observed and should be excluded from the list of key intermediates under normal conditions. Instead, intermediate $[Ir(H)_2(IMes)(\kappa^1\text{-pyr})(DMSO)_2]$ (**2**), forms in the system. Consequently, the corresponding $^{1}H$ and $^{13}C$ NMR resonances previously associated with **4**, in fact, belong to complex **2**. Additionally, we observe a pronounced *intramolecular* hydride ligand exchange in complexes **1**-**3**, which was previously unreported but can significantly limit the observable $^{13}C$ hyperpolarization levels. Furthermore, we present a Density Functional Theory (DFT) computational analysis of the proposed mechanism, revealing a potential role for $Na^+$ interactions in influencing relative stabilities and offering an explanation for the previous misassignment.

Our experimental approach leverages the fact that the hydride ligand protons (IrHH) in species **1**-**3** do not overlap, and can therefore be targeted separately by frequency-selective polarization transfer in exchange NMR experiments.[44,45] The resulting data enable the determination of the kinetic parameters for ligand exchange in **Fig. 2** with high sensitivity by using $pH_2$. Previously, this sensitization method enabled the measurement of *J*-coupling constants between hydride ligands and $^{13}C$ nuclei of bound pyruvate on SABRE complexes.[43]

Overall, we present several crucial results that necessitate reinterpretation of earlier data and suggest that further studies are needed to predict the quantitative outcomes of SABRE if an optimal solution is to be achieved.

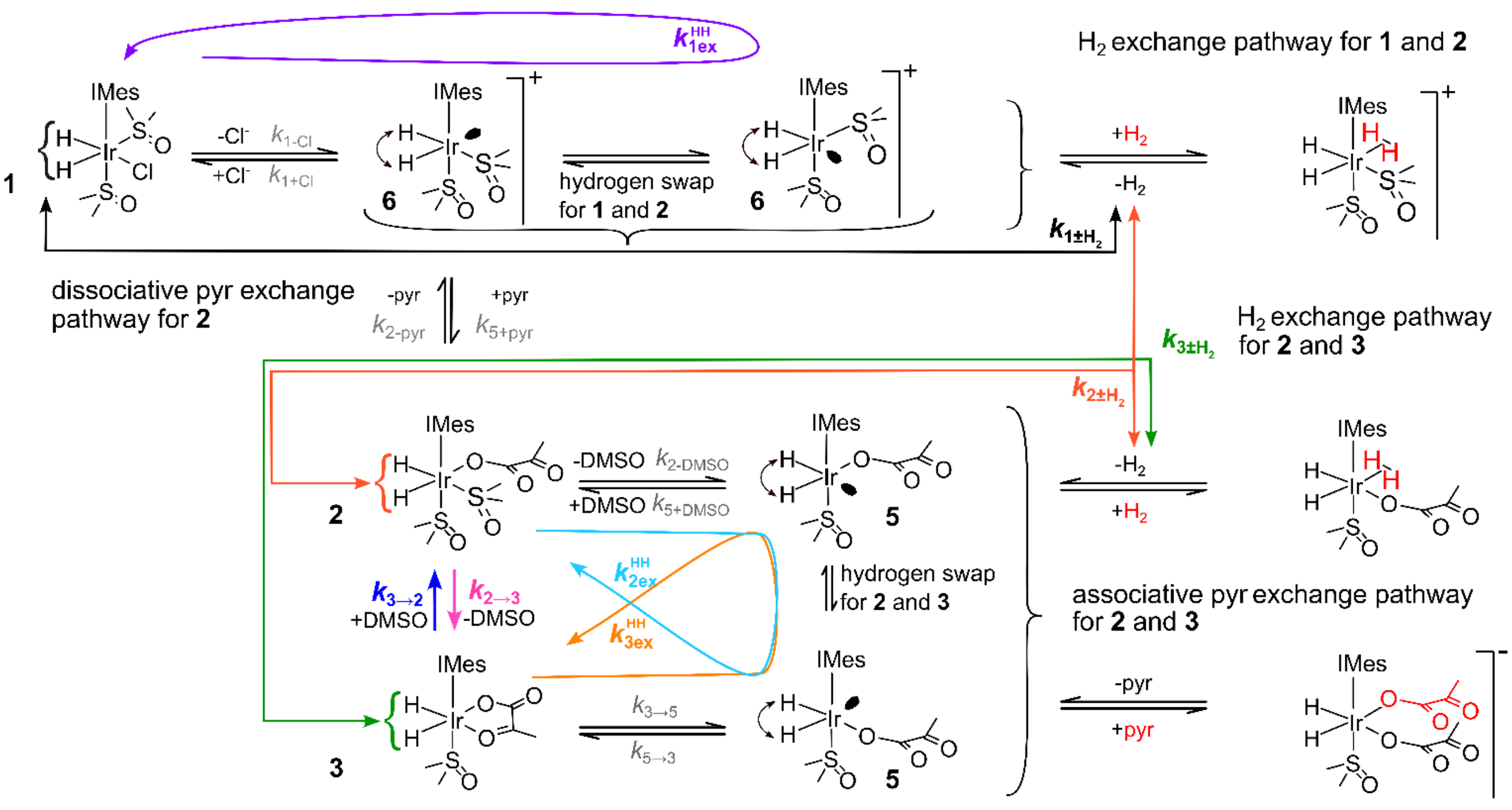

**Fig. 2. Schematic of pyruvate, DMSO, and $H_2$ exchange, which explains the observed experimental kinetic dependencies.** Critical elements of the exchange: (1) $H_2$ and pyruvate exchange on a transition between **1** and **2**, where "+pyr" and "–pyr" indicate coordination from or dissociation to the solution phase; (2) intramolecular H-H exchange happens during rearrangement between **2** and **3** and **1** and **2**; (3) transformation from **3** to **2** is [DMSO] dependent. Elementary reaction steps are indicated with grey colored rates and single-barbed arrows. In contrast, transmission rates are indicated with colors and regular arrowheads. For example, transmission $k_{3ex}^{HH}$ involves multiple transformations going from **3** to **5** and back to **3**. Note that for the sake of simplicity $k_{2ex}^{HH}$ is indicated only in the cycle **2**-**5**-**2**, while it could also occur in **2**-**6**-**2**. The transmission $k_{3\text{-}pyr}$ is not indicated, but it goes from 3 to 5 to the associative pyr exchange. Transmission $k_{2\text{-}pyr}$ consists of two pathways, one through complex **5** (associative) and the other one through complex **6** (dissociative).

## RESULTS AND DISCUSSION

**Reevaluation of NMR signal assignments and structures in pyruvate SABRE:** Our experiments discussed here and below indicate that a revised assignment of SABRE complexes is necessary. In particular, we believe that $^{1}$H NMR signals of hydride ligands in complex **2** at δ -14.98 and -24.09 ppm (**Fig. 3A**) were incorrectly ascribed to complex **4** in the literature[22,27,43,46–49].

The first key piece of evidence comes from NOE 2D NMR. We investigated a sample containing IrCl(COD)(IMes), DMSO-$h_6$, and [2-$^{13}$C]pyruvate in methanol-$d_4$ at 253 K after exposure to $H_2$. $^{1}$H-$^{1}$H NOE revealed cross peaks between hydride protons of **2** (δ -14.98 and -24.09 ppm) and three methyl groups of DMSO ($^{1H}\delta$ 3.37, 2.98, and 2.92 ppm, **ESI**, **Fig. S19**). Due to the low hydride-ligand symmetry, a single DMSO ligand would produce two connections to the hydrides, whereas our observation of three peaks was indicative of two DMSO ligands.

To confirm this finding, we carried out two alternative experiments using [U-$^{13}$C]DMSO and nonlabelled pyruvate: HMQC (heteronuclear multiple-quantum correlation)[50] $^{1}$H-$^{13}$C correlation 2D NMR (**ESI, Fig. S20**) and $^{1}$H-$^{13}$C SEPP-SPINEPT (frequency-selective excitation of parahydrogen-derived PASADENA polarization followed by frequency-selectively pulsed insensitive nuclei enhanced by polarization transfer) that transfers polarization from p$H_2$ to neighboring carbons[43] (**ESI, Fig. S21**). Here, PASADENA refers to the high-field p$H_2$ experiment: p$H_2$ and synthesis allow dramatic enhancement of nuclear alignment[51] (**Fig. 3A**).

The SEPP-SPINEPT applied to hydrides of **2** yielded four hyperpolarized carbons of this isotopomer of DMSO (**ESI, Fig. S21**). HMQC, on the other hand, provided resonances for the neighboring methyl protons ($^{1H}\delta$-$^{13C}\delta$ pairs were 3.37-48.27, 2.98-58.98, 2.92-44.19, and 3.14-50.20 ppm, **ESI, Table S17**). With this we confirmed the presence of two DMSO ligands and assigned four $^{13}$C and $^{1}$H resonance pairs, establishing that the correct identity of the complex is [Ir(H)$_2$(IMes)(κ$^1$-pyr)(DMSO)$_2$] (**2**) rather than the previously proposed [Ir(H)$_2$(IMes)(κ$^1$-κ$^2$-pyr)(DMSO)], (**4**)[22].

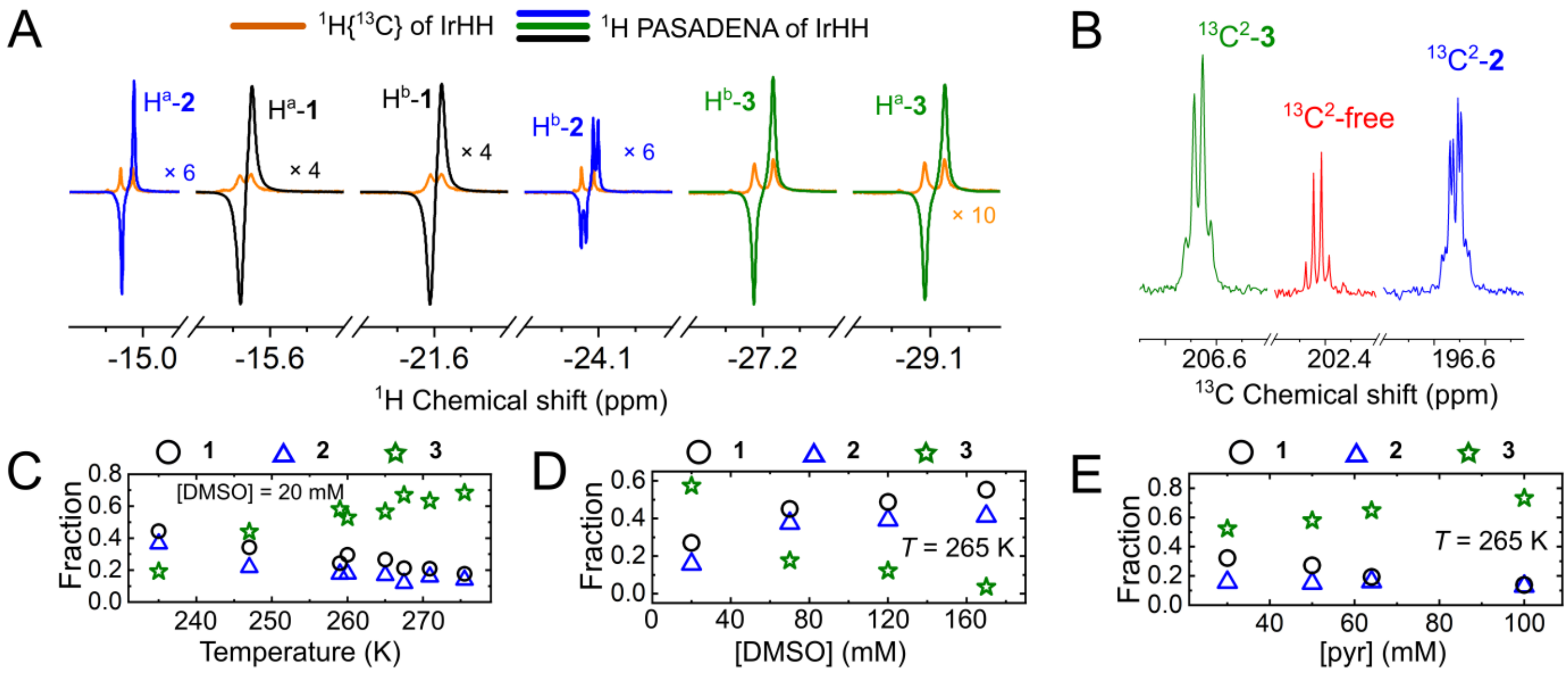

**Fig. 3. Overview of hydride spectra and concentration dependencies of dominant Ir complexes present in an IrIMes, DMSO, pyruvate, and $H_2$ in methanol reaction mixture**. (A) $^{1}$H{$^{13}$C} NMR spectrum of thermally polarized (vermilion: multiplied by 10, ×10) and $^{1}$H NMR of hyperpolarized PASADENA spectrum of IrHH proton resonances of **1**-**3** (blue: ×6, black: ×4, and green). The thermal spectrum was acquired with 100 scans with a receiver gain (RG) of 101, while the hyperpolarized spectra were recorded in a single scan with an RG of 28.5. (B) Representative $^{13}$C spectra of free [2-$^{13}$C]pyruvate in methanol (red), coordinated to **2** (blue) and **3** (green), obtained at 265 K. The spectra were hyperpolarized using the SEPP-SPINEPT-SABRE sequence. (C, D, E) Relative concentrations of hydride protons of **3** (green star), **2** (blue triangle), and **1** (black circle) as a function of temperature (C, [DMSO] = 20 mM), [DMSO] (D, T = 265 K), and [pyr] (E, T = 265 K, [DMSO] = 20 mM). Note that the changes in [**3**] and [**2**] as a function of [DMSO] have different dependencies, indicating different numbers of DMSO ligands. Experimental parameters: $B_0$ = 9.4 T, 40 mM sodium pyruvate, and 4 mM [Ir(IMes)(COD)(Cl)] precatalyst were dissolved in methanol-$d_3$. Concentrations were measured using $^{1}$H NMR at 9.4 T, with 2 scans for averaging; the tube was constantly maintained under 6.9 bar (100 psi) $H_2$ pressure ([$H_2$] = 23.73 mM)[52].

**Proposed chemical exchange model and the key features of the pyruvate SABRE system:** To rationalize our experimental findings, we propose a general scheme of pyruvate SABRE to provide the reader with a basic picture of the involved processes.

As noted in the introduction, a relatively well-studied SABRE mechanism for pyridyl-type systems indicates dissociative loss of pyridine and associative $H_2$ exchange. Hence, by analogy, dissociative loss of DMSO and chloride in **1** and **2** (**Fig. 1**) is likely. However, pyruvate dissociation will be more complicated, as the reversible formation of a κ$^1$-pyr intermediate from the κ$^1$-κ$^2$-pyr counterpart would be expected, for instance, during the interconversion of **3** and **2**. This intermediate will correspond to 16-electron $[Ir(H)_2$(IMes)(κ$^1$-pyr)(DMSO)] **(5)**, and likely have an octahedral geometry with inequivalent hydride ligands, where one is *trans* to a vacant site. Any $H_2$ exchange is then expected to be associative, via the analogous dihydrogen-dihydride intermediate (**Fig. 2**).

**Fig. 2** shows such a hypothesized reaction pathway that allows the equilibration of **1**, **2,** and **3** through a series of dissociative reaction steps involving loss of $Cl^-$ ($k_{1\text{-Cl}}$), DMSO ($k_{2\text{-DMSO}}$), and pyr$^-$ ($k_{2\text{-pyr}}$), or pyr transition from bidentate to monodentate binding ($k_{3\rightarrow6}$) and back ($k_{6\rightarrow3}$). These steps involve the 16-electron intermediates **5** ($k_{2\text{-DMSO}}$, $k_{3\rightarrow5}$) and $[Ir(H)_2(IMes)(DMSO)_2]^+$ (**6**, $k_{2\text{-pyr}}$ and $k_{1\text{-Cl}}$). These intermediates are expected to bind DMSO, methanol, $Cl^-$, pyruvate, and water. Unless complexes **1**-**3** are formed, the resulting species are unobservable in NMR because they are energetically very unfavorable. The associative bimolecular $H_2$ loss occurs via binding of $H_2$ to **5** and **6** intermediates (**Fig. 2**). We propose that these two intermediates also enable the hydride ligands to interchange positions. The corresponding, now first-order, rearrangement steps are indicated as “hydrogen swap” in **Fig. 2**. The observed hydrogen dissociation transmission rate for **1** is displayed as $k_{1\text{-H2}}$ and involves several steps, while only the initial reagent and final product are observable.

When viewing this figure, note that elementary reaction steps (grey rates with single-barbed arrows) and transmission rates (not grey rates with regular arrowheads) are shown. Care should be taken since only overall transmission rates between reagents and products are observed, not each intermediate step.

As was noted above, based on the experimental and theoretical evidence supported by DFT calculations (**ESI, Fig. S24-29**), we have found that complex **4** (**Fig. 1**) does not play a significant role in pyruvate SABRE and rule it out from the list of key intermediates of this system in **Fig. 2:** Instead, we prove the existence of complex **2** as one of the key species for pyruvate SABRE, along with complexes **1** and **3**.

In the sections below, we evaluate the validity of the model by performing a series of exchange spectroscopy (EXSY) NMR measurements (data and experimental schemes are in **Figs. 3, 5, and 6**). EXSY allowed us to monitor the chemical exchange of free pyruvate, DMSO, and $H_2$ with their counterparts bound to the iridium complexes. We repeated experiments with varying concentrations of $[H_2]$, [pyr], and [DMSO] to quantitatively probe the exchange dynamics. Then we discuss how concentrations affect the rates of chemical transformations in the context of the proposed reaction mechanisms here or previously. These observations and the resulting hypothesis of pyruvate SABRE kinetics (**Fig. 2**) are supported by a computational study detailed below.

**Evaluation of thermodynamic properties of the pyruvate SABRE system:** Temperature strongly influences the kinetic parameters and the equilibrium populations of the complexes, affecting the hyperpolarization process[26]. Moreover, it provides insights into the energetics of the intersystem conversion, which can then be compared with DFT simulations to verify the SABRE mechanism. Therefore, we first measured the fractions of complexes **1**-**3** as a function of temperature by using $^1$H NMR signals of hydride ligands (**Fig. 3C**). These temperature dependencies were probed multiple times with freshly prepared samples and at two different research centers to ensure reproducibility. Notably, the fraction of **3** increased with increasing temperature, while the amounts of **2** and **1** decreased. This observation implies that **3** is thermodynamically less favorable than **2** and **1,** being higher in relative energy. Considering equilibrium reaction conditions for these complexes, the Gibbs free energies of **1**⟵**3** and **2**⟵**3** transformations were estimated using the concentration ratios $\frac{[2]}{[3]} = [DMSO]e^{\frac{-\Delta G^0_{2\leftarrow3}}{RT}}$ and $\frac{[1]}{[3]} = [DMSO]$, as detailed in (**ESI**, **Eqs. S41** and **S44**). These analyses yielded $\Delta G^0_{2\leftarrow3} = \Delta G^0_2 - \Delta G^0_3$ = -(5.3 ± 0.14) kJ/mol and $\Delta G^0_{1\leftarrow3}$ = -(11.85 ± 0.51) kJ/mol at $T$ = 270.9 K (**ESI**, **Section 6**).

**DMSO concentration dependency:** We found that when the concentration of DMSO was increased, the relative proportions of the two-DMSO complexes, **2** and **1**, increased, while that of the one-DMSO complex, **3**, decreased (**Fig. 3D**). This behavior cannot be rationalized with the previous assignment of complexes **1**, **3,** and **4**, since in this case, only [**1**] would increase, while [**3**] and [**4**] would relatively decrease. Thus, we further confirmed the revised identity of **2**, whose hydride ligand chemical shift signals were found not to change with increasing [DMSO] (**ESI**, **Fig. S19**). A related DMSO-dependent change in observed $^{13}C$ hyperpolarized signal amplitudes was independently reported by Mamone et al.[46], although without analysis of equilibrium populations or reassignment of complex identities. Given the modified structures and associated chemical exchange diagrams (**Fig. 2**), the dependence on [DMSO] becomes evident (**ESI, Section 6**).

**Pyruvate concentration dependency:** A further study was then undertaken where the [pyr] was increased (**Fig. 3E**). The proportions of the three iridium complexes were found to depend on the pyruvate concentration, with their total concentration decreasing to approximately 40% of the total iridium concentration as [pyr] increased. This is in line with previous observations in which polarization yield is negatively correlated with [pyr][49,53].

Surprisingly, the concentration of **3** also increases relative to **2** as a function of [pyr], which was not expected based on the presented chemical model (**Fig. 2**). This is because, according to their chemical structures, both **2** and **3** contain one pyr ligand; hence, changes of [pyr] should not favor one of them. The observed trend, therefore, indicates that the system behaves more intricately than the simple model predicts.

**Computational study:**

Electronic structure calculations at the B3LYP-D4/def2-TZVP CPCM (Methanol) level of theory[54–62] using the ORCA 6 program package[63,64] yielded relative Gibbs free energy differences ($\Delta G_{50\%}$,[65–67] see **ESI, Section 18** for details, **Table S18**) of +9.1 and +15.8 kJ/mol for **3** and **4** with respect to **2** (**Fig. 4**). Within the error bars of DFT, this is consistent with the experimentally evaluated thermodynamic parameters presented above, which indicated that **2** is energetically favorable over **3** by about 5 kJ/mol.

Upon studying the relative energies and electronic structures of the complexes, we found that the electrostatic potentials of **2**, **3**, and **4** (**Fig. 4**) are particularly indicative of potentially strong counterion interactions. We found that their relative stabilities are strongly influenced by the explicit addition of species such as $Na^+$. Note that implicit solvation models are known to exhibit larger errors in the case of strong specific interactions with solvent molecules or counterions. We also checked the impact of $Na^+$ because SABRE presumably works with sodium pyruvate, as originally proposed, rather than with pyruvic acid, for which SABRE has not yet been shown to work, providing another experimental indication of the possible role $Na^+$-pyr pairing plays in complex stabilization. Here, we found that the complex-counterion interaction with $Na^+$ is strongest for **2** ($\Delta E$ = 144.1 kJ/mol), followed by **3** ($\Delta E$ = 116.8 kJ/mol) and **4** ($\Delta E$ = 113.0 kJ/mol), in agreement with the electrostatic potential assessment. For a series of simple adduct models (complex with $Na^+$, methanol, or DMSO, **ESI**, **Fig. S26**), this can lead to changes in the relative free energies of more than 10 kJ/mol. While these preliminary results should not be regarded as quantitative, they hint at the great potential of this influence, and further detailed experimental and computational studies of the ions' effects are underway in our group.

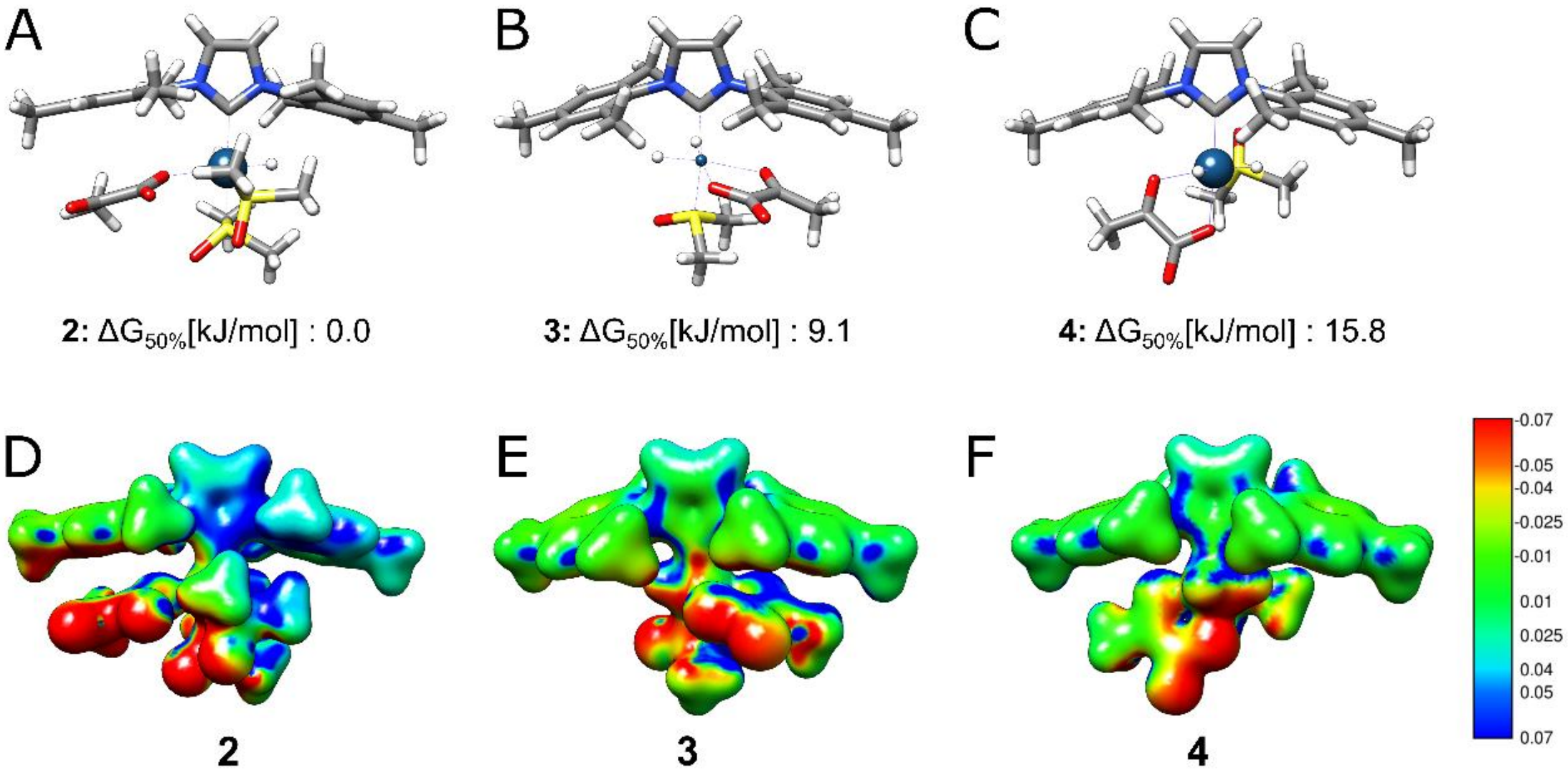

**Fig. 4. Optimized structures of complex 2 (A, D) and 3 (B, E) and 4 (C, F), and corresponding electrostatic potential surfaces (C, D) calculated at the B3LYP-D4/def2-tzvp CPCM(methanol) level of theory**. $\Delta G_{50\%}$ refers to relative Gibbs free energies for which the entropy contribution has been scaled by 0.5 to compensate for deficiencies in the description of the translational entropy correction upon DMSO dissociation.

**Hyperpolarization-enhanced observation of proton exchange:** To validate the equilibrium data, we investigated the kinetics of chemical exchange, focusing on $H_2$ and pyruvate. These measurements allowed us to test the proposed mechanism and quantify the rates of exchange under varying experimental conditions.

To study reaction kinetics, $pH_2$-derived hyperpolarization was used to sensitize the associated measurements according to the PASADENA protocol. As PASADENA yields antiphase signals after a single hard pulse[51], we used the frequency-selective excitation of PASADENA (SEPP)[44,45] to create longitudinal polarization for the selectively excited proton (e.g., $\hat{I}_z^{H^a-(1)}$ for $H^a$ of **1**) to monitor its subsequent chemical exchange. Utilization of $pH_2$ and SEPP allowed us to enhance, for example, the $H^a$ or $H^b$ proton signals of **1** at 265 K by ~8060-fold at 9.4 T ([DMSO]= 75 mM, [Ir] = 4 mM, [pyr] = 40 mM), which dramatically accelerated data collection and proved critical for the low stability of **3** (the samples can degrade after preparation[68]). To improve data quality, after the SEPP pulse element (90°-$\tau_2$-180°-$\tau_2$, **Fig. 5A**), the magnetization associated with the selectively excited proton was rotated into the z-axis, and a pulsed field gradient was applied to suppress nonzero-order quantum coherences arising during the exchange encoding time $\tau_e$ (**Fig. 5B, C**). We refer to this experiment as SEPP-SABRE (**Fig. 5A**).

A simple two-step phase cycle and dephasing gradients were used to improve background signal suppression, with a 180° phase shift of the first pulse and the receiver to suppress signals unaffected by the frequency-selective first pulse (**Fig. S2B**). No additional filters[69] were used to enable the observation of the early exchange, because remaining zero-quantum terms do not contribute to the total integral[37] (**Fig. 5B)**. Relaxation was taken into account in the data fitting process, as the $T_1$'s of the interchanging sites may significantly differ.

Using this technique, the hydride site interchange rate, isomer interconversion rate, and any ligand loss rates were assessed as a function of the concentrations of $H_2$, DMSO, chloride, and pyruvate (**Fig. 5D-K**).

It is worth noting some limitations of this approach. As with all EXSY-type experiments, it is not always possible to distinguish between nuclear Overhauser effect (NOE) cross-relaxation and actual chemical exchange; however, in some cases, NOE can lead to a transfer of polarization with the opposite sign, which is equivalent to a negative flux rate constant. Despite this limitation, the exchange rates obtained from the data fitting are of the correct order of magnitude, since chemical exchange processes are faster than NOE cross-relaxation in most of the cases considered here. Therefore, we did not impose any constraints on the flux rates during the fitting process. We observed a consistently negative flux-rate constant (indicative of NOE-mediated polarization transfer) during measurements of slow intramolecular magnetization exchange between two hydrides in complex **2** (**Fig. 5E, G, I, K**). In all other cases, the flux rates were positive in sign despite an expected but apparently weak negative cross-relaxation contribution.

***Intercomplex* exchange:** When the two hydride protons of **3** were selectively excited at 265 K in different experiments, at DMSO concentrations of less than 20 mM under 8.5 bar of $H_2$, no exchange with **2** was observed. However, when DMSO concentrations were greater than 20 mM, the conversion of **3** to **2** became visible. Moreover, the conversion from **3** to **2** was linearly [DMSO] dependent. This flux $k_{3\to2}$ is therefore approximately equal to [DMSO]×$k_{2+DMSO}$, further confirming the proposed reaction scheme and indicating that we did not reach the saturation of **3** to **2** conversion and [DMSO]×$k_{2+DMSO} \ll k_{3\to5}$ for [DMSO] below 300 mM and T of 265 K (**Fig. 5H, I**).

**Hydrogen site *intramolecular* exchange in 3:** The additional process of hydride site exchange in **3**, flux $k_{3ex}^{HH}$, was readily evident in all of the $^1$H NMR spectra shown in **Fig. 5B**, with directly analogous behavior seen regardless of which hydride ligand was selected, alongside the slightly slower transfer of the magnetization into that of free $H_2$ (integrals shown in **Fig. 5C**). Moreover, the observed $k_{3ex}^{HH}$ flux decreased gradually with increasing [DMSO], which is consistent with possible interception in the κ$^1$ intermediate **5** during $k_{2+DMSO}$ step. Notably, the symmetric conversion observed from an excited hydride (e.g., $H^a$) to both $H^a$ and $H^b$ within the same complex **3** or **2** indicates that the intramolecular hydride exchange is much faster than competing processes, such as DMSO binding ([DMSO]·$k_{5+DMSO}$), or pyruvate rearrangement ($k_{5\to3}$). Furthermore, the dissociation of $H_2$ via **3** occurs much faster than via **2**, and can be accelerated by increasing [$H_2$] (**Fig. 5F, G**) or inhibited by increasing [DMSO] (**Fig. 5H, I**) through competitive trapping of **5.**

**Hydrogen-observed intramolecular exchange in 2:** In contrast to intramolecular exchange in **3**, when the two hydride ligand signals of **2** were excited separately in two separate experiments, no hydride site interchange is evident on the same reaction timescale. Instead, we observed the buildup of negative polarization of the non-excited hydride, indicating a cross-polarization effect within the complex. Accordingly, the fitted conversion rate appears negative (**Fig. 5E, G, I, K**). This shows that polarization is first transferred to the other hydride (with opposite sign) before dissociation to free $H_2$. A similar effect is likely present for complexes **1** and **3**, indicating that the observed exchange rates are underestimating actual exchange rates due to the intramolecular cross-relaxation rate. At the same time, the slow production of $H_2$ and slightly faster conversion of **2** into **3** were evident. Under these conditions, $k_{2\text{-}DMSO}$ controls access to **5** proving $k_{5\to3}$ to be faster than $H_2$ dissociation, $k_{2\text{-}H2}$, and *intracomplex* exchange $k_{2ex}^{HH}$.

**$H_2$ exchange:** The effect of [$H_2$] on these processes was evaluated by changing p$H_2$ pressure over the range of 1 to 8.5 bar. This pressure change, and as a result [$H_2$] change, led to a progressive increase in the degree of visible $H_2$ loss, flux $k_{3\text{-}H2}$ from **3** (**Fig. 5F**), thereby confirming that these exchange processes occur by $H_2$ addition to **5** (i.e. also $S_N2$ mechanism as with [Ir(H)$_2$(IMes)(pyridine)$_3$]Cl-type systems), accessed by κ$^2$-pyr to κ$^1$-pyr rearrangement. The loss of $H_2$ from **2**, flux $k_{2\text{-}H2}$, however, was always much lower than flux $k_{3\text{-}H2}$ (**Fig. 5G**). The dissociation of $H_2$ from **1**, flux $k_{1\text{-}H2}$, however, also slightly increased with [$H_2$] and was an order of magnitude faster than that of **3** (**ESI, Fig. S5.1, Table S5.1)**. When the hydride resonances of **2** were examined, the rates of formation of free $H_2$ and of conversion to **1** were found to decrease as [$H_2$] increased, indicating that suppressing the flux to **1** impacted relayed $H_2$ loss via **1**.

The [DMSO]-dependency data at 8.5 bar of $H_2$ revealed that the **2** to **3** conversion, flux $k_{2\to3}$, was decelerated while the **3** to **2** conversion, flux $k_{3\to2}$, was accelerated when [DMSO] was increased (**Fig. 5H,I**). There were minimal changes seen in the behavior of **1** in this case (**ESI, Fig. S6**). The impact of [pyr] on these changes was also evaluated for concentrations between 40 and 174 mM with 8.5 bar p$H_2$. This led to a slight decrease in the rate of $H_2$ liberation from **2** and **3** (**Fig. 5 J,K**).

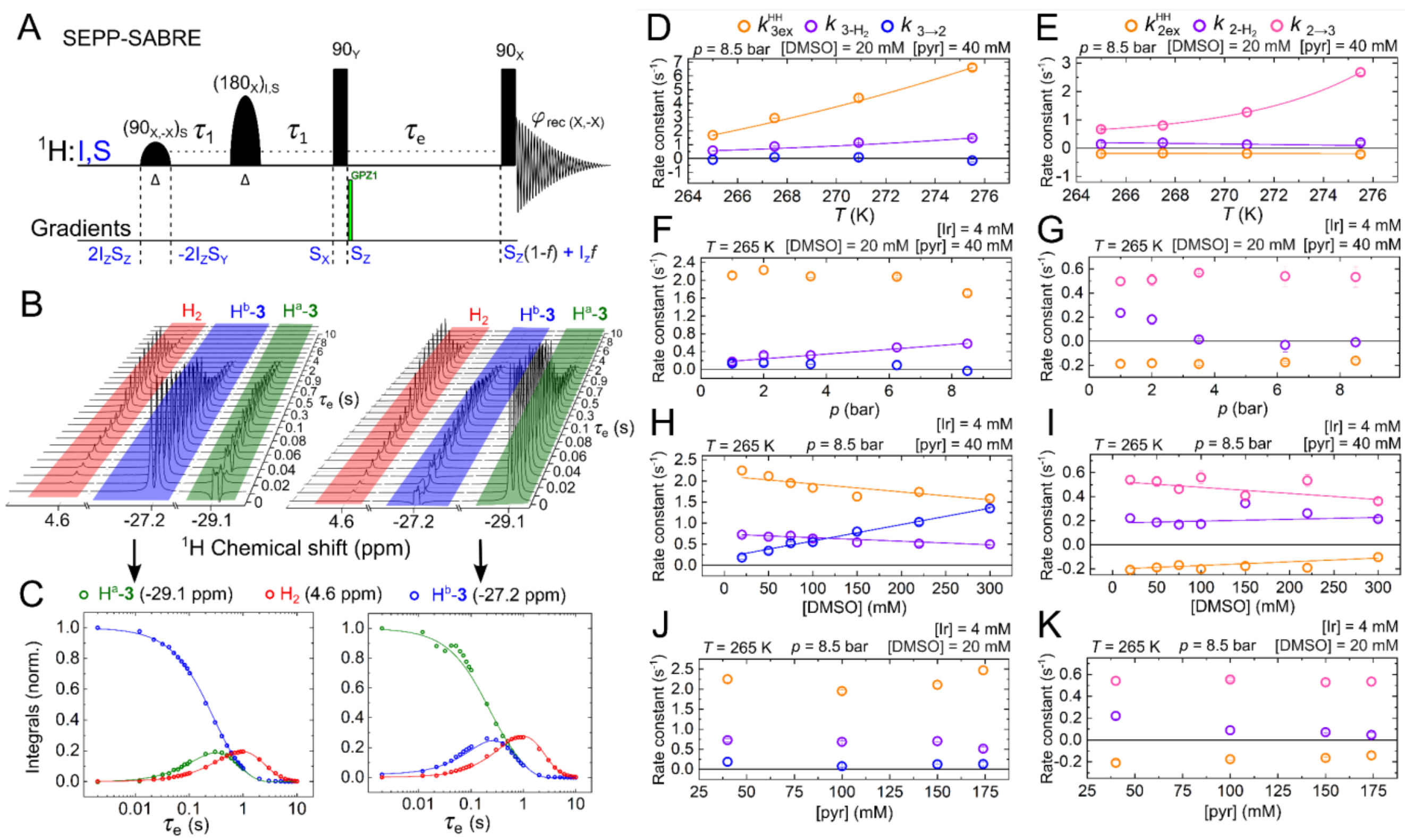

**Fig. 5. Observation of hydride chemical exchange.** Using a SEPP-SABRE pulse sequence (A) enabled us to enhance the signals of hydride protons selectively (B, left: H$^b$-**3** and right: H$^a$-**3**) and observe the intra- and intermolecular $H_2$ exchange (C) and extract exchange rates of **3** and **2** as a function of temperature (D, E), $pH_2$ pressure (F, G), [DMSO] (H, I), and [pyr] (J, K) using (**3**↔**2**)↔$H_2$ model as described in SI, S6-8. Note the increase of $k_{3\to2}$ as a function of [DMSO] and large $k^{HH}_{3ex}$ compared to other rates. The round pulses were frequency-selective (a Gaussian shape was used) with different amplitudes, and the exact duration $\Delta$ = 10 ms. Additional indices for selective pulses indicate the excited nuclear spins. Rectangle pulses indicate the hard pulses. Several delays were applied: $\tau_1 \cong 1/4J_{IS}$ = 10 ms to achieve the refocusing of the spins after the selective pulse; $\tau_e$ stands for the time of longitudinal magnetization evolution, which was varied to measure the exchange. During $\tau_e$, a fraction $f$ of longitudinal magnetization of the spins S chemically (or through cross-relaxation) exchanges with spin-*I*. We repeated the experiment twice, changing the phase of the first selective RF pulse, $\varphi_1$ = [X, -X], and the phase of the receiver $\phi_{rec}$ = [X, -X], which suppressed the background signal of spins not affected directly by the first selective pulse or indirectly via chemical exchange with this spin. A gradient pulse was applied for 2 ms with SMSQ10.100 shape at 31% amplitude.

**Hyperpolarization-enhanced tracking of pyruvate exchange:** To investigate pyruvate exchange, we selectively hyperpolarized the 2-$^{13}$C nucleus of bound pyruvate in **2** or **3** and followed the label over time. To do so, we transferred the polarization first to one of the two hydride protons H$^a$ or H$^b$ (using SEPP), and then to the 2-$^{13}$C of bound pyruvate (using selective INEPT, SPINEPT, **Fig. 6A**)[45]. By adding two more 90° pulses and a variable evolution time $\tau_e$, we measured the exchange of pyruvate using the resulting SEPP-SPINEPTplus-SABRE sequence as a function of [DMSO], *T*, [pyr], and hydrogen pressure for complexes **2** and **3** (**Fig. 6**). The kinetics (exemplary **Fig. 6B, C**) were analyzed using a global fit with the model ($^{13}$C$^2$-**3**↔$^{13}$C$^2$-**2**)↔$^{13}$C$^2$-free (**ESI**), with the primary parameters plotted in **Fig. 6:** the dissociation flux rate $k_{2\text{-pyr}}$, and intramolecular exchange flux rates $k_{3\to2}$, and $k_{2\to3}$ were defined in **Fig. 2**. Previously, a somewhat similar approach was proposed to transfer $pH_2$ spin order to $^1$H nuclei of ligands[35], which we cannot employ due to the too narrow chemical shift dispersion of $^1$H pyruvate signals and rapid relaxation of protons.

**Pyruvate-observed *inter*complex exchange:** For the low [DMSO] of 20 mM, the flux $k_{3\to2}$ was 6-8 times lower than that of flux $k_{2\to3}$. At high [DMSO] (hundreds of mM), $k_{3\to2} \approx k_{2\to3}$ (**Fig. 6H,I**). Both fluxes proved to be temperature-dependent, as expected. Interestingly, flux $k_{3\to2}$ grew proportionally to [DMSO], while flux $k_{2\to3}$ decayed slowly because high [DMSO] prevented pyruvate from binding back to form **3**. Hence, increased [DMSO] proved to inhibit $2 \rightarrow 3$ conversion as expected through the [DMSO] dependent $k_{5+DMSO}$ step. This aligned with the observations made by SEPP-SABRE applied to hydrides and supported the identity of **2**: $k_{2\rightarrow3}$ and $k_{3\rightarrow2}$ flux values measured by probing hydride were within error of those observed here by probing pyruvate.

The increased temperature promoted complex reactivity and hence pyruvate unhooking dissociation in **3**, or DMSO loss from **2**, thereby facilitating *intra*-complex exchange. The concentrations of [$H_2$] and [pyr] did not significantly influence the flux of these processes: the slight decline in $k_{3\rightarrow2}$ flux as a function of increased [pyr] aligned with the observed stabilization of **3** relative to **1** and **2** with an excess of pyr (**Fig. 3E**).

**Pyruvate dissociation rate:** The observed pyruvate dissociation flux at studied temperatures was always slower than the flux of *inter*complex exchange between **2** and **3** mediated by **5**. For example, at [Ir] = 4 mM, [DMSO] = 20 mM, [pyr] = 40 mM at 265 K, the fluxes $k_{2-pyr}$ and $k_{3-pyr}$ were (0.042 ± 0.004) $s^{-1}$, and (0.016 ± 0.001) $s^{-1}$, respectively, whereas the interconversion fluxes were $k_{3\rightarrow2}$ = (0.121 ± 0.006) $s^{-1}$ and $k_{2\rightarrow3}$= (0.630 ± 0.007) $s^{-1}$.

Pyruvate exchange can go via dissociative exchange and complex **6**, or via associative exchange and complex **5**. When it goes through **5**, the complex [Ir(H)$_2$(IMes)($\kappa^1$-pyr)$_2$(DMSO)] is generated as a subsequent step; thus, the observed exchange should be [pyr]-dependent. This is in accordance with the observation of the acceleration of $k_{3-pyr}$ with the increase of [pyr] (**ESI, Fig. S17G**). Hence, the addition of pyr could accelerate pyr exchange. At the same time, $k_{2-pyr}$ is less impacted by [pyr] (**ESI, Fig. S17H**) than $k_{3-pyr}$, indicating that for **2**, the dissociative exchange has a greater contribution than the associative exchange at 265 K and [pyr] up to 175 mM.

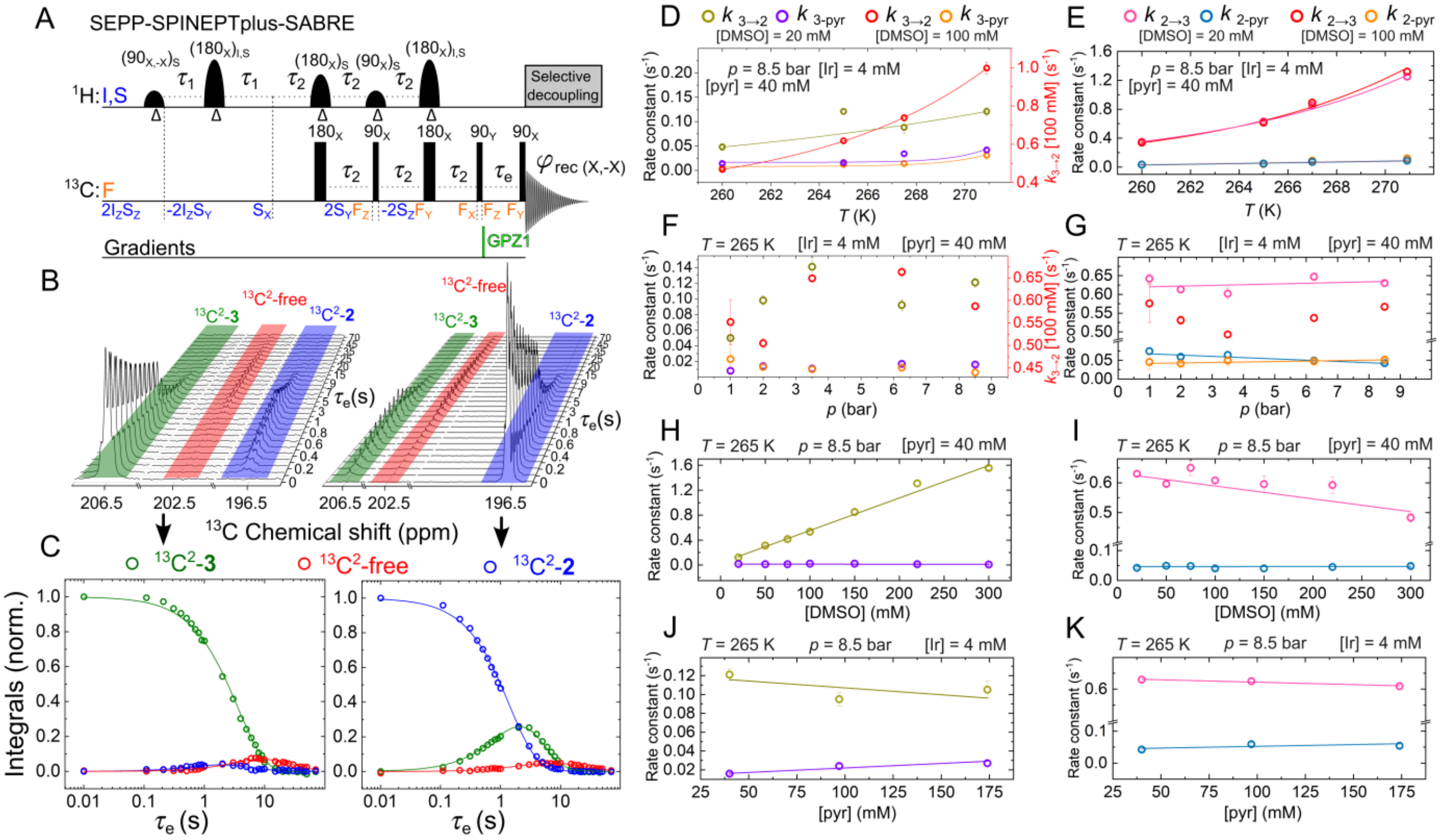

**Fig. 6. Observation of pyruvate chemical exchange.** Scheme of SEPP-SPINEPTplus-SABRE spin order transfer sequence (A), corresponding $^{13}C$ NMR spectra at 265 K, where selective $^{13}C^2$ carbon of **3** (B, left) and **2** (B, right), corresponding integral kinetics (C), and extracted from such kinetics pyruvate dissociation rates $k_{3-pyr}$, and $k_{2-pyr}$, exchange flux between complexes $k_{3\rightarrow2}$, and $k_{2\rightarrow3}$as a function of temperature (D, E), p$H_2$ pressure (F, G), [DMSO] (H, I), and [pyr] (J, K) by performing a global fit using a model ($^{13}C^2$-**3**↔$^{13}C^2$-**2)**↔$^{13}C^2$-free described in SI (SI, S10,11). The round pulses were frequency-selective (a Gaussian shape was used) with different amplitudes, and the exact duration $\Delta$ = 10 ms. Additional indices for selective

pulses indicate which nuclear spins are excited. Rectangle pulses indicate the hard pulses. Several delays were applied: $\tau_1 \cong 1/4\,J_{IS}$ = 10 ms to achieve the refocusing of the spins after the selective pulse, $\tau_2 \cong 1/4\,J_{SF}$ = 70 ms in **3**[43], and 80 ms in **2**[43], $\tau_e$ stands for the time of longitudinal magnetization evolution, which was varied to measure the exchange. We repeated the experiment twice, changing the phase of the first selective RF pulse, $\varphi_1$ = [X, -X], and the phase of the receiver $\phi_{rec}$ = [X, -X], which suppressed the background signal of spins not affected directly by the first selective pulse or via chemical or indirectly via chemical exchange with this spin. A gradient pulse was applied for 2 ms with an SMSQ10.100 shape at 31% amplitude.

## CONCLUSION

**Rationalization of SABRE: implications of $H_2$ and pyruvate exchange:** A comprehensive mechanism for the complex transformations during SABRE was developed (**Fig. 2**), based on both experimental observation and DFT calculations. While complex **3** remains critical to this process, it is notable that the identity of the other critical complex **2** differs from the previously suggested complex **4**.[22,27,43,47–49] Electronic structure calculations indicate a strong influence of $Na^+$ on the relative stability of these complexes, which might play an important role in the observed populations of pyruvate SABRE complexes.

Exchange within a complex is detrimental to SABRE polarization yield. While **3** was considered optimally suited for polarization transfer due to the associated *J*-coupling network,[43] and was assumed to be an energetically favorable state, the loss of the oxo-$\kappa^2$-bonding interaction, which occurs during the intramolecular hydride exchange, disrupts the critical spin-spin coupling network, thereby hindering polarization transfer. Moreover, the rapid exchange of the hydride ligands averages different *J*-coupling constants. In the extreme, the two become equal; the $pH_2$ spin order cannot be converted into polarization of the target nucleus (**ESI**, **Section 16**, **Fig. S22**).

The need for stable complex **2**, which can be accessed via complex **3** or **1**, for optimal SABRE performance is therefore questionable. Our results suggest that both **1** and **2** negatively affect polarization transfer. Consequently, stabilizing **3**, while maintaining sufficient $H_2$ exchange, appears promising for boosting pyruvate hyperpolarization. This finding suggests that designing a ligand sphere that favors complex **3**, and suppresses complexes **1** and **2,** could maximize the concentration of the active catalyst, thereby improving the target polarization. Alternatively, populating only **2** could also be potentially a remedy.

Similarly, reducing the intramolecular hydrogen exchange would also increase polarization. In fact, reduced intramolecular hydride exchange may be the reason why the highest polarization levels of SABRE were achieved at low temperatures.[19,26] In addition, higher temperatures are required for pyruvate dissociation, in accordance with the temperature-cycling process used to deliver polarized free material.[26]

These measurements have also revealed that the **2**–**3** interconversion process also facilitates indirect $H_2$ loss. This situation arises because $H_2$ dissociates from **3** much faster than from **2**, and this indirect pathway contributes to the experimentally observed $H_2$ loss when **2** was probed. Furthermore, the fastest observed process involved hydride site exchange in **3**, thereby confirming that the prerequisite $k_{5\rightarrow3}$ is the fastest individual step. As stated, this process leads to bond rupture and, therefore, interruption of the spin-spin pathway associated with $^{13}C$-SABRE transfer. Thus, while rapid $H_2$ exchange will be limiting SABRE, *intra*molecular hydride site exchange must also be considered when evaluating the potential activity of a catalyst. However, the slow process of pyruvate exchange must also be optimized. At -8 °C, there is no significant pyruvate exchange[26,70], as shown by the fact that the observed flux $k_{3-pyr}$ is (0.005 ± .004) $s^{-1}$ and $k_{2-pyr}$ = (0.031 ± 0.005) $s^{-1}$.

Our study indicates that the current system is not yet suitable for polarizing at high pyruvate concentrations. Moreover, the refined structure for **2**, when considered alongside previous experimental observations[43], raises the question of why the four-bond $^4J_{CH}$ is greater than the three-bond $^3J_{CH}$. The accuracy of DFT-calculated *J*-coupling constants to hydrides is low, making it difficult to use this metric to explain this trend.[43,47] Future studies should therefore consider matching DFT-derived reaction paths to reaction flux to optimize outcomes. The potentially significant influence of the cation counterion also suggests that solvation will be a key parameter in improving the SABRE outcome.

## METHODS

**Chemicals.** Perdeuterated Ir precatalyst [Ir-$d_{22}$] = [IrCl(COD)(IMes-$d_{22}$)] was synthesized according to Ref.[30] (IMes = 1,3-bis(2,4,6-trimethylphenyl)imidazol-2-ylidene, COD = 1,5-cyclooctadiene), sodium pyruvate-2-$^{13}$C ($^{13}$C-pyr, 490725, Sigma-Aldrich), dimethylsulfoxide-$d_6$ (DMSO, 00905-25, Deutero GmbH), methanol-$d_4$ (441384, Sigma-Aldrich), were used here.

**Sample preparation.** The samples were prepared by mixing 40 mM of $^{13}$C-pyr with 4 mM of [Ir-$d_{22}$] and 20 mM of DMSO in 0.6 mL of methanol-$d_4$ unless otherwise stated.

**Parahydrogen enrichment.** $pH_2$-enriched hydrogen gas with a 93% fraction of $pH_2$ was prepared by passing high-purity hydrogen over hydrated iron(III) oxide at 25 K using a $pH_2$ generator similar to the one in Ref.[71].

**SABRE experiments.** All hyperpolarization experiments were carried out on a Bruker 400 MHz WB Avance NEO spectrometer using a 5 mm BBFO probe. The NMR tube was then attached to a bubbling system (similar to the one used in Ref.[72]) and placed inside the spectrometer. The solution was bubbled with parahydrogen-enriched $H_2$ at various specified pressures, e.g., 8.5 bar. First, the sample was bubbled with $pH_2$ for 30 minutes to activate the catalyst. One experiment consisted of a 5-second relaxation delay followed by 2 seconds of WALTZ-16 on the $^1$H channel, then 10 seconds of $pH_2$ bubbling through the sample, followed by a 1.5-second delay for the bubbles to dissipate, and finally, the SOT sequence was applied. Two SOT sequences were used: SEPP-SABRE (**Fig. 5A**) and SEPP-SPINEPTplus-SABRE (**Fig. 6A**).

**Acquisition of thermally polarized NMR spectra.** To acquire reference spectra for signal enhancement estimation of the $^1$H NMR spectra at 9.4 T, we prepared a sample containing 4 mM of [Ir-$d_{22}$], 40 mM of $^{13}C^2$-pyr, and 20 mM of DMSO in 500 µL of methanol-$d_3$. We performed a zgesgp $^1$H NMR experiment that suppresses the water peak 6.58 ppm due to using a $D_2O$ lock, which shifts the water signal. We used 100 scans with a 90° flip angle, with RG = 101 using TR = 9s.

For the $^{13}$C NMR thermally polarized NMR spectra, we used a 40 mM sample of $^{13}$C-pyr containing 4 mM of [Ir-$d_{22}$] and 20 mM of DMSO in 0.6 mL of methanol-$d_4$. We performed 600 acquisitions after a 90° flip angle at 9.4 T using TR = 130 s.

**Data processing and fitting.** All NMR spectra were analyzed using the spectral data analyzing software Bruker TopSpin (4.1.4), MestReNova (14.2.2), and Origin (2021). All the kinetics profiles were analyzed using MATLAB scripts (available as supporting materials). All error margins for the fitted values are standard deviations estimated using the MATLAB nonlinear regression "*nlinfit*" function.

**Thermally polarized exchange spectroscopy:** Kinetic measurements using thermal polarization were performed on an Avance III Bruker 600 MHz spectrometer equipped with a variable temperature unit. A standard EXSY NMR experiment (selnogpzs.2 Bruker experiment) was used in the measurements. The following mixing times were used in the experiments (s): 0.010, 0.020, 0.050, 0.075, 0.100, 0.150, 0.200, 0.300, 0.400, 0.500, 0.600, 0.800, 1.200, 1.600, 2.400, 3.200. The number of scans was 256. The measured EXSY curves, as well as the curves obtained from the data fitting procedures, are shown in the **Fig S8, Section 7.5, ESI**. For the details of the fitting procedure, see **Section 1.1, ESI**.

## ASSOCIATED CONTENT

### Supporting Information

The authors have cited additional references within the Supporting Information.[73–84] The supporting information contains a detailed method section, a description of kinetic fitting models, all SEPP-SABRE and SEPP-SPINEPTplus-SABRE kinetics of IrHH hydrogens, and $^{13}C^2$ of 1, 2 and 3, a calibration of the variable temperature NMR unit, signal enhancement and polarization calculations, $T_1$ measurements for $H_2$ and free-$^{13}C^2$ as a function of temperature, and additional computational analysis of SABRE. Raw data, simulation Matlab scripts, and DFT calculated geometries of complexes and dissociation geometries can be accessed via Zenodo DOI: https://doi.org/10.5281/zenodo.18449765.

## AUTHOR INFORMATION

### Corresponding Authors

Vladimir V. Zhivonitko, ORCID 0000-0003-2919-8690;
Email: valdimir.zhivonitko@oulu.fi

Andrey N. Pravdivtsev, ORCID 0000-0002-8763-617X;
Email: andrey.pravdivtsev@rad.uni-kiel.de

**Authors**

Charbel D. Assaf, ORCID 0000-0003-1968-2112;
Amaia Vicario, ORCID 0009-0006-7075-1952;
Alexander. A. Auer, ORCID 0000-0001-6012-30027;
Simon B. Duckett, ORCID 0000-0002-9788-6615;
Jan-Bernd Hövener, ORCID 0000-0001-7255-7252;

**Author Contributions**

[†]C.D.A., V.V.Z. contributed equally. A.N.P., V.V.Z., C.D.A.: conceptualization, C.D.A., A.V., V.V.Z., S.B.D.: experiments, A.N.P.: spin dynamics simulations, A.N.P., C.D.A., V.V.Z., S.B.D.: investigation, A.N.P., C.D.A.: development of exchange models, A.A.A. all quantum chemistry simulations, A.N.P, V.V.Z, and J.B.H.: supervision, funding acquisition. All authors contributed to discussions and interpretation of the results, writing the original draft, and have approved the final version of the manuscript.

**Notes**

None of the authors has any conflicts of interest to declare.

## ACKNOWLEDGMENT

A.N.P., C.D.A., J.B.H. acknowledge funding from the German Federal Ministry of Education and Research (BMBF, 03WIR6208A hyperquant), Heisenberg DFG grant (565789098), DFG (562203308, 555951950, 527469039, HO-4602/2-2, HO-4602/3, HO-4604/6-1, HO-4604/8-1, Inst 257/747-1, EXC2167/2, FOR5042, TRR287). MOIN CC was founded by a grant from the European Regional Development Fund (ERDF) and the Zukunftsprogramm Wirtschaft of Schleswig-Holstein (Project no. 122-09-053). S.B.D is grateful to the UK Research and Innovation (UKRI), under the UK government's Horizon Europe funding guarantee [grant number EP/X023672/1], for funding. V.V.Z. is grateful for the support from the Research Council of Finland (grant number 362959) and the University of Oulu (Kvantum Institute). A.A.A. acknowledges support by the Max-Planck Society and the Max-Planck Institut für Kohlenforschung. A.V. was participating in this research during his internship in Kiel [a] in 2024, supported by the Deutsche Akademische Austauschdienst DAAD-RISE program.

**Keywords:** Hyperpolarization, SABRE, intracomplex exchange, dissociation rate, parahydrogen.

**Figure 1:**

# Figure 2:

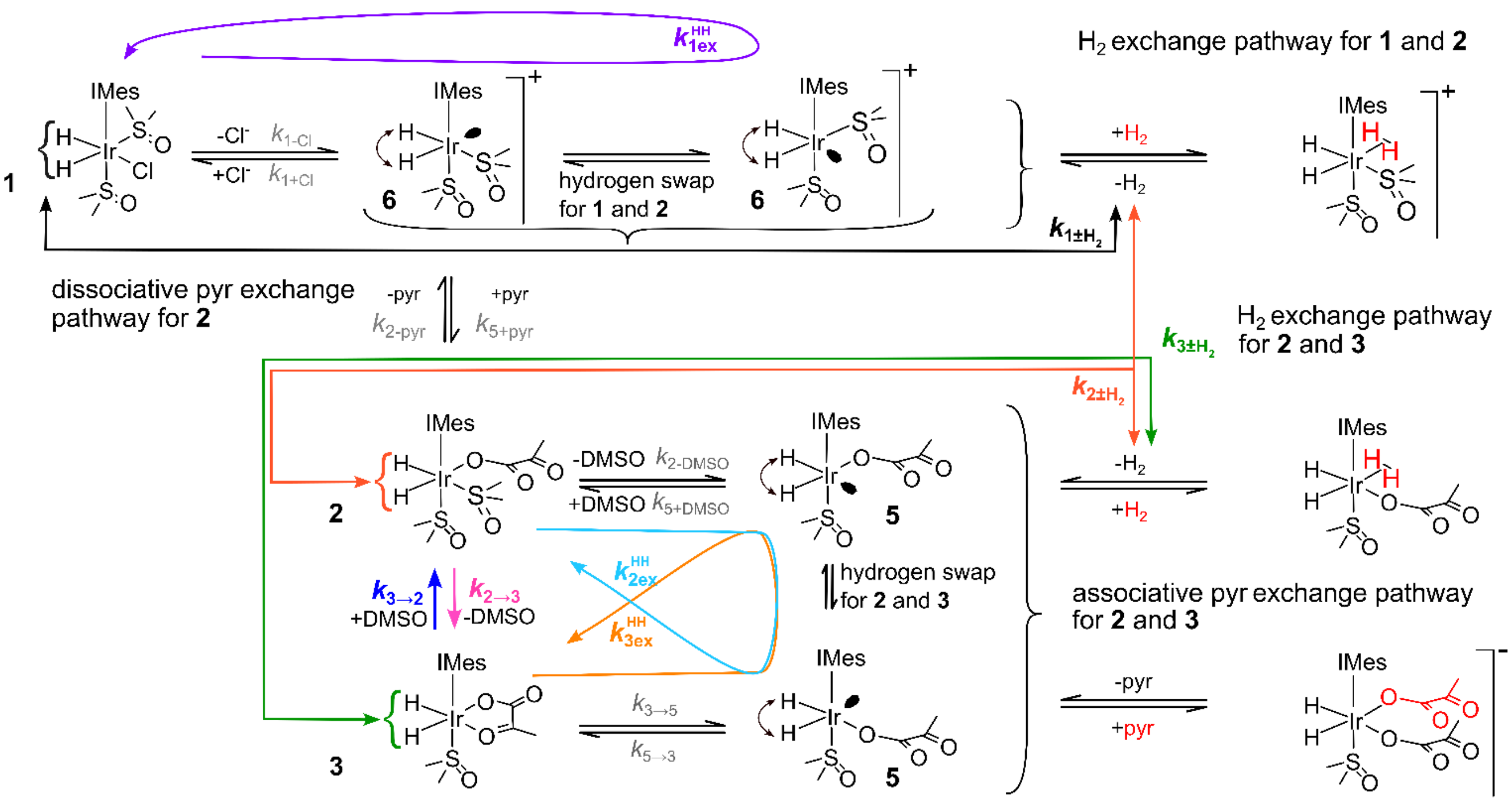

# Figure 3:

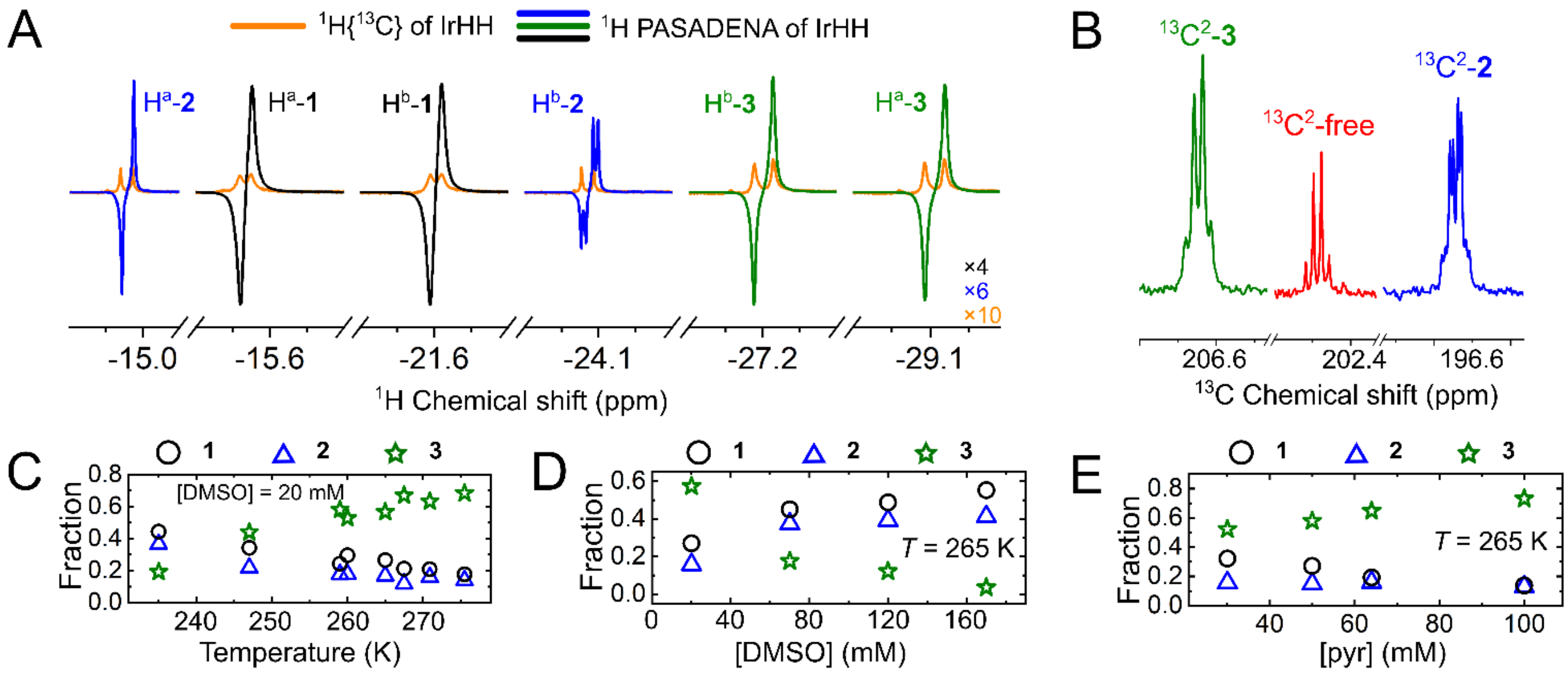

# Figure 4:

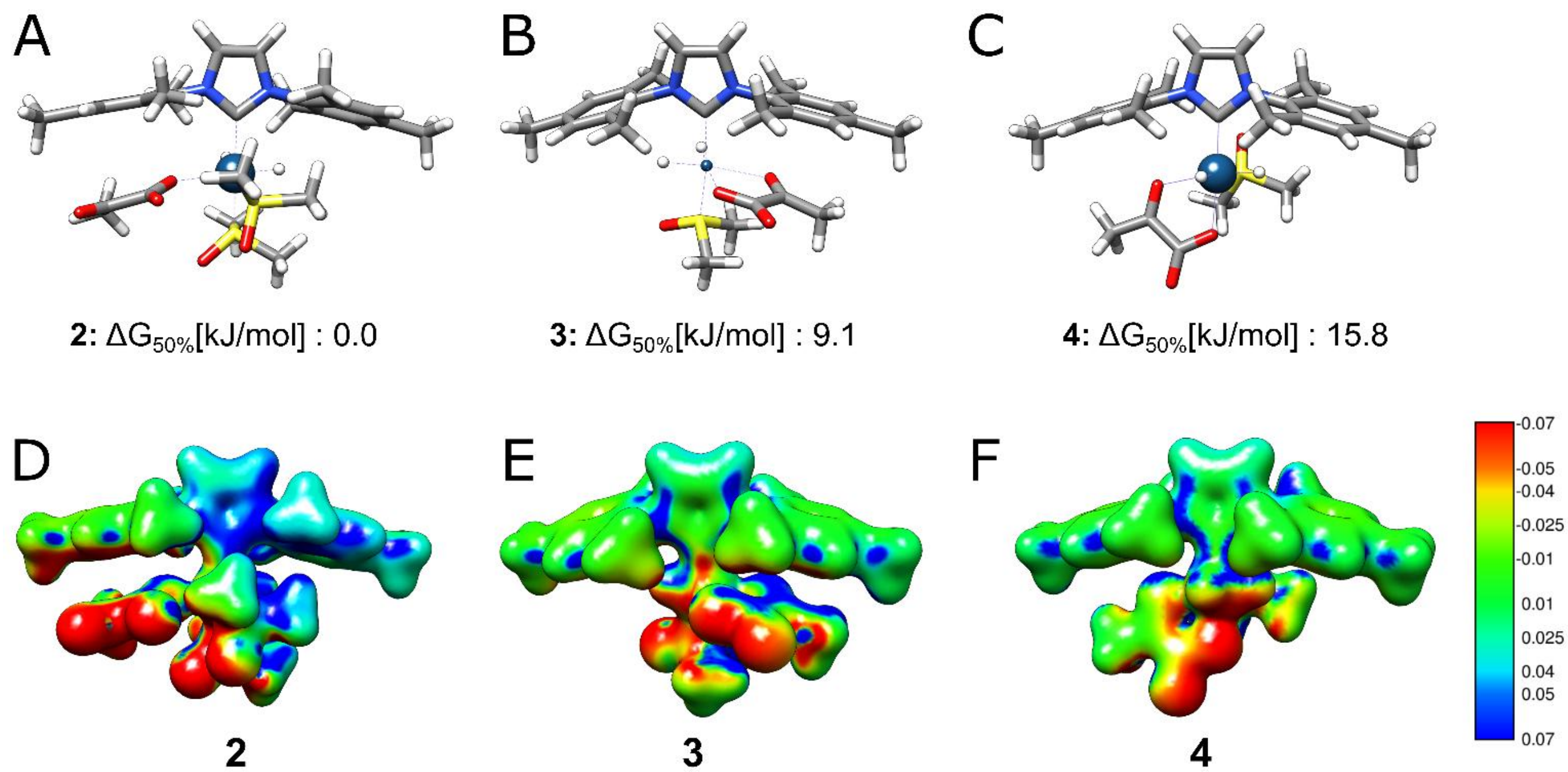

# Figure 5:

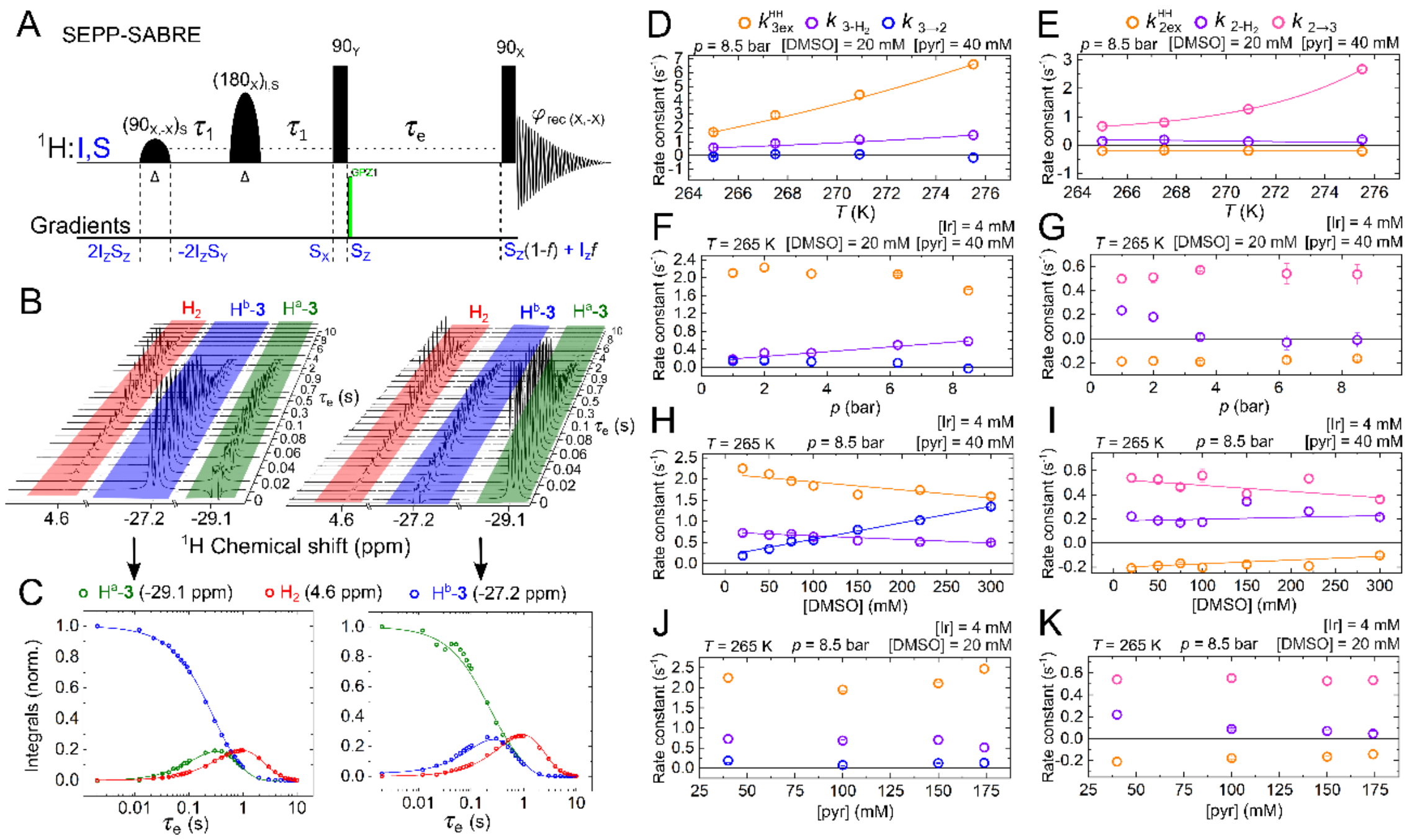

# Figure 6:

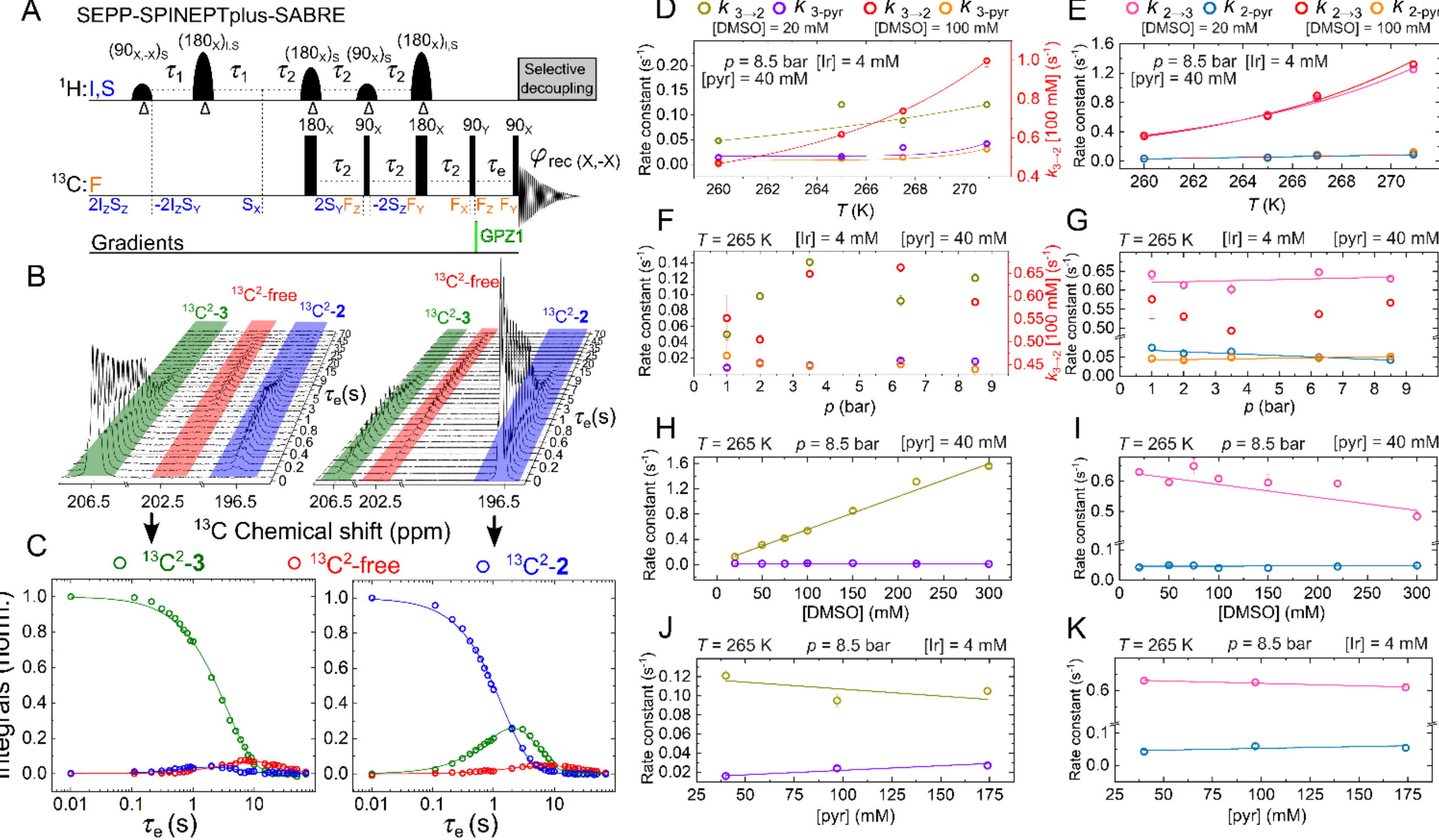

# Graphical Table of content:

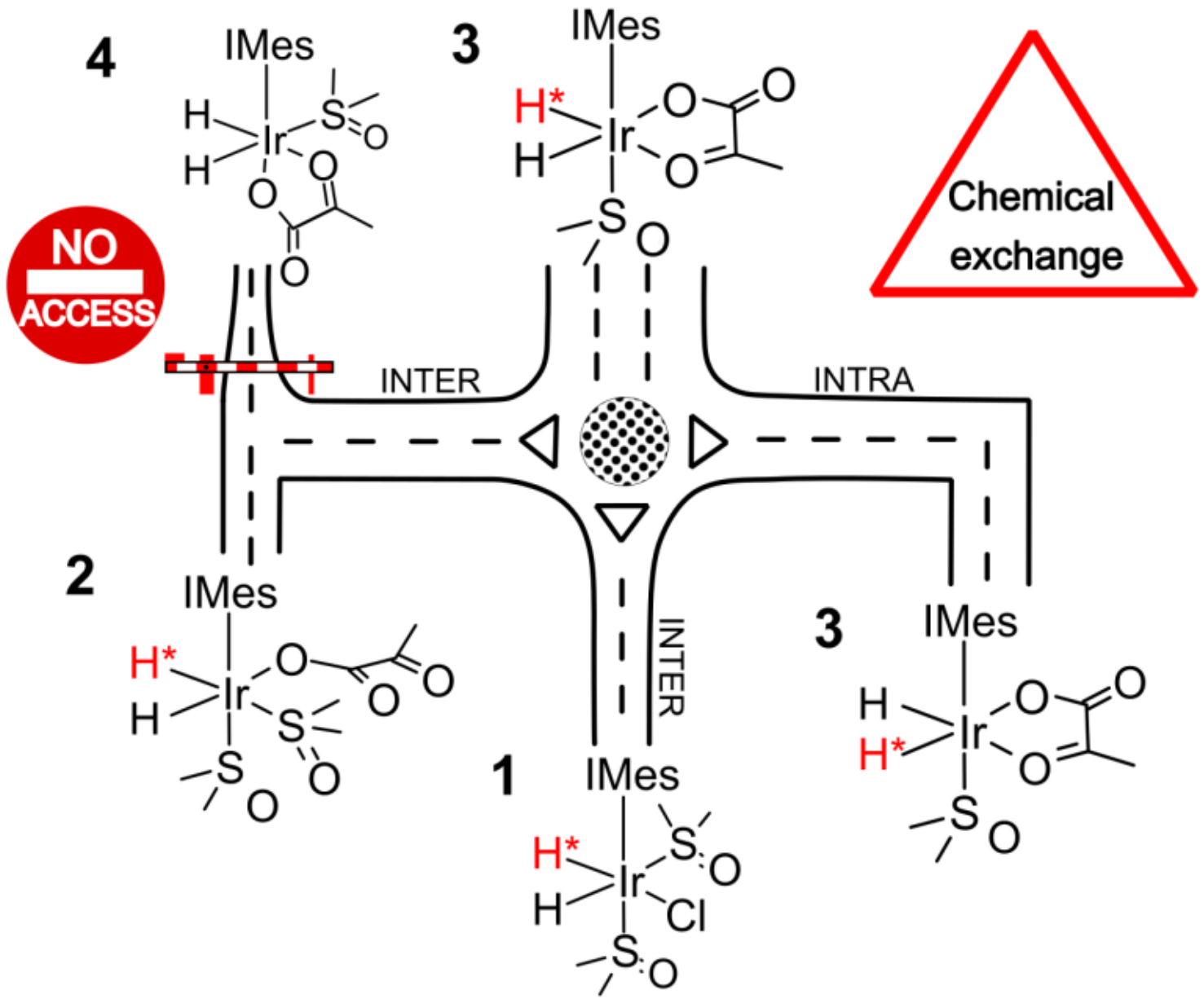

# Mechanistic Insights into Chemical Exchange during the Signal Amplification by Reversible Exchange Sensitization of Pyruvate

Charbel D. Assaf†[a], Vladimir V. Zhivonitko*†[b], Amaia Vicario[a,c], Alexander A. Auer[d], Simon B. Duckett[e], Jan-Bernd Hövener[a] and Andrey N. Pravdivtsev*[a]

†equal contribution
*corresponding authors

[a] Section Biomedical Imaging (SBMI), Molecular Imaging North Competence Center (MOINCC), Department of Radiology and Neuroradiology, University Hospital Schleswig-Holstein, Kiel University, *Am Botanischen Garten 14/18*, 24118 Kiel, Germany. E-mail: andrey.pravdivtsev@rad.uni-kiel.de

[b] NMR Research Unit, Faculty of Science, University of Oulu, P.O. Box 3000, Oulu, Finland. E-mail: vladimir.Zhivonitko@oulu.fi

[c] University of Bath, Claverton Down, Bath, BA2 7AY, UK

[d] Max-Planck-Institut für Kohlenforschung, Kaiser-Wilhelm-Platz 1, 45470 Mülheim an der Ruhr.

[e] Centre for Hyperpolarization in Magnetic Resonance (CHyM), Department of Chemistry, University of York, Heslington YO10 5NY, UK

# 1. Models of hydrogen exchange during SABRE

This section describes the kinetic models used to model ligand exchange, as described in the main text.

The first model, **IrHH↔H₂**, considers only exchange with one complex with free $H_2$ hydrogen.

The second model, **(3↔2)↔H₂,** considers how exchange between complexes affects H2 exchange by reference to complexes **2** and **3**.

## 1.1. Model IrHH↔H₂

**Chemical exchange.** The iridium complexes discussed feature two inequivalent hydride protons represented by $\mathrm{IrHH}$. These two hydrogens can exchange within the complex, at an exchange rate $k_{\mathrm{IrH'H \to IrHH'}}$, and dissociate from the complex at an exchange rate $k_{\mathrm{IrHH-H2}}$, both processed forming an intermediate Ir-complex. Finally, this intermediate can react with free $\mathrm{H_2}$ in the sample according to the pseudo-first-order rate constant $k_{\mathrm{a}}'^{\mathrm{H2}}$(valid if [$H_2$] is high and therefore invariant). Such chemical system can be represented by the following chemical equations:

$$\begin{aligned} \mathrm{H_2} &\xrightarrow{k_{\mathrm{a}}'^{\mathrm{H2}}} \mathrm{IrHH} \\ \mathrm{IrH'H} &\xrightarrow{k_{\mathrm{IrH'H \to IrHH'}}} \mathrm{IrHH'} \\ \mathrm{IrHH} &\xrightarrow{k_{\mathrm{IrHH-H2}}} \mathrm{H_2} \end{aligned} \qquad (Eq.\ S1)$$

The apostrophe is used to highlight the exchange of the two chemically inequivalent hydrogens. Based on preserving molecularity, we consider that both the hydrides from the same complex dissociate at the same rate. In this form $1/k_{\mathrm{IrHH-H2}}$ is therefore the lifetime of the active Ir complex ($\mathrm{IrHH}$) and $1/k_{\mathrm{a}}'^{\mathrm{H2}}$ is a characteristic time of $\mathrm{H_2}$ exchange. The chemical kinetics are given by the following kinetic equation:

$$\frac{d[\mathrm{H_2}]}{dt} = -\frac{d[\mathrm{IrHH}]}{dt} = +k_{\mathrm{IrHH-H2}}[\mathrm{IrHH}] - k_{\mathrm{a}}'^{\mathrm{H2}}[\mathrm{H_2}] \qquad (Eq.\ S2)$$

Based on this equation, the steady-state equilibrium ratio of $[\mathrm{IrHH}]$, and $[\mathrm{H_2}]$ can be found:

$$\frac{k_{\mathrm{a}}'^{\mathrm{H2}}}{k_{\mathrm{IrHH-H2}}} = \frac{[\mathrm{IrHH}]}{[\mathrm{H_2}]} \qquad (Eq.\ S3)$$

where [Ir] and [$H_2$] are the concentrations of the Ir precatalyst and natural hydrogen $H_2$.

**Magnetization exchange (Inter and intramolecular exchange).** By using Eq. S2 it is possible to derive how the evolution of magnetization, driven by relaxation and chemical exchange proceeds. The following approach is based on the Bloch-McConnell equation (73McConnellJCP). We define magnetization, for simplicity, as the product of polarization and concentration, which is therefore to a constant value, that does not affect to output of the calculations. The magnetization values for the constituents are therefore:

$$\begin{cases} M_{\mathrm{H_2}} = 2P_{\mathrm{H_2}}[\mathrm{H_2}] \\ M_{\mathrm{IrH^a}} = P_{\mathrm{IrH^a}}[\mathrm{IrHH}] \\ M_{\mathrm{IrH^b}} = P_{\mathrm{IrH^b}}[\mathrm{IrHH}] \\ M_{\mathrm{IrHH}} = M_{\mathrm{IrH^a}} + M_{\mathrm{IrH^b}} = \left(P_{\mathrm{IrH^a}} + P_{\mathrm{IrH^b}}\right)[\mathrm{IrHH}] \end{cases} \qquad (Eq.\ S4)$$

$P_{\mathrm{H_2}}$ is the average polarization of two protons. The corresponding thermal equilibrium magnetization terms are $M^0_{\mathrm{IrH^a}} = P^0[\mathrm{IrHH}]$, $M^0_{\mathrm{IrH^b}} = P^0[\mathrm{IrHH}]$, $M^0_{\mathrm{H_2}} = 2P^0[\mathrm{H_2}]$ or $M^0_{\mathrm{IrHH}} = M^0_{\mathrm{IrH^a}} + M^0_{\mathrm{IrH^b}}$,where $P^0$ is the thermal polarization of a $^1$H nucleus.

The corresponding kinetic equation for magnetization within this sytem for $\mathrm{IrHH}$, $\mathrm{IrH^a}$, $\mathrm{IrH^b}$, and $\mathrm{H_2}$ is therefore as follows:

$$\begin{cases} \dfrac{dM_{\mathrm{H_2}}}{dt} = -k_{\mathrm{a}}'^{\mathrm{H2}} M_{\mathrm{H_2}} + k_{\mathrm{IrHH-H2}}\left(M_{\mathrm{IrH^a}} + M_{\mathrm{IrH^b}}\right) - \left(M_{\mathrm{H_2}} - M^0_{\mathrm{H_2}}\right)R_{\mathrm{H_2}} \\ \dfrac{dM_{\mathrm{IrHH}}}{dt} = +k_{\mathrm{a}}'^{\mathrm{H2}} M_{\mathrm{H_2}} - k_{\mathrm{IrHH-H2}} M_{\mathrm{IrHH}} - \left(M_{\mathrm{IrHH}} - M^0_{\mathrm{IrHH}}\right)R_{\mathrm{IrHH}} \\ \dfrac{dM_{\mathrm{IrH^a}}}{dt} = +\dfrac{k_{\mathrm{a}}'^{\mathrm{H2}}}{2} M_{\mathrm{H_2}} - k_{\mathrm{IrHH-H2}} M_{\mathrm{IrH^a}} - k_{\mathrm{IrH'H \rightarrow IrHH'}}\left(M_{\mathrm{IrH^a}} - M_{\mathrm{IrH^b}}\right) - \left(M_{\mathrm{IrH^a}} - M^0_{\mathrm{IrH^a}}\right)R_{\mathrm{IrH^a}} \\ \dfrac{dM_{\mathrm{IrH^b}}}{dt} = +\dfrac{k_{\mathrm{a}}'^{\mathrm{H2}}}{2} M_{\mathrm{H_2}} - k_{\mathrm{IrHH-H2}} M_{\mathrm{IrH^b}} + k_{\mathrm{IrH'H \rightarrow IrHH'}}\left(M_{\mathrm{IrH^a}} - M_{\mathrm{IrH^b}}\right) - \left(M_{\mathrm{IrH^b}} - M^0_{\mathrm{IrH^b}}\right)R_{\mathrm{IrH^b}} \end{cases}$$

(*Eq. S5*)

An internal check is provided by the fact $\frac{dM_{\mathrm{IrHH}}}{dt} = \frac{dM_{\mathrm{IrH^a}}}{dt} + \frac{dM_{\mathrm{IrH^b}}}{dt}$. Here we assumed $R_{\mathrm{IrH^a}} \cong R_{\mathrm{IrH^b}} \cong \frac{R_{\mathrm{IrH^a}} + R_{\mathrm{IrH^b}}}{2} = R_{\mathrm{IrHH}}$, an assumption that provides a high order of accuracy. The three independent components, can be written in a matrix form as:

$$\frac{d}{dt}\begin{pmatrix} M_{\mathrm{IrH^a}} \\ M_{\mathrm{IrH^b}} \\ M_{\mathrm{H_2}} \end{pmatrix} = \hat{L}\begin{pmatrix} M_{\mathrm{IrH^a}} \\ M_{\mathrm{IrH^b}} \\ M_{\mathrm{H_2}} \end{pmatrix} + \begin{pmatrix} R_{\mathrm{IrHH}} M^0_{\mathrm{IrH^a}} \\ R_{\mathrm{IrHH}} M^0_{\mathrm{IrH^b}} \\ R_{\mathrm{H_2}} M^0_{\mathrm{H_2}} \end{pmatrix}$$

$$\hat{L} = \begin{pmatrix} -(R_{\mathrm{IrHH}} + k_{\mathrm{IrHH-H2}} + k_{\mathrm{ex}}) & k_{\mathrm{IrH'H \rightarrow IrHH'}} & \dfrac{k_{\mathrm{a}}'^{\mathrm{H2}}}{2} \\ k_{\mathrm{IrH'H \rightarrow IrHH'}} & -(R_{\mathrm{IrHH}} + k_{\mathrm{IrHH-H2}} + k_{\mathrm{IrH'H \rightarrow IrHH'}}) & \dfrac{k_{\mathrm{a}}'^{\mathrm{H2}}}{2} \\ k_{\mathrm{IrHH-H2}} & k_{\mathrm{IrHH-H2}} & -\left(R_{\mathrm{H_2}} + k_{\mathrm{a}}'^{\mathrm{H2}}\right) \end{pmatrix}$$

(*Eq. S6*)

Hence, there are only five unknown parameters: $k_{\mathrm{IrHH-H2}}$, $k_{\mathrm{a}}'^{\mathrm{H2}}$, $k_{\mathrm{IrH'H \rightarrow IrHH'}}$, $R_{\mathrm{IrHH}}$ and $R_{\mathrm{H_2}}$.

The general solution of the Bloch-McConnell equation, like (Eq. S6), is a superposition of exponentially decaying functions with decay rates equal to the eigenvalues of $\hat{L}$ plus the thermal magnetization $\vec{M}^0$ that can be calculated numerically using the following equation:

$$\vec{M}(d_{\mathrm{mix}}) = e^{\hat{L} d_{\mathrm{mix}}}\left[\vec{M}(d_{\mathrm{mix}} = 0) - \vec{M}^0\right] + \vec{M}^0 \qquad (\textit{Eq. S7})$$

as a function of mixing time, $d_{\mathrm{mix}}$. Here, $\vec{M}(d_{\mathrm{mix}} = 0)$ is the initial magnetization of the system at $d_{\mathrm{mix}} = 0$, which is, for example a result of the NMR pulse sequence preparation or spin labeling step. In case of hyperpolarization, $\vec{M}^0$ can be ignored, and the simplified solution can be calculated as follows:

$$\vec{M}(d_{\mathrm{mix}}) = e^{\hat{L} d_{\mathrm{mix}}}\vec{M}(d_{\mathrm{mix}} = 0) \qquad (\textit{Eq. S8})$$

All hyperpolarization experiments were fitted using this solution with the appropriate $\vec{M}$ and $\hat{L}$ parameters as described in the corresponding model.

**Magnetization exchange (Intermolecular exchange only, $\boldsymbol{M_{\mathrm{IrHH}} = M_{\mathrm{IrH^a}} + M_{\mathrm{IrH^b}}}$).** The equations above can be significantly simplified by considering the sum of the magnetization

IrHH. Consequently, we need to consider only $\frac{dM_{\mathrm{H2}}}{dt}$ and $\frac{dM_{\mathrm{IrHH}}}{dt}$ of Eq. S5. The resulting equations in the matrix form are:

$$\begin{cases} \mathrm{H_2} \xrightarrow{k'^{\mathrm{H2}}_{\mathrm{a}}} \mathrm{IrHH} \\ \mathrm{IrHH} \xrightarrow{k_{\mathrm{IrHH-H2}}} \mathrm{H_2} \end{cases}$$

$$\frac{d}{dt}\begin{pmatrix} M_{\mathrm{IrHH}} \\ M_{\mathrm{H_2}} \end{pmatrix} = \hat{L}\begin{pmatrix} M_{\mathrm{IrHH}} \\ M_{\mathrm{H_2}} \end{pmatrix} + \begin{pmatrix} R_{\mathrm{IrHH}} M^0_{\mathrm{IrHH}} \\ R_{\mathrm{H_2}} M^0_{\mathrm{H_2}} \end{pmatrix} \quad (Eq.\ S9)$$

$$\hat{L} = \begin{pmatrix} -(R_{\mathrm{IrHH}} + k_{\mathrm{IrHH-H2}}) & k'^{\mathrm{H2}}_{\mathrm{a}} \\ k_{\mathrm{IrHH-H2}} & -(R_{\mathrm{H_2}} + k'^{\mathrm{H2}}_{\mathrm{a}}) \end{pmatrix}$$

This approach is convenient as it simplifies the system and allows us to readily obtain key $H_2$ exchange parameters, and then use them to find the rest using the model that includes inter- and intramolecular exchange.

This equation has 3 free parameters: $R_{\mathrm{IrHH}}$, $k_{\mathrm{IrHH-H2}}$ and $k'^{\mathrm{H2}}_{\mathrm{a}}$ considering $R_{\mathrm{H_2}}$ is measured. When the ratio $\frac{k'^{\mathrm{H2}}_{\mathrm{a}}}{k_{\mathrm{IrHH-H2}}} = \frac{[\mathrm{IrHH}]}{[\mathrm{H_2}]}$ is measured by 1D spectroscopy at thermal equilibrium, then only 2 free parameters are left.

### 1.2. Model (3↔2)↔H₂

**Chemical exchange.** Experimentally, we observed chemical exchange between complexes **2** and **3** (**Figure 1**).

Complex **1** does not directly interconvert into **2** and **3**, but does produce free $H_2$; therefore, it can be treated separately using the previous model. The following model accounts for the chemical exchange of hydrogens in these two complexes.

$$\begin{cases} \mathrm{H_2} = \mathrm{H^xH^y} \xrightarrow{k'^{\mathrm{H_2}}_{3\mathrm{a}}} \frac{1}{2}\mathbf{3}^{\mathrm{xy}} + \frac{1}{2}\mathbf{3}^{\mathrm{yx}} \\ \mathrm{H_2} = \mathrm{H^xH^y} \xrightarrow{k'^{\mathrm{H_2}}_{2\mathrm{a}}} \frac{1}{2}\mathbf{2}^{\mathrm{xy}} + \frac{1}{2}\mathbf{2}^{\mathrm{yx}} \\ \mathbf{3}^{\mathrm{xy}} \xrightarrow{k^{\mathrm{H}}_{3-\mathrm{ex}}} \mathbf{3}^{\mathrm{yx}} \\ \mathbf{2}^{\mathrm{xy}} \xrightarrow{k^{\mathrm{H}}_{2-\mathrm{ex}}} \mathbf{2}^{\mathrm{yx}} \\ \mathbf{3}^{\mathrm{xy}} \xrightarrow{k_{3\to2}} \frac{1}{2}\mathbf{2}^{\mathrm{xy}} + \frac{1}{2}\mathbf{2}^{\mathrm{yx}} \\ \mathbf{2}^{\mathrm{xy}} \xrightarrow{k_{2\to3}} \frac{1}{2}\mathbf{3}^{\mathrm{xy}} + \frac{1}{2}\mathbf{3}^{\mathrm{yx}} \\ \mathbf{3}^{\mathrm{xy}} \xrightarrow{k_{3-\mathrm{H2}}} \mathrm{H_2} \\ \mathbf{2}^{\mathrm{xy}} \xrightarrow{k_{2-\mathrm{H2}}} \mathrm{H_2} \end{cases} \quad (Eq.\ S10)$$

The model involves 8 exchange rate constants: pseudo-first-order association rate constants $k'^{\mathrm{H_2}}_{3\mathrm{a}}$, $k'^{\mathrm{H_2}}_{2\mathrm{a}}$, dissociation rate constants $k_{3-\mathrm{H_2}}$, $k_{2-\mathrm{H_2}}$ intra-complex exchange rate constants $k^{\mathrm{H}}_{3-\mathrm{ex}}$, $k^{\mathrm{H}}_{2-\mathrm{ex}}$ and inter-complex exchange rate constants $k_{3\to2}$, $k_{2\to3}$. The superscript "H" means this was measured using a $^1$H experiment of hydride exchange, and **2** and **3** refer to the complex association, dissociation, or exchange it is related to. Here, we assume that both hydrides dissociate at the same rate under the same conditions. Finally, we also added superscripts to the complexes "xy". These labels represent the positions of two hydrides and are used here to indicate the changes in the positions of two protons. For example, the first position refers to a proton, with chemical shift $H^a$, and the second to a proton with chemical shift $H^b$ in the respective complex. For $H_2$, both protons have the same chemical shift.

Using these chemical equations, the kinetic behavior is governed by the following equation:

$$\frac{d[\mathrm{H_2}]}{dt} = +k_{3-\mathrm{H_2}}[3] + k_{2-\mathrm{H_2}}[2] - \left({k'}_{3a}^{\mathrm{H_2}} + {k'}_{2a}^{\mathrm{H_2}}\right)[\mathrm{H_2}] \qquad \textit{(Eq. S11)}$$

From which a steady-state equilibrium can be obtained:

$$k_{3-\mathrm{H_2}}[3] + k_{2-\mathrm{H_2}}[2] = \left({k'}_{3a}^{\mathrm{H_2}} + {k'}_{2a}^{\mathrm{H_2}}\right)[\mathrm{H_2}] \qquad \textit{(Eq. S12)}$$

**Magnetization exchange (inter- and intramolecular exchange).** As before, magnetization is proportional to concentration and polarization and can be defined for the constituents:

$$\begin{cases} M_{\mathrm{H_2}} = 2P_{\mathrm{H}}[\mathrm{H_2}] \\ M_{\mathbf{3}^{a}} = P_{\mathbf{3}^{a}}[3] \\ M_{\mathbf{3}^{b}} = P_{\mathbf{3}^{b}}[3] \\ M_{\mathbf{2}^{a}} = P_{\mathbf{2}^{a}}[2] \\ M_{\mathbf{2}^{b}} = P_{\mathbf{2}^{b}}[2] \\ M_{\mathbf{3}} = M_{\mathbf{3}^{a}} + M_{\mathbf{3}^{b}} = \left(P_{\mathbf{3}^{a}} + P_{\mathbf{3}^{b}}\right)[3] \\ M_{\mathbf{2}} = M_{\mathbf{2}^{a}} + M_{\mathbf{2}^{b}} = \left(P_{\mathbf{2}^{a}} + P_{\mathbf{2}^{b}}\right)[2] \end{cases} \qquad \textit{(Eq. S13)}$$

The corresponding thermal equilibrium magnetization $M^0_{\mathrm{H_2}}$, $M^0_{\mathbf{3}^{a}}$, $M^0_{\mathbf{3}^{b}}$, $M^0_{\mathbf{2}^{a}}$, $M^0_{\mathbf{2}^{b}}$, can again be obtained by substituting the polarization values with the thermal polarization $P^0$. Then, the corresponding kinetic equation for the magnetization of $\mathbf{3}$, $\mathbf{2}$, $\mathbf{3}^{a}$, $\mathbf{3}^{b}$, $\mathbf{2}^{a}$, $\mathbf{2}^{b}$, and $H_2$ is as follows:

$$\begin{cases} \dfrac{dM_{\mathbf{3}}}{dt} = {k'}_{3a}^{\mathrm{H_2}} M_{\mathrm{H_2}} - k_{3-\mathrm{H_2}}M_{\mathbf{3}} - k_{3\to2}M_{\mathbf{3}} + k_{2\to3}M_{\mathbf{2}} - R_{\mathbf{3}}\left(M_{\mathbf{3}} - M^0_{\mathbf{3}}\right) \\ \dfrac{dM_{\mathbf{2}}}{dt} = {k'}_{2a}^{\mathrm{H_2}} M_{\mathrm{H_2}} - k_{2-\mathrm{H_2}}M_{\mathbf{2}} - k_{2\to3}M_{\mathbf{2}} + k_{3\to2}M_{\mathbf{3}} - R_{\mathbf{2}}\left(M_{\mathbf{2}} - M^0_{\mathbf{2}}\right) \\ \dfrac{dM_{\mathbf{3}^{a}}}{dt} = \dfrac{{k'}_{3a}^{\mathrm{H_2}}}{2} M_{\mathrm{H_2}} - k_{3-\mathrm{H_2}}M_{\mathbf{3}^{a}} - k^{\mathrm{H}}_{3-\mathrm{ex}}\left(M_{\mathbf{3}^{a}} - M_{\mathbf{3}^{b}}\right) - k_{3\to2}M_{\mathbf{3}^{a}} + \dfrac{k_{2\to3}}{2}\left(M_{\mathbf{2}^{a}} + M_{\mathbf{2}^{b}}\right) - R_{\mathbf{3}}\left(M_{\mathbf{3}^{a}} - M^0_{\mathbf{3}^{a}}\right) \\ \dfrac{dM_{\mathbf{3}^{b}}}{dt} = \dfrac{{k'}_{3a}^{\mathrm{H_2}}}{2} M_{\mathrm{H_2}} - k_{3-\mathrm{H_2}}M_{\mathbf{3}^{b}} + k^{\mathrm{H}}_{3-\mathrm{ex}}\left(M_{\mathbf{3}^{a}} - M_{\mathbf{3}^{b}}\right) - k_{3\to2}M_{\mathbf{3}^{b}} + \dfrac{k_{2\to3}}{2}\left(M_{\mathbf{2}^{a}} + M_{\mathbf{2}^{b}}\right) - R_{\mathbf{3}}\left(M_{\mathbf{3}^{b}} - M^0_{\mathbf{3}^{b}}\right) \\ \dfrac{dM_{\mathbf{2}^{a}}}{dt} = \dfrac{{k'}_{2a}^{\mathrm{H_2}}}{2} M_{\mathrm{H_2}} - k_{2-\mathrm{H_2}}M_{[2]^{a}} - k^{\mathrm{H}}_{2-\mathrm{ex}}\left(M_{\mathbf{2}^{a}} - M_{\mathbf{2}^{b}}\right) - k_{2\to3}M_{\mathbf{2}^{a}} + \dfrac{k_{3\to2}}{2}\left(M_{\mathbf{3}^{a}} + M_{\mathbf{3}^{b}}\right) - R_{\mathbf{2}}\left(M_{\mathbf{2}^{a}} - M^0_{\mathbf{2}^{a}}\right) \\ \dfrac{dM_{\mathbf{2}^{b}}}{dt} = \dfrac{{k'}_{2a}^{\mathrm{H_2}}}{2} M_{\mathrm{H_2}} - k_{2-\mathrm{H_2}}M_{\mathbf{2}^{b}} + k^{\mathrm{H}}_{2-\mathrm{ex}}\left(M_{\mathbf{2}^{a}} - M_{\mathbf{2}^{b}}\right) - k_{2\to3}M_{\mathbf{2}^{b}} + \dfrac{k_{3\to2}}{2}\left(M_{\mathbf{3}^{a}} + M_{\mathbf{3}^{b}}\right) - R_{\mathbf{2}}\left(M_{\mathbf{2}^{b}} - M^0_{\mathbf{2}^{b}}\right) \\ \dfrac{dM_{\mathrm{H_2}}}{dt} = -\left({k'}_{3a}^{\mathrm{H2}} + {k'}_{2a}^{\mathrm{H2}}\right) M_{\mathrm{H_2}} + k_{3-\mathrm{H_2}}\left(M_{\mathbf{3}^{a}} + M_{\mathbf{3}^{b}}\right) + k_{2-\mathrm{H_2}}\left(M_{\mathbf{2}^{a}} + M_{\mathbf{2}^{b}}\right) - R_{\mathrm{H_2}}\left(M_{\mathrm{H_2}} - M^0_{\mathrm{H_2}}\right) \end{cases}$$

(*Eq. S14*)

The relaxation rates of the hydrides ligands are assumed to be the same, $R_{\mathbf{3}^{a}} \cong R_{\mathbf{3}^{b}} \cong \frac{R_{3}a+R_{3}b}{2} = R_{\mathbf{3}}$ and $R_{\mathbf{2}^{a}} \cong R_{\mathbf{2}^{b}} \cong \frac{R_{\mathbf{2}}a+R_{\mathbf{2}}b}{2} = R_{\mathbf{2}}$ for each complex. Now, the sum of $\frac{dM_{\mathbf{2}}a}{dt} + \frac{dM_{\mathbf{2}}b}{dt}$ is $\frac{dM_{\mathbf{3}}}{dt}$, and $\frac{dM_{\mathbf{2}}a}{dt} + \frac{dM_{\mathbf{2}}b}{dt}$ is $\frac{dM_{\mathbf{2}}}{dt}$. These equations can then be written in a matrix form as:

$$\frac{d}{dt}\begin{pmatrix} M_{\mathbf{3}^{\mathrm{a}}} \\ M_{\mathbf{3}^{\mathrm{b}}} \\ M_{\mathbf{2}^{\mathrm{a}}} \\ M_{\mathbf{2}^{\mathrm{b}}} \\ M_{\mathrm{H}_2} \end{pmatrix} = \hat{L}\begin{pmatrix} M_{\mathbf{3}^{\mathrm{a}}} \\ M_{\mathbf{3}^{\mathrm{b}}} \\ M_{\mathbf{2}^{\mathrm{a}}} \\ M_{\mathbf{2}^{\mathrm{b}}} \\ M_{\mathrm{H}_2} \end{pmatrix} + \begin{pmatrix} R_{\mathbf{3}^{\mathrm{a}}} M^0_{\mathbf{3}^{\mathrm{a}}} \\ R_{\mathbf{3}^{\mathrm{b}}} M^0_{\mathbf{3}^{\mathrm{b}}} \\ R_{\mathbf{2}^{\mathrm{a}}} M^0_{\mathbf{2}^{\mathrm{a}}} \\ R_{\mathbf{2}^{\mathrm{b}}} M^0_{\mathbf{2}^{\mathrm{b}}} \\ R_{\mathrm{H}_2} M^0_{\mathrm{H}_2} \end{pmatrix} \qquad \textit{(Eq. S15)}$$

$$\hat{L} =$$

$$\begin{pmatrix} -\left(R_3 + k_{3-\mathrm{H}_2} + k^{\mathrm{H}}_{3-\mathrm{ex}} + k_{3\to 2}\right) & k^{\mathrm{H}}_{3-\mathrm{ex}} & \frac{k_{2\to 3}}{2} & \frac{k_{2\to 3}}{2} & \frac{{k'_{3\mathrm{a}}}^{\mathrm{H2}}}{2} \\ k^{\mathrm{H}}_{3-\mathrm{ex}} & -\left(R_{\mathbf{3}} + k_{3-\mathrm{H}_2} + k^{\mathrm{H}}_{3-\mathrm{ex}} + k_{3\to 2}\right) & \frac{k_{2\to 3}}{2} & \frac{k_{2\to 3}}{2} & \frac{{k'_{3\mathrm{a}}}^{\mathrm{H2}}}{2} \\ \frac{k_{3\to 2}}{2} & \frac{k_{3\to 2}}{2} & -\left(R_{\mathbf{2}} + k_{2-\mathrm{H}_2} + k^{\mathrm{H}}_{2-\mathrm{ex}} + k_{2\to 3}\right) & k^{\mathrm{H}}_{2-\mathrm{ex}} & \frac{{k'_{2\mathrm{a}}}^{\mathrm{H2}}}{2} \\ \frac{k_{3\to 2}}{2} & \frac{k_{3\to 2}}{2} & k^{\mathrm{H}}_{2-\mathrm{ex}} & -\left(R_{\mathbf{2}} + k_{2-\mathrm{H}_2} + k^{\mathrm{H}}_{2-\mathrm{ex}} + k_{2\to 3}\right) & \frac{{k'_{2\mathrm{a}}}^{\mathrm{H2}}}{2} \\ k_{3-\mathrm{H}_2} & k_{3-\mathrm{H}_2} & k_{2-\mathrm{H}_2} & k_{2-\mathrm{H}_2} & -\left(R_{\mathrm{H}_2} + {k'_{3\mathrm{a}}}^{\mathrm{H2}} + {k'_{2\mathrm{a}}}^{\mathrm{H2}}\right) \end{pmatrix}$$

Consequently, there are unknown parameters: $k_{3-\mathrm{H}_2}$, $k_{2-\mathrm{H}_2}$,${k'_{3\mathrm{a}}}^{\mathrm{H}_2}$, ${k'_{2\mathrm{a}}}^{\mathrm{H}_2}$, $k^{\mathrm{H}}_{3-\mathrm{ex}}$, $k^{\mathrm{H}}_{2-\mathrm{ex}}$, $R_{\mathbf{3}}$, $R_{\mathbf{2}}$, and $R_{\mathrm{H}_2}$. The general solution to this equation is Eq. S8.

**Magnetization exchange (Intermolecular exchange only, $M_{\mathbf{3}} = M_{\mathbf{3}^{\mathrm{a}}} + M_{\mathbf{3}^{\mathrm{b}}}$, $M_{\mathbf{2}} = M_{\mathbf{2}^{\mathrm{a}}} + M_{\mathbf{2}^{\mathrm{b}}}$).** The equation above can be significantly simplified when we consider the magnetization of IrHH protons together. In this case, we need to consider only $\frac{dM_{\mathrm{H}_2}}{dt}$, $\frac{dM_{\mathbf{3}}}{dt}$ and $\frac{dM_{\mathbf{2}}}{dt}$ from Eq. S14. The resulting equations in the matrix form are:

$$\begin{cases} \mathrm{H}_2 \xrightarrow{{k'_{3\mathrm{a}}}^{\mathrm{H}_2}} \mathbf{3} \\ \mathrm{H}_2 \xrightarrow{{k'_{2\mathrm{a}}}^{\mathrm{H}_2}} \mathbf{2} \\ \mathbf{3} \xrightarrow{k_{3\to 2}} \mathbf{2} \\ \mathbf{2} \xrightarrow{k_{2\to 3}} \mathbf{3} \\ \mathbf{3} \xrightarrow{k_{3-\mathrm{H}_2}} \mathrm{H}_2 \\ \mathbf{2} \xrightarrow{k_{2-\mathrm{H}_2}} \mathrm{H}_2 \end{cases}$$

$$\frac{d}{dt}\begin{pmatrix} M_{\mathbf{3}} \\ M_{\mathbf{2}} \\ M_{\mathrm{H}_2} \end{pmatrix} = \hat{L}\begin{pmatrix} M_{\mathbf{3}} \\ M_{\mathbf{2}} \\ M_{\mathrm{H}_2} \end{pmatrix} + \begin{pmatrix} R_{\mathbf{3}} M^0_{\mathbf{3}} \\ R_{\mathbf{2}} M^0_{\mathbf{2}} \\ R_{\mathrm{H}_2} M^0_{\mathrm{H}_2} \end{pmatrix}$$

$$\hat{L} = \begin{pmatrix} -\left(k_{3-\mathrm{H}_2} + k_{3\to 2} + R^{\mathrm{inter}}_{\mathbf{3}}\right) & k_{2\to 3} & {k'_{3\mathrm{a}}}^{\mathrm{H}_2} \\ k_{3\to 2} & -\left(k_{2-\mathrm{H}_2} + k_{2\to 3} + R^{\mathrm{inter}}_{\mathbf{2}}\right) & {k'_{2\mathrm{a}}}^{\mathrm{H}_2} \\ k_{3-\mathrm{H}_2} & k_{2-\mathrm{H}_2} & -\left({k'_{3\mathrm{a}}}^{\mathrm{H}_2} + {k'_{2\mathrm{a}}}^{\mathrm{H}_2} + R_{\mathrm{H}_2}\right) \end{pmatrix}$$

(*Eq. S16*)

This is convenient as it simplifies the system, meaning some exchange parameters can be defined and used in a expanded model including inter- and intramolecular exchange. Setting $R^{\mathrm{inter}}_{\mathbf{3}} = R^{\mathrm{inter}}_{\mathbf{2}}$, and measuring $R_{\mathrm{H}_2}$, we have 7 free parameters: $k_{3\to 2}$, $k_{2\to 3}$,$k_{3-\mathrm{H}_2}$, $k_{2-\mathrm{H}_2}$, ${k'_{3\mathrm{a}}}^{\mathrm{H2}}$, ${k'_{2\mathrm{a}}}^{\mathrm{H2}}$ and $R^{\mathrm{inter}}_{\mathbf{3}} = R^{\mathrm{inter}}_{\mathbf{2}}$. When these parameters are obtained, we can estimate $k^{\mathrm{H}}_{3-\mathrm{ex}}$ and $k^{\mathrm{H}}_{2-\mathrm{ex}}$ using inter- and intramolecular exchange models (Eq. S15). The measured $R_{\mathrm{H}_2}$ values at different temperatures and pressures are presented in Table S13, and prove to be of the order 1.9 s and decrease with high temperature, reaching 1.5 s at 300 K as in (74 SchmidtCPC).

Because of observable cross-relaxation effects between protons of the same complex, the obtained relaxation rates $R^{\mathrm{inter}} = R^{\mathrm{inter+intra}} + \sigma$ for complexes **3** and **2** (see Solomon equations for the explanation). The $\sigma_3$ and $\sigma_2$ in the inter- and intramolecular exchange model therefore is included in the observed $k^{\mathrm{H}}_{3-\mathrm{ex}}$ and $k^{\mathrm{H}}_{2-\mathrm{ex}}$ rate constants. Different $R$ parameters were therefore used in the two models.

Additional constraints on the parameters can be set according to data from the thermal equilibrium experiments. The stationary conditions from (Eq. S29) are

$$\begin{cases} k'^{\;\mathrm{H2}}_{3\mathrm{a}}[\mathrm{H}_2] = \left(k_{3-H_2} + k_{3\to 2}\right)[3] - k_{2\to 3}[2] \\ k'^{\;\mathrm{H2}}_{2\mathrm{a}}[\mathrm{H}_2] = \left(k_{2-H_2} + k_{2\to 3}\right)[2] - k_{3\to 2}[3] \\ k_{3-\mathrm{H}_2}[3] + k_{2-\mathrm{H}_2}[2] = \left(k'^{\;\mathrm{H2}}_{3\mathrm{a}} + k'^{\;\mathrm{H2}}_{2\mathrm{a}}\right)[\mathrm{H}_2] \end{cases} \qquad \textit{(Eq. S17)}$$

For $[3] = f_{3,2}[2]$, the first two equations can be divided and one obtains:

$$k'^{\;\mathrm{H}_2}_{3\mathrm{a}} = k'^{\;\mathrm{H}_2}_{2\mathrm{a}} \frac{\left(k_{3-\mathrm{H}_2} + k_{3\to 2}\right) f_{3,2} - k_{2\to 3}}{\left(k_{2-\mathrm{H}_2} + k_{2\to 3}\right) - k_{3\to 2} f_{3,2}} \qquad \textit{(Eq. S18)}$$

Hence, when $f_{3,2} = [3]/[2]$ ratio is known, the number of parameters goes down to 6 (see **Table S13**): $k_{3\to 2}$, $k_{2\to 3}$, $k_{3-\mathrm{H}_2}$, $k_{2-\mathrm{H}_2}$, $k'^{\;\mathrm{H}_2}_{2\mathrm{a}}$, and $R^{\mathrm{inter}}_{\mathbf{3}} = R^{\mathrm{inter}}_{\mathbf{2}}$ (or to 7 if $R^{\mathrm{inter}}_{\mathbf{3}} \neq R^{\mathrm{inter}}_{\mathbf{2}}$). Using the parameters estimated this way, one can apply them to the inter and intramolecular exchange model, which will have four free parameters: $R_{\mathbf{3}}$, $R_{\mathbf{2}}$, $k^{\mathrm{H}}_{3-\mathrm{ex}}$, $k^{\mathrm{H}}_{2-\mathrm{ex}}$.

# 2. Models of pyruvate exchange during SABRE

Since we observed exchange between complexes **2** and **3**, it is not reasonable to model the exchange of only one complex without the other; hence, only a model with two Ir complexes will be considered here.

## 2.1. Model ($^{13}C^{2}$-3↔$^{13}C^{2}$-2)↔$^{13}C^{2}$-free

The model is analogous to Model (**2**↔**3**)↔$H_2$ with only a few modifications:

$$\begin{pmatrix} \mathrm{pyr} \xrightarrow{{k'_{3\mathrm{a}}}^{\mathrm{P}}} \mathbf{3} \\ \mathrm{pyr} \xrightarrow{{k'_{2\mathrm{a}}}^{\mathrm{P}}} \mathbf{2} \\ \mathbf{3} \xrightarrow{k_{3\to 2}} \mathbf{2} \\ \mathbf{2} \xrightarrow{k_{2\to 3}} \mathbf{3} \\ \mathbf{3} \xrightarrow{k_{3\to \mathrm{pyr}}} \mathrm{pyr} \\ \mathbf{2} \xrightarrow{k_{2\to \mathrm{pyr}}} \mathrm{pyr} \end{pmatrix} \qquad \textit{(Eq. S19)}$$

Here, superscript "P" refers to the values measured by observing pyruvate exchange. The meaning of all exchange rate constants is as before. The corresponding chemical kinetic equations are

$$\begin{cases} \frac{d[1]}{dt} = -\left(k_{3\to\mathrm{pyr}} + k_{3\to 2}\right)[3] + k_{2\to 3}[2] + {k'_{3\mathrm{a}}}^{\mathrm{P}}[\mathrm{pyr}] \\ \frac{d[2]}{dt} = +k_{3\to 2}[3] - \left(k_{2\to\mathrm{pyr}} + k_{2\to 3}\right)[2] + {k'_{2\mathrm{a}}}^{\mathrm{P}}[\mathrm{pyr}] \\ \frac{d[\mathrm{pyr}]}{dt} = +k_{3\to\mathrm{pyr}}[3] + k_{2\to\mathrm{pyr}}[2] - \left({k'_{3\mathrm{a}}}^{\mathrm{P}} + {k'_{2\mathrm{a}}}^{\mathrm{P}}\right)[\mathrm{pyr}] \end{cases} \qquad \textit{(Eq. S20)}$$

Corresponding steady-state concentrations of $[\mathrm{pyr}]$ and catalysts are:

$$\begin{cases} {k'_{3\mathrm{a}}}^{\mathrm{P}}[\mathrm{pyr}] = \left(k_{3\to\mathrm{pyr}} + k_{3\to 2}\right)[3] - k_{2\to 3}[2] \\ {k'_{2\mathrm{a}}}^{\mathrm{P}}[\mathrm{pyr}] = \left(k_{2\to\mathrm{pyr}} + k_{2\to 3}\right)[2] - k_{3\to 2}[3] \\ k_{3\to\mathrm{pyr}}[3] + k_{2\to\mathrm{pyr}}[2] = \left({k'_{3\mathrm{a}}}^{\mathrm{P}} + {k'_{2\mathrm{a}}}^{\mathrm{P}}\right)[\mathrm{pyr}] \end{cases} \qquad \textit{(Eq. S21)}$$

For $[3] = f_{3,2}[2]$, the first two equations can be divided, and one obtains:

$${k'_{3\mathrm{a}}}^{\mathrm{P}} = {k'_{2\mathrm{a}}}^{\mathrm{P}} \frac{\left(k_{3\to\mathrm{pyr}} + k_{3\to 2}\right) f_{3,2} - k_{2\to 3}}{\left(k_{2\to\mathrm{pyr}} + k_{2\to 3}\right) - k_{3\to 2} f_{3,2}} \qquad \textit{(Eq. S22)}$$

The magnetization is proportional to concentration and polarization, and for the discussed compounds is:

$$\begin{cases} M_{\mathbf{3}} = P_{\mathbf{3}}[\mathbf{3}] \\ M_{\mathbf{2}} = P_{\mathbf{2}}[\mathbf{2}] \\ M_{\mathrm{pyr}} = P_{\mathrm{pyr}}[\mathrm{pyr}] \end{cases} \qquad \textit{(Eq. S23)}$$

The corresponding thermal equilibrium magnetization $M_{\mathbf{3}}^{0}$, $M_{\mathbf{2}}^{0}$, or $M_{[\mathrm{pyr}]}^{0}$ can be obtained by substituting the polarization values with the thermal polarization $P^{0}$. Then, the corresponding kinetic equations for magnetization of **3**, **2**, and pyr are as follows:

$$\begin{cases} \frac{dM_3}{dt} = +{k'_{3a}}^{\mathrm{P}} M_{\mathrm{pyr}} - (k_{3\to\mathrm{pyr}} + k_{3\to 2}) M_{\mathbf{3}} + k_{2\to 3} M_{\mathbf{2}} - (M_{\mathbf{3}} - M_{\mathbf{3}}^0) R_{\mathbf{3}} \\ \frac{dM_2}{dt} = +{k'_{2a}}^{\mathrm{P}} M_{\mathrm{pyr}} + k_{3\to 2} M_{\mathbf{3}} - (k_{2\to\mathrm{pyr}} + k_{2\to 3}) M_{\mathbf{2}} - (M_{\mathbf{2}} - M_{\mathbf{2}}^0) R_{\mathbf{2}} \\ \frac{dM_{\mathrm{pyr}}}{dt} = -({k'_{3a}}^{\mathrm{P}} + {k'_{2a}}^{\mathrm{P}}) M_{\mathrm{pyr}} + k_{3\to\mathrm{pyr}} M_{\mathbf{3}} + k_{2\to\mathrm{pyr}} M_{\mathbf{2}} - (M_{\mathrm{pyr}} - M_{\mathrm{pyr}}^0) R_{\mathrm{pyr}} \end{cases} \qquad \textit{(Eq S24)}$$

In a matrix form, this can be written as follows:

$$\frac{\mathrm{d}}{\mathrm{dt}} \begin{pmatrix} M_{\mathbf{3}} \\ M_{\mathbf{2}} \\ M_{\mathrm{pyr}} \end{pmatrix} = \hat{L} \begin{pmatrix} M_{\mathbf{3}} \\ M_{\mathbf{2}} \\ M_{\mathrm{pyr}} \end{pmatrix} + \begin{pmatrix} R_{\mathbf{3}} M_{\mathbf{3}}^0 \\ R_{\mathbf{2}} M_{\mathbf{2}}^0 \\ R_{\mathrm{pyr}} M_{\mathrm{pyr}}^0 \end{pmatrix}$$

$$\hat{L} = \begin{pmatrix} -(R_{\mathbf{3}} + k_{3\to\mathrm{pyr}} + k_{3\to 2}) & k_{2\to 3} & {k'_{3a}}^{\mathrm{P}} \\ k_{3\to 2} & -(R_{\mathbf{2}} + k_{2\to\mathrm{pyr}} + k_{2\to 3}) & {k'_{2a}}^{\mathrm{P}} \\ k_{3\to\mathrm{pyr}} & k_{2\to\mathrm{pyr}} & -(R_{\mathrm{pyr}} + {k'_{3a}}^{\mathrm{P}} + {k'_{2a}}^{\mathrm{P}}) \end{pmatrix} \qquad \textit{(Eq S25)}$$

Consequently, there are unknown parameters: $k_{3\to\mathrm{pyr}}$, $k_{2\to\mathrm{pyr}}$, ${k'_{3a}}^{\mathrm{P}}$, ${k'_{2a}}^{\mathrm{P}}$, $k_{2\to 3}$, $k_{3\to 2}$, $R_{\mathbf{3}}$, $R_{\mathbf{2}}$, and $R_{\mathrm{pyr}}$. Setting $R_{\mathbf{3}} = R_{\mathbf{2}}$ and measuring $R_{\mathrm{pyr}}$ for pyruvate in $H_2$ bubbled methanol without the catalyst (**Table S15**), leaves only 7 independent parameters. When the ratio $f_{3,2} = [3]/[2]$, the number of parameters goes down to 6 using Eq. S25. This ratio can be measured from the IrHH signals provided independently by **3** and **2** (see **Table S13**).

## 3. Calibration of variable temperature NMR unit

We calibrated the variable temperature unit (VTU) of the NMR spectrometer using a method sample, of observing the chemical shift separation Δ between its OH and $CH_3$ resonances (calibration sample temperature method provided by Brucker). We prepared a 600 µL of 4:96 mixture of isotopically unlabeled methanol and methanol-$d_4$ and measured its $^1$H NMR spectra at different temperatures. Then, we used the following two equations to calculate the actual temperature of the sample, as in the Bruker Instruments Manual for VT-Calibration:

1) For values between 230 – 270 K:

$$T = (3.92 - \Delta)/0.008 \qquad (Eq.\ S31)$$

2) For more accurate values between 270 – 300 K, we used this equation:

$$T = (4.109 - \Delta)/0.008708 \qquad (Eq.\ S32)$$

**Table S1.** The user set the nominal NMR temperature, and the calibrated temperature was calculated using Eqs. S31 and S32. Deviation indicates the relative deviation between the calibrated and the nominal temperatures.

| Nominal NMR *T* (K) | Calibrated *T* (K) | Deviation (%) |
|---|---|---|
| 260 | 257.5 | -0.96 |
| 263 | 260 | -1.14 |
| 267 | 265 | -0.75 |
| 270 | 267.5 | -0.92 |
| 273 | 270.9 | -0.77 |
| 277 | 275.5 | -0.54 |
| 280 | 277.8 | -0.79 |
| 283 | 281.2 | -0.62 |
| 288 | 287 | -0.35 |

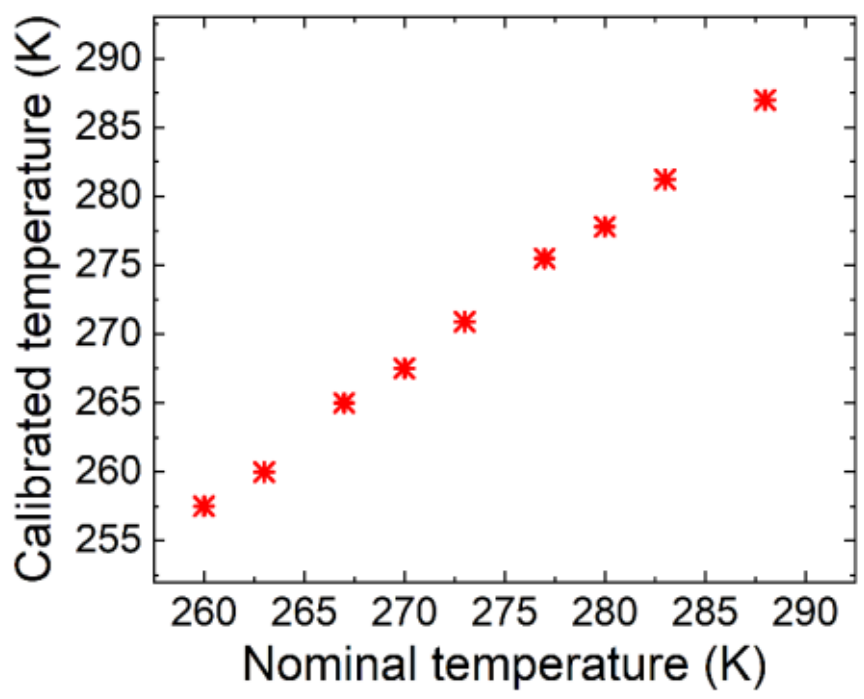

**Figure S1:** Calibrated temperature measured from chemical shift difference of methanol peaks, Δ, as a function of nominal temperature.

# 4. Signal enhancement and polarization calculations

**Signal enhancement and polarization**

The enhancement factor $\varepsilon$ of substrates was calculated using the signal intensities of the spectra of thermally polarized and hyperpolarized solutions, taking into account the differences in acquisition parameters. The following formula was used when the signals had net magnetization:

$$\varepsilon = \frac{I^{\mathrm{HP}}}{I^{\mathrm{TP}}} \times \frac{\sin(\alpha^{\mathrm{TP}})}{\sin(\alpha^{\mathrm{HP}})} \times \frac{RG^{\mathrm{TP}}}{RG^{\mathrm{HP}}} \times \frac{NS^{\mathrm{TP}}}{NS^{\mathrm{HP}}} \qquad \textit{(Eq. S33)}$$

where $I^{\mathrm{HP}}$ is the integral of the hyperpolarized signal and $I^{\mathrm{TP}}$ is the integral of the thermally polarized signal of the same species. $NS^{\mathrm{TP}}$ and $NS^{\mathrm{HP}}$ are the numbers of scans for the thermally polarized and hyperpolarized samples, and $\alpha^{\mathrm{TP}}$ and $\alpha^{\mathrm{HP}}$ are the excitation angles of RF pulses used to acquire the corresponding spectra. $RG^{\mathrm{TP}}$ and $RG^{\mathrm{HP}}$ are values of linear receiver gain.

In the case of $^{1}$H PASADENA polarization, the integral of only one $^{1}$H proton was considered after the application of the SEPP SOT; the corresponding excitation angle is $\sin(\alpha^{\mathrm{HP}}) = 1$.

To compare the enhancements at 9.4 T more easily, the enhancement factors $\varepsilon$ were converted to a polarization level, P, using Eq. S33, where, $\gamma$ is the gyromagnetic ratio, $B_0$ is the detection field, $T$ is the temperature, $\hbar$ is the reduced Planck's constant, and $k_B$ is Boltzmann's constant.

$$P = \varepsilon \frac{\gamma B_0 \hbar}{2\, k_{\mathrm{B}} T} = \varepsilon P^{\mathrm{TP}} \qquad \textit{(Eq. S34)}$$

The example thermal polarization values at 9.4 T and 300 K are $P^{\mathrm{TP}}(^{1}\mathrm{H}) \cong 3.6 \cdot 10^{-5}$ and $P^{\mathrm{TP}}(^{13}\mathrm{C}) \cong 8.05 \cdot 10^{-6}$.

**Line-enhancement in PASADENA:** In the case of the $^{1}$H PASADENA experiment, we integrated separately two lines of the antiphase PASADENA spectrum of a given $^{1}$H resonance and then subtracted them (or added together absolute values) which gave us an integral over the PASADENA line, $I^{\mathrm{PASADENA}}$, of one $^{1}$H spin. This value is very close to the integral over the spectral line when it is given in amplitude mode. Since two IrHH protons are polarized, we used an average integral value for each complex: $I_{\mathrm{avg}}^{\mathrm{PASADENA}} = 0.5(I_{\mathrm{H^a}}^{\mathrm{PASADENA}} + I_{\mathrm{H^b}}^{\mathrm{PASADENA}})$. We used the modification of Eq. S33 to calculate the signal enhancement as

$$\varepsilon^{PASADENA} = \frac{I_{\mathrm{avg}}^{\mathrm{PASADENA}}}{I^{\mathrm{TP}}} \times \frac{\sin(\alpha^{\mathrm{TP}})}{\sin(2\alpha^{\mathrm{HP}})} \times \frac{RG^{\mathrm{TP}}}{RG^{\mathrm{HP}}} \times \frac{NS^{\mathrm{TP}}}{NS^{\mathrm{HP}}} \qquad \textit{(Eq. S35)}$$

Note that here, the angle dependence is $\sin(2\alpha^{\mathrm{HP}})$, hence for $\alpha^{\mathrm{HP}} = 45°$, it is 1 (maximum signal in the PASADENA experiment). To calculate the polarization of PASADENA two spin order, $\hat{I}_{\mathrm{z}}^{\mathrm{H^a}} \hat{I}_{\mathrm{z}}^{\mathrm{H^b}}$, one should use the following equation:

$$P_{\mathrm{zz}} = 2\varepsilon^{PASADENA} P^{\mathrm{TP}} \qquad \textit{(Eq. S36)}$$

The additional factor of 2 reflects that the PASADENA signal is measured with a 45° pulse so ½ of the signal results relative to a spin with 100% net magnetization. Derivation of the net and PASADENA signals is discussed in the SI of (75PravdivtsevJPC). When $P_{\mathrm{zz}} = 1$ is reached, it means that 100% $pH_2$ was used, and there were no losses in spin order during the experiment.

## 5. $^{1}$H NMR Spectra of IrHH

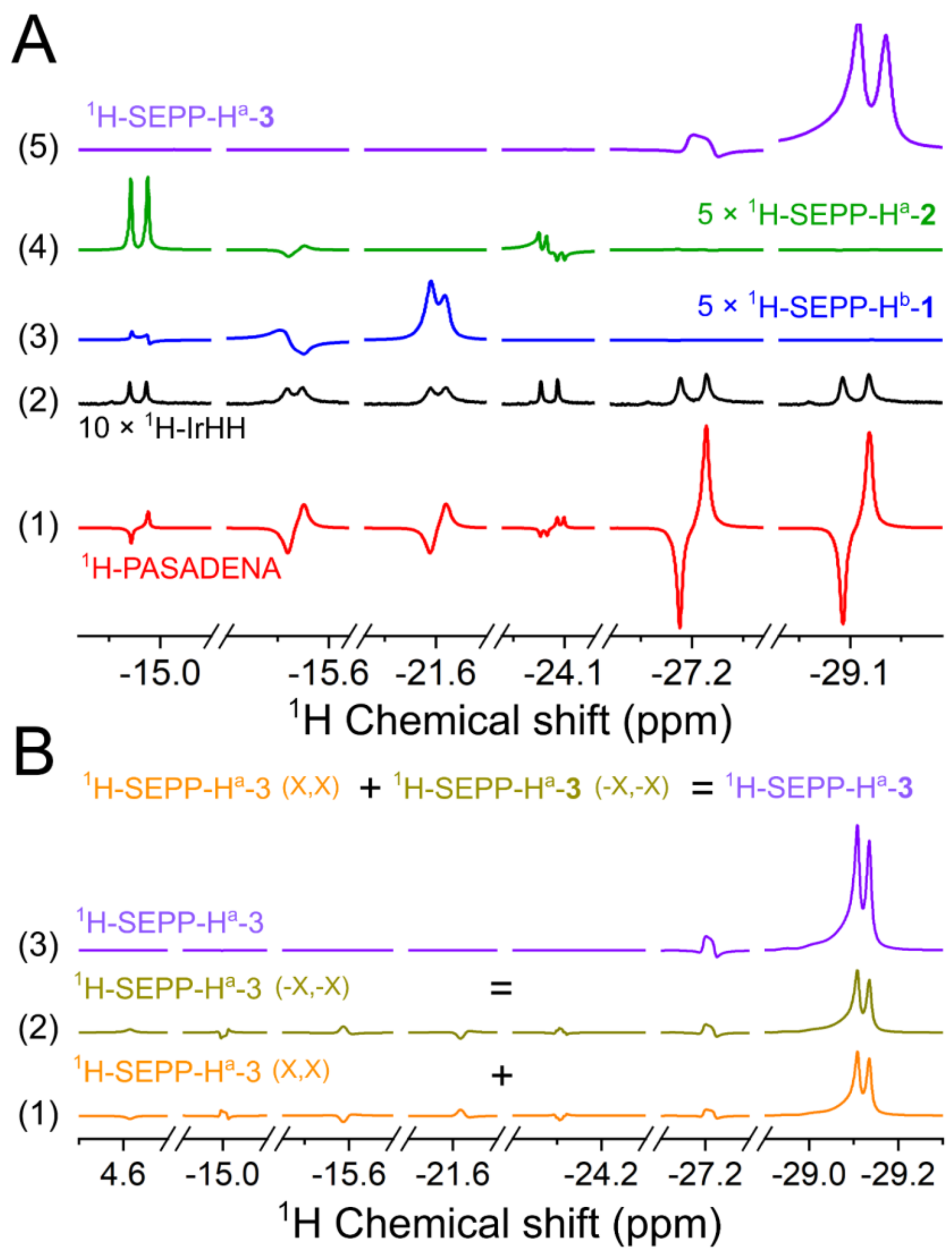

**Figure S2: $^{1}$H NMR spectra of IrHH protons**. (A.1)$^{1}$H-PASADENA (NS = 1, $\alpha = 45^{o}$) (red),(A. 2) $^{1}$H thermally polarized spectrum (NS = 100, TR = 9 s, $\alpha = 90^{o}$) (black), (A. 3) $^{1}$H-SEPP of H$^{b}$-**1** (-21.59 ppm, $\tau_1$ = 26 ms, $\tau_2$ = 0 ms, $\Delta$ = 10 ms) (blue), (A. 4) $^{1}$H-SEPP of H$^{a}$-**2** (-14.98 ppm, $\tau_1$ = 26 ms, $\tau_2$ = 0 ms, $\Delta$ = 10 ms) (green), and (A. 5) $^{1}$H-SEPP of H$^{a}$-**3** (-29.1 ppm, $\tau_1$ = 23.8 ms, $\tau_2$ = 0 ms, $\Delta$ = 10 ms) (purple). (B. 1) $^{1}$H-SEPP of H$^{a}$-**3** (-29.1 ppm, $\tau_1$ = 23.8 ms, $\tau_2$ = 0 ms, $\Delta$ = 10 ms) with phase (X, X) of the first 90$^{o}$ pulse and recording phase (orange). (B. 2) $^{1}$H-SEPP of H$^{a}$-**3** (-29.1 ppm, $\tau_1$ = 23.8 ms, $\tau_2$ = 0 ms, $\Delta$ = 10 ms) with phase (-X, -X) of the first 90$^{o}$ pulse and recording phase (dark yellow). (B. 1) $^{1}$H-SEPP of H$^{a}$-**3** (-29.1 ppm, $\tau_1$ = 23.8 ms, $\tau_2$ = 0 ms, $\Delta$ = 10 ms) after 2 scans (B.1) + (B.2) (purple). Sample compositions and conditions: [Ir] = 4mM, [2-$^{13}$C-pyr] = 40mM, [DMSO] = 20 mM, in methanol-$d_4$ at 8.5 bar and 265 K, with 10 s bubbling time, and a settling time of 1.5 s. Achieved signal enhancements and polarization for SEPP spectra are 1995, 8.28% (A. 3), 1053, 4.36% (A. 4), 2723, 11.3% (A. 5). Achieved signal enhancements and polarization in PASADENA spectrum are 8058, 66.89% **3**, 2320, 19.26% **2**, 5717, 47.45% **1**.

## 6. Analysis of chemical equilibria for 1-4

**2-3 and 3-4:** To assess the connection between equilibrium concentrations of complexes **2**, **3**, and **4**, it is convenient to consider the following thermodynamic equilibria with the corresponding equilibrium constants $K_{\{2\leftarrow3\}}$ and $K_{\{4\leftarrow3\}}$:

$$\mathbf{3} + \mathrm{DMSO} \overset{K_{\{2\leftarrow3\}}}{\rightleftharpoons} \mathbf{2} \qquad (\textit{Eq. S37})$$

where

$$K_{\{2\leftarrow3\}} = \frac{[\mathbf{2}]}{[\mathbf{3}][\mathrm{DMSO}]} \Rightarrow \frac{[\mathbf{2}]}{[\mathbf{3}]} = K_{\{2\leftarrow3\}}[\mathrm{DMSO}] \qquad (\textit{Eq. S38})$$

And

$$\mathbf{3} \overset{K_{\{4\leftarrow3\}}}{\rightleftharpoons} \mathbf{4} \qquad (\textit{Eq. S39})$$

Where

$$K_{\{4\leftarrow3\}} = \frac{[\mathbf{4}]}{[\mathbf{3}]} \qquad (\textit{Eq. S40})$$

The obtained ratios of equilibrium concentrations reveal a clear difference since the [**2**]/[**3**] ratio is dependent on the equilibrium concentration of DMSO (see Eq. S38), whereas the [**4**]/[**3**] ratio must be DMSO concentration independent (see Eq. S40). The latter conclusion comes from the fact that complexes **3** and **4** are structural isomers, and their equilibrium concentration ratio is defined solely by the thermodynamic chemical potential of only these two species. In other words, since there are no other species involved, [**4**]/[**3**] becomes concentration independent.

Since the analysis of concentration dependencies for **2** and **3** shown in **Figure 3D** of the main text demonstrates an apparent increase of [**2**] and decrease of [**3**] with the total DMSO concentration (**3** → **2** transformation), it provides additional evidence that the hydride signals at -14.93 and -24.02 ppm in $^{1}$H NMR belong to complex **2** because it matches expectation on the DMSO concentration dependence for the [**2**]/[**3**] ratio (Eq. S38). If those resonances were to belong to **4**, the complex concentration ratio would remain unchanged, as discussed above.

Moreover, the dependence of the kinetic rate for the **3**→**2** transformation on [DMSO] shown in **Figure 5H** supports the hydride signal assignment of **2,** since this process will also be DMSO concentration dependent according to the equilibrium Eq. S37, whereas such a kinetic rate would be concentration independent in the case of equilibrium Eq. S39.

To determine the standard Gibbs free energy change for the **2**←**3** transition, we assumed that

$$K_{\{2\leftarrow3\}} = \frac{[\mathbf{2}]}{[\mathbf{3}][\mathrm{DMSO}]} = \mathrm{e}^{-\frac{\Delta G^{0}_{2\leftarrow3}}{RT}}. \qquad (\textit{Eq. S41})$$

where, $\Delta G^{0}_{\mathbf{2}\leftarrow\mathbf{3}} = \Delta G^{0}_{\mathbf{2}} - \Delta G^{0}_{\mathbf{3}}$. Rearranging this equation and assuming $\Delta G^{0}_{2\leftarrow3} = \Delta H^{0}_{2\leftarrow3} - T\Delta S^{0}_{2\leftarrow3}$ gives:

$$\ln\left(\frac{[\mathbf{2}]}{[\mathbf{3}][\mathrm{DMSO}]}\right) = -\frac{\Delta H^{0}_{2\leftarrow3}}{R} \times \frac{1}{T} + \frac{\Delta S^{0}_{2\leftarrow3}}{R} \qquad (\textit{Eq. S42})$$

A plot of $\ln\left(\frac{[\mathbf{2}]}{[\mathbf{3}][\mathrm{DMSO}]}\right)$ as a function of $\frac{1}{T}$ yields a straight line, where the slope corresponds to $-\frac{\Delta H^0_{2\leftarrow 3}}{R}$ and the constant term to $\frac{\Delta S^0_{2\leftarrow 3}}{R}$. Utilizing the equilibrium concentrations from **Table S2,** this linearization equation was used to fit experimental dependency (**Figure S3B**) and obtain $\Delta H^0_{2\leftarrow 3}$, $\Delta S^0_{2\leftarrow 3}$, and $\Delta G^0_{2\leftarrow 3}$ in **Table S3**.

**1-3:** Similarly, the transformation from **3** to **1** can be analyzed using the following equation,

$$\mathbf{3} + \mathrm{DMSO} + \mathrm{Cl}^- \overset{K_{\{\mathbf{1}\leftarrow\mathbf{3}\}}}{\rightleftharpoons} \mathbf{1} + \mathrm{pyr} \qquad (Eq.\ S43)$$

leading to

$$K_{\{\mathbf{1}\leftarrow\mathbf{3}\}} = \frac{[\mathbf{1}][\mathrm{pyr}]}{[\mathbf{3}][\mathrm{DMSO}][\mathrm{Cl}^-]} \Rightarrow \frac{[\mathbf{1}]}{[\mathbf{3}]} = K_{\{\mathbf{1}\leftarrow\mathbf{3}\}}\frac{[\mathrm{DMSO}][\mathrm{Cl}^-]}{[\mathrm{pyr}]} \qquad (Eq.\ S44)$$

Performing the necessary rearrangements and assuming $\Delta G^0_{1\leftarrow 3} = \Delta H^0_{1\leftarrow 3} - T\Delta S^0_{1\leftarrow 3}$ gives:

$$\ln\left(\frac{[\mathbf{1}][\mathrm{pyr}]}{[\mathbf{3}][\mathrm{DMSO}][\mathrm{Cl}^-]}\right) = -\frac{\Delta H^0_{1\leftarrow 3}}{R} \times \frac{1}{T} + \frac{\Delta S^0_{1\leftarrow 3}}{R}. \qquad (Eq.\ S45)$$

Utilizing the equilibrium concentrations from **Table S2,** this linearization equation was used to fit experimental dependency (**Figure S3C**) and obtain $\Delta H^0_{1\leftarrow 3}$, $\Delta S^0_{1\leftarrow 3}$, and $\Delta G^0_{1\leftarrow 3}$ in **Table S3**.

**Table S2**: Equilibrium concentrations of complexes 1-9 were estimated from the NMR spectra (**Figure S3A**) measured at different temperatures. Nominal concentrations of constituents $[\mathrm{DMSO}]^0$ = 20 mM, $[\mathrm{pyr}]^0$ = 40 mM, $[\mathrm{Ir}]^0$ = 4 mM, and $[\mathrm{Cl}^-]^0$ = 4 mM, and their corresponding estimated equilibrium concentrations [DMSO], [pyr] and $[\mathrm{Cl}^-]$.

| C, mM | **1** | **2** | **3** | **7** | **8** | **9** | -25.84 | -26.47 | [DMSO] | [pyr] | $[\mathrm{Cl}^-]$ |
|---|---|---|---|---|---|---|---|---|---|---|---|
| DMSO | 2 | 2 | 1 | 1 | 2 | 2 | X | X | | | |
| Cl | 1 | 0 | 0 | 0 | 0 | 0 | X | X | | | |
| Pyr | 0 | 1 | 1 | 0 | 0 | 0 | X | X | | | |
| 220 K | 0.79 | 1.51 | 0.1 | 0.27 | 0.83 | 0.05 | 0.15 | 0.24 | 12.76 | 38.3 | 3.20 |
| 235 K | 1.16 | 1.02 | 0.51 | 0.08 | 0.77 | 0.06 | 0.18 | 0.26 | 12.92 | 38.4 | 2.84 |
| 247 K | 1.13 | 0.81 | 0.98 | 0.03 | 0.72 | 0.02 | 0.10 | 0.23 | 13.30 | 38.2 | 2.87 |
| 259 K | 0.77 | 0.5 | 1.48 | 0.13 | 0.78 | 0.02 | 0.11 | 0.21 | 13.89 | 38.1 | 3.22 |
| 260 K | 0.58 | 0.31 | 1.19 | 0.90 | 0.77 | 0.05 | 0.08 | 0.18 | 14.24 | 38.5 | 3.42 |
| 265 K | 0.45 | 0.24 | 1.06 | 1.19 | 0.75 | 0.08 | 0.09 | 0.20 | 14.38 | 38.6 | 3.54 |
| 267.5 K | 0.35 | 0.21 | 1.13 | 1.27 | 0.72 | 0.05 | 0.087 | 0.23 | 14.63 | 38.7 | 3.65 |
| 270.9 K | 0.30 | 0.20 | 1.13 | 1.42 | 0.66 | 0.10 | 0.07 | 0.22 | 14.64 | 38.7 | 3.70 |
| 275.5 K | 0.22 | 0.15 | 1.16 | 1.50 | 0.64 | 0.12 | 0.09 | 0.24 | 14.77 | 38.7 | 3.78 |

The equilibrium concentrations (**Table S2**) were used together with Eq S42 and S45 to estimate $\Delta G^0_{2\leftarrow 3}$ and $\Delta G^0_{1\leftarrow 3}$ (**Figure S3B,C**).

**Table S3**: Fitted from temperature dependences, $\Delta H^0_{2\leftarrow 3}$ = -(39 ± 2) kJ/mol, $\Delta S^0_{2\leftarrow 3}$ = -(126 ± 9) J/(mol K) for transition **2-3** and $\Delta H^0_{1\leftarrow 3}$ = -(39 ± 2) kJ/mol $\Delta S^0_{1\leftarrow 3}$ = -(99 ± 9) J/(mol K) for transition **1-3**. Here for example $\Delta G^0_{2\leftarrow 3} = \Delta G^0_{\mathbf{2}} - \Delta G^0_{\mathbf{3}}$.

| | Experimental | |
|---|---|---|
| Temperature (K) | $\Delta G^0_{1\leftarrow 3}$ (J/mol) | $\Delta G^0_{2\leftarrow 3}$ (J/mol) |
| 235 | -15430 | -9864 |
| 247 | -14236 | -8351 |
| 259 | -13042 | -6839 |
| 260 | -12942 | -6713 |
| 265 | -12445 | -6083 |
| 267.5 | -12196 | -5767 |
| 270.9 | -11858 | -5339 |
| 275.5 | -11400 | -4759 |

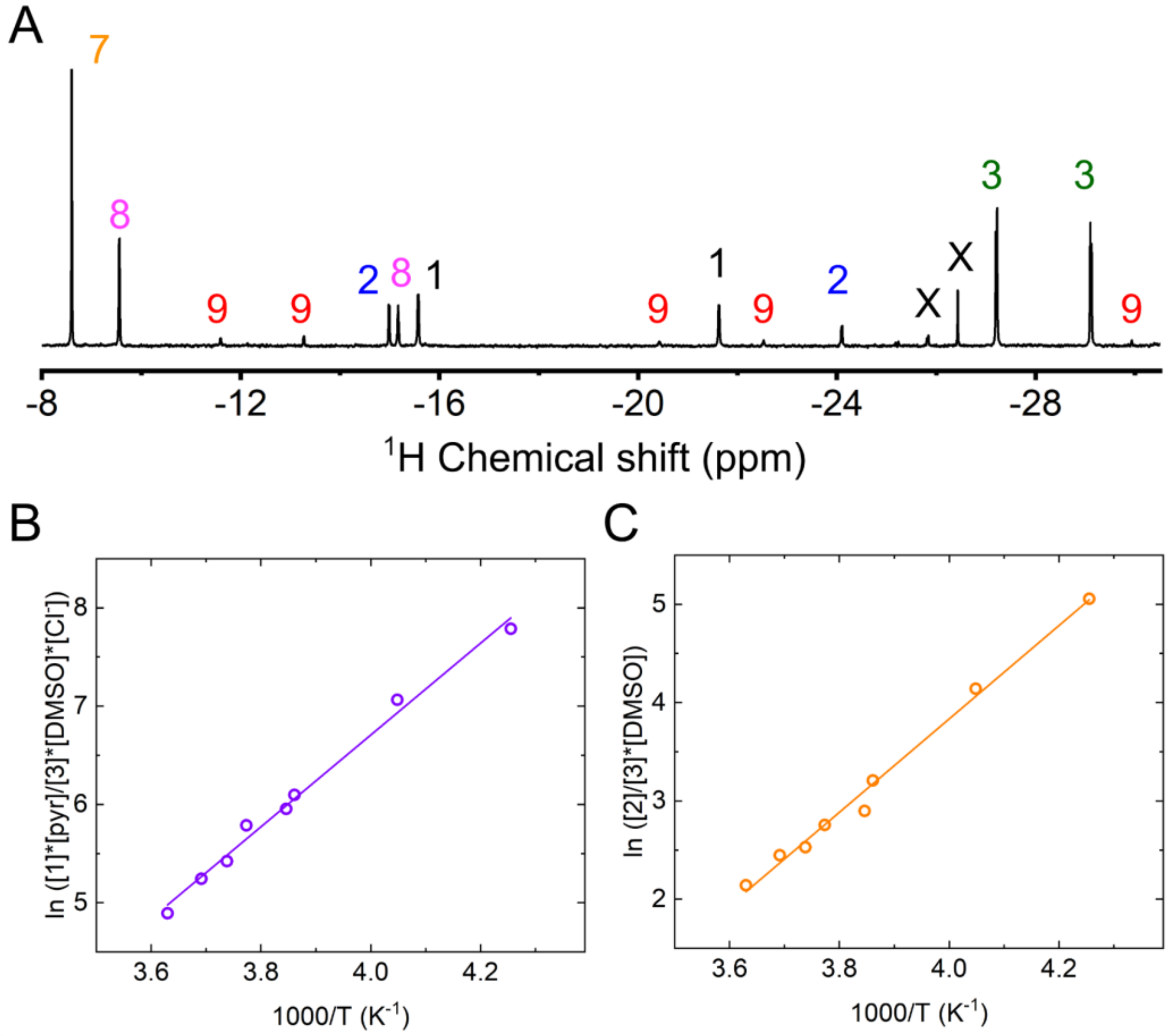

**Figure S3:** (A) $^1$H NMR spectrum of the hydride region (negative chemical shifts) for the SABRE-pyruvate system showing peak assignments corresponding to iridium complexes **1-3** and **7-9**. These complexes were assigned in (32CowleyJACS). The peaks at -25.84 ppm and -26.47 ppm have not been structurally assigned but play no role in the observed kinetic exchange profile. (B) The Van't Hoff plot of ln([1][pyr]/[3][DMSO][$Cl^-$]) versus 1000/T based on the thermodynamic equilibrium between complexes [1] and [3] (Eq. S45). The slope and intercept of the linear fit yield the enthalpy ($\Delta H^0_{1\leftarrow 3}$) and entropy ($\Delta S^0_{1\leftarrow 3}$) changes, respectively. (C) The Van't Hoff plot of ln([2]/[3][DMSO]) versus 1000/T, corresponding to the equilibrium between complexes [2] and [3] (Eq. S42). $\Delta H^0_{2\leftarrow 3}$ and $\Delta S^0_{2\leftarrow 3}$ were determined from the slope and intercept of the linear fit.

# 7. IrHH hydrogen exchange

## 7.1. SEPP-SABRE kinetics of IrHH hydrogens as a function of temperature

**Sample:** 4 mM of [Ir-$d_{22}$], 40 mM of $^{13}$C-pyr, and 20 mM of DMSO in 600 µL of methanol-$d_4$.

**Experiment:** In all the experiments below, we used a 9.4 T NMR system, $pH_2$ at 8.5 bar pressure, and a SEPP-SABRE sequence for $^1$H hyperpolarization of the selected proton. The temperature varied between 260 K and 275.5 K. We performed SEPP-SABRE on 6 hydrogens $H^a$ and $H^b$ of all 3 complexes. Consequently, 2 sets of data were measured for each complex. Data showed an exchange of hydrogens within complexes **3** and **2**. The hyperpolarized selected proton of **3** and **2** showed no exchange with the hydrogens of **1**; selection of 1 also did not show conversion to **3** and **2**.

**Simulations:** To streamline the analysis, we fitted **3** and **2** together using model (**3**↔**2**)↔$H_2$, and **1** separately using model IrHH↔$H_2$. The scripts were written in MATLAB, and global fitting was used when possible.

### 7.1.1. Complex 3 & 2

**Experiment:** For each temperature, 4 sets of data were measured. Set 1 - hyperpolarization of $H^a$-**3** at -29.1 ppm using SEPP-SABRE and following exchanging with $H^a$-**3** at -27.2 ppm, $H^a$-**2** at -14.98 ppm, $H^b$-**2** at -24.08 ppm, and free $H_2$ at 4.6 ppm (**Figure S4-A, E**). Set 2 - hyperpolarization of $H^b$-**3** using SEPP-SABRE and following exchanging with $H^a$-**3** , $H^a$-**2** , $H^b$-**2** , and $H_2$ (**Figure S4-B, F**). Set 3 - hyperpolarization of $H^a$-**2** using SEPP-SABRE and following exchanging with $H^b$-**2**, $H^a$-**3** , $H^b$-**3** , and $H_2$ (**Figure S4-C, G**). Set 4 - hyperpolarization of $H^b$-**2** using SEPP-SABRE and following exchanging with $H^a$-**2**, $H^a$-**3**, $H^b$-**3** , and $H_2$ (**Figure S4-D, H**). $R_{H2}$ was measured for different temperatures (**Table S15**).

**Simulations:**

To streamline the analysis, we employed a global fitting approach for model (**3**↔**2**)↔$H_2$, consisting of two steps of fitting, using MATLAB scripts to simultaneously fit all parameters across 4 sets of data.

Across all 4 sets and for each temperature, we share ten parameters: 2 dissociation rate constants $k_{3-\mathrm{H_2}}$, and $k_{2-\mathrm{H_2}}$, two inter-complex exchange rate constants $k_{3\to2}$, $k_{2\to3}$, two exchange constants $k_{3-\mathrm{ex}}^{\mathrm{HH}}$, and $k_{2-\mathrm{ex}}^{\mathrm{HH}}$, two association rate constants ${k'_{3\mathrm{a}}}^{\mathrm{H}}$ and ${k'_{2\mathrm{a}}}^{\mathrm{H}}$, and two distinct relaxation exchanges $R_3$, $R_2$. 2 steps of fitting were followed to get these parameters. Step 1: kinetics of $H^a$-**3**, $H^b$-**3**, and $H^a$-**2**, $H^b$-**2** were summed together having 3 kinetics within 1 set: $H^a$-**3** + $H^b$-**3**, $H^a$-**2** + $H^b$-**2**, and $H_2$. Then, a global fit was applied using Eq. S16 of model (**3**↔**2**)↔$H_2$ (Intermolecular exchange) allows us to get, as explained in **section 1.2**, 6 out of 10 parameters $k_{3\to2}$, $k_{2\to3}$, $k_{3-H_2}$, $k_{2-\mathrm{H_2}}$, ${k'_{3\mathrm{a}}}^{\mathrm{H}}$ and ${k'_{2\mathrm{a}}}^{\mathrm{H}}$. Step 2: We used 4 sets of data, each set containing 5 kinetics: $H^a$-**3**, $H^b$-**3**, $H^a$-**2**, $H^b$-**2**, and $H_2$. Then, a global fit was applied using Eq. S15 (inter and intramolecular exchange) to find the rest of the parameters $k_{3-\mathrm{ex}}^{\mathrm{HH}}$, $k_{2-\mathrm{ex}}^{\mathrm{HH}}$, $R_3$, and $R_2$. The reported errors are the results of such fits using MATLAB's nonlinear regression function “nlinfit”.

The fitting script and integrals are available in Zenodo https://doi.org/10.5281/zenodo.18449765.

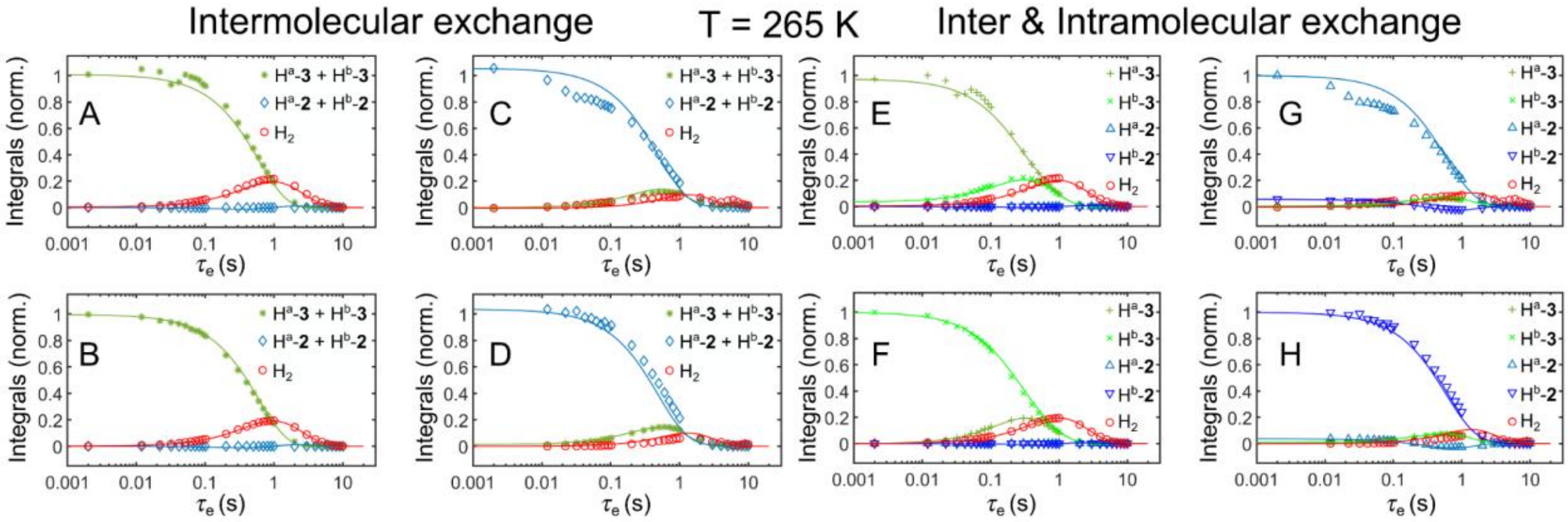

**Figure S4.265 K: Normalized integrals of 3 and 2 SEPP-SABRE kinetics at 265 K.** $^1$H NMR integrals of the hydride at -29.2 ppm ($H^a$-**3**, green plus), at -27.2 ppm ($H^b$-**3**, green cross), and 4.6 ppm ($H_2$, red circle) at 265 K. The summed integrals of **3**'s hydrogens ($H^a$-**3** + $H^b$-**3**, green star), and of **2**'s hydrogens ($H^a$-**2** + $H^b$-**2**, blue diamond) are used for the intermolecular exchange. The curves are fitted using a global fit of the model (**3**↔**2**)↔$H_2$ (A, B, C, D, Intermolecular exchange), and (E, F, G, H, Inter and Intramolecular exchange). Estimated exchange rate constants, dissociations, associations, and relaxation-exchanges are given in **Table S4.**

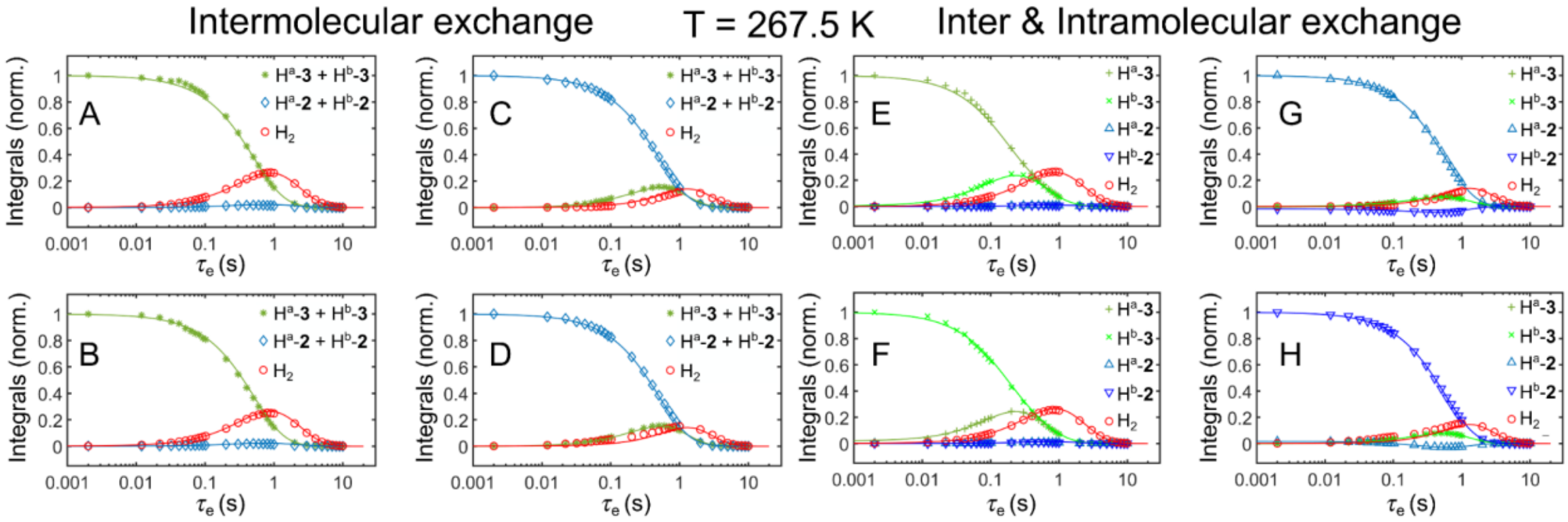

**Figure S4.267.5 K: Normalized integrals of 3 and 2 SEPP-SABRE kinetics at 267.5 K.** $^1$H NMR integrals of the hydride at -29.2 ppm ($H^a$-**3**, green plus), at -27.2 ppm ($H^b$-**3**, green cross), and 4.6 ppm ($H_2$, red circle) at 267.5 K. The summed integrals of **3**'s hydrogens ($H^a$-**3** + $H^b$-**3**, green star), and of **2**'s hydrogens ($H^a$-**2** + $H^b$-**2**, blue diamond) are used for the intermolecular exchange. The curves are fitted using a global fit the model (**3**↔**2**)↔$H_2$ (A, B, C, D, Intermolecular exchange), and (E, F, G, H, Inter and Intramolecular exchange). Estimated exchange rate constants, dissociations, associations, and relaxation-exchanges are given in **Table S4.**

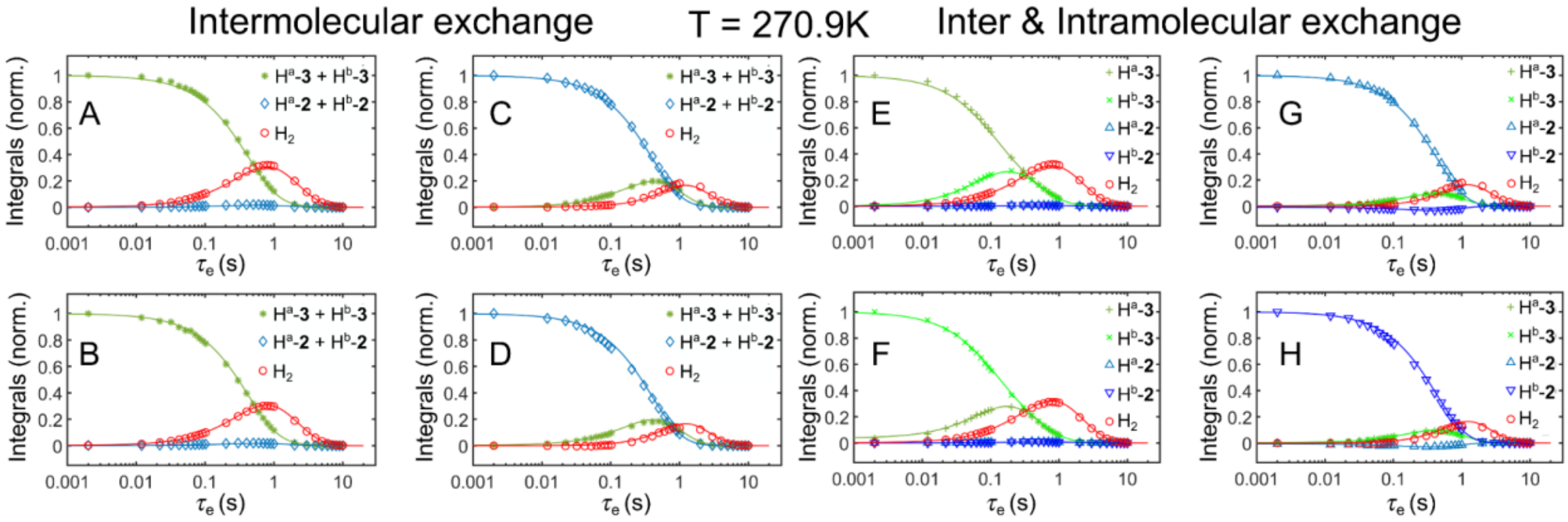

**Figure S4.270.9 K: Normalized integrals of 3 and 2 SEPP-SABRE kinetics at 270.9 K.** $^1$H NMR integrals of the hydride at -29.2 ppm ($H^a$-**3**, green plus), at -27.2 ppm ($H^b$-**3**, green cross), and 4.6 ppm ($H_2$, red circle) at 270.9 K. The summed integrals of **3**'s hydrogens ($H^a$-**3** + $H^b$-**3**, green star), and of **2**'s hydrogens ($H^a$-**2** + $H^b$-**2**, blue diamond) are used for the intermolecular exchange. The curves are fitted using a global fit of the model (**3**↔**2**)↔$H_2$ (A, B, C, D, Intermolecular exchange), and (E, F, G, H, Inter and Intramolecular exchange). Estimated exchange rate constants, dissociations, associations, and relaxation-exchanges are given in **Table S4.**

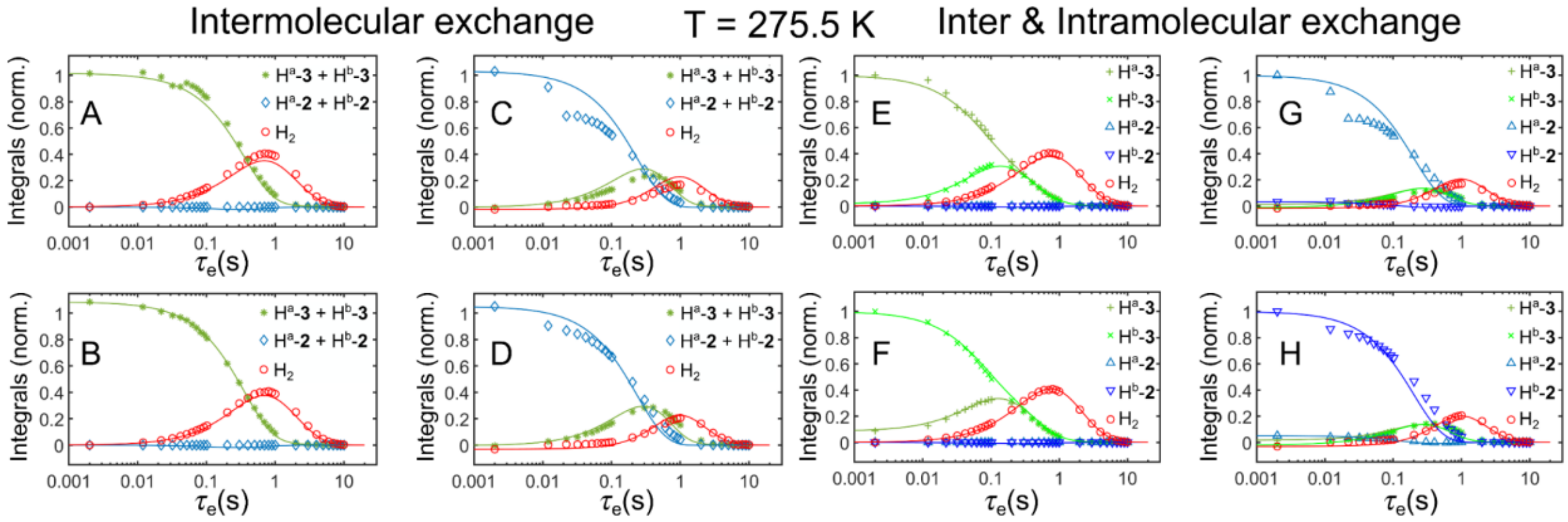

**Figure S4.275.5 K: Normalized integrals of 3 and 2 SEPP-SABRE kinetics at 275.5 K.** $^{1}$H NMR integrals of the hydride at -29.2 ppm (H$^{a}$-**3**, green plus), at -27.2 ppm (H$^{b}$-**3**, green cross), and 4.6 ppm ($H_2$, red circle) at 275.5 K. The summed integrals of **3**'s hydrogens (H$^{a}$-**3** + H$^{b}$-**3**, green star), and of **2**'s hydrogens (H$^{a}$-**2** + H$^{b}$-**2**, blue diamond) are used for the intermolecular exchange. The curves are fitted using a global fit of the model (**3**↔**2**)↔$H_2$ (A, B, C, D, Intermolecular exchange), and (E, F, G, H, Inter and Intramolecular exchange). Estimated exchange rate constants, dissociations, associations, and relaxation-exchanges are given in **Table S4.**

**Table S4.** Fitting parameters of **3** and **2** SEPP-SABRE kinetics for each temperature using a 2-step fitting process of model (**3**↔**2**)↔$H_2$.Parameters are defined in section 2.

| **Temp (K)** | $k_{3-H_2}$ (s$^{-1}$) | $k_{2-H_2}$ (s$^{-1}$) | $k_{3\to2}$ (s$^{-1}$) | $k_{2\to3}$ (s$^{-1}$) | $k_{3-\mathrm{ex}}^{\mathrm{HH}}$ (s$^{-1}$) | $k_{2-\mathrm{ex}}^{\mathrm{HH}}$ (s$^{-1}$) | $k'_{3\mathrm{a}}{}^{\mathrm{H}}$ (s$^{-1}$) | $k'_{2\mathrm{a}}{}^{\mathrm{H}}$ (s$^{-1}$) | $R_3$ (s$^{-1}$) | $R_2$ (s$^{-1}$) | $R_{\mathrm{H2}}$ (s$^{-1}$) |
|---|---|---|---|---|---|---|---|---|---|---|---|
| **265** | 0.562 ± 0.031 | 0.141 ± 0.024 | -0.091 ± 0.041 | 0.665 ± 0.033 | 1.695 ± 0.064 | -0.202 ± 0.024 | -0.162 ± 0.12 | 0.191 ± 0.115 | 1.136 ± 0.031 | 1.054 ± 0.036 | 0.518 ± 0.02 |
| **267.5** | 0.868 ± 0.012 | 0.189 ± 0.009 | 0.090 ± 0.014 | 0.798 ± 0.012 | 2.928 ± 0.029 | -0.182 ± 0.008 | 0.033 ± 0.03 | 0.094 ± 0.028 | 1.014 ± 0.010 | 0.948 ± 0.012 | 0.531 ± 0.02 |
| **270.9** | 1.148 ± 0.013 | 0.126 ± 0.010 | 0.083 ± 0.016 | 1.270 ± 0.015 | 4.401 ± 0.036 | -0.191 ± 0.010 | 0.089 ± 0.03 | 0.053 ± 0.025 | 1.065 ± 0.010 | 1.131 ± 0.014 | 0.526 ± 0.02 |
| **275.5** | 1.484 ± 0.062 | 0.196 ± 0.065 | -0.163 ± 0.089 | 2.679 ± 0.103 | 6.607 ± 0.204 | -0.219 ± 0.09 | 0.117 ± 0.14 | -0.021 ± 0.12 | 1.122 ± 0.038 | 2.330 ± 0.106 | 0.549 ± 0.02 |

*The negative values of $k_{2-\mathrm{ex}}^{\mathrm{HH}}$ are reflective of the NOE contribution dominating under limited/zero exchange.

### 7.1.2. Complex 1

**Sample:** 4 mM of [Ir-*d*$_{22}$], 40 mM of $^{13}$C-pyr, and 20 mM of DMSO in 600 μL of methanol-*d*$_4$.

**Experiment:** In all the experiments below, we used a 9.4 T NMR system, p$H_2$ at 8.5 bar pressure, and a SEPP-SABRE sequence for $^{1}$H hyperpolarization of the selected proton. For each temperature, 2 sets of data were measured. Set 1 - hyperpolarization of H$^{a}$-**1** at -15.56 ppm using SEPP-SABRE and following exchanging with H$^{b}$-**1** at -21.59 ppm, and free $H_2$ at 4.6 ppm (**Figure S4.1-top**). Set 2 - hyperpolarization of H$^{b}$-**1** using SEPP-SABRE and following exchanging with H$^{a}$-**1**, and $H_2$ (**Figure S4.1-bottom**). $R_{\mathrm{H2}}$ was measured for different temperatures (**Table S15**).

**Simulations:**

To streamline the analysis, we employed a global fitting approach for model IrHH↔$H_2$ using MATLAB scripts to fit all parameters simultaneously using 2 sets of kinetics. Across both kinetics and for each temperature, we share 5 parameters: the dissociation rate constant $k_{1-\mathrm{H}_2}$, exchange constant $k_{1-\mathrm{ex}}^{\mathrm{HH}}$, association rate constant $k'_{1\mathrm{a}}{}^{\mathrm{H}}$, natural $H_2$ relaxation exchange $R_{\mathrm{H}_2}$ and relaxation exchange $R_{\mathrm{IrHH}}$. The reported errors are the results of such fits using MATLAB's nonlinear regression function “nlinfit”. The fitting script, together with integrals, is available in Zenodo DOI: https://doi.org/10.5281/zenodo.18449765.

**Figure S4.1: Normalized integrals of 1 SEPP-SABRE kinetics.** [1]H NMR integrals of the hydride at -14.98 ppm ($H^a$-**1**, black square), at -21.59 ppm ($H^b$-**1**, grey star), and 4.6 ppm ($H_2$, red circle), at the nominal temperatures of 260 K (A, B), 265 K (C, D), and 267.5 K (E, F). The curves are fitted using the model IrHH↔$H_2$ (Inter and Intramolecular exchange). Estimated exchange rate constants, dissociations, associations, and relaxation-exchanges are given in Table S2.1.

**Table S4.1**. Fitting parameters of **1** SEPP-SABRE kinetics for each temperature using model IrHH↔$H_2$. Parameters are defined in section 2

| **Temp (K)** | $k_{1-H_2}$ ($s^{-1}$) | $k_{1-\mathrm{ex}}^{\mathrm{HH}}$ ($s^{-1}$) | ${k'_{1a}}^{\mathrm{H}}$ ($s^{-1}$) | $R_{\mathrm{IrHH}}$ ($s^{-1}$) | $R_{\mathrm{H2}}$ ($s^{-1}$) |
|---|---|---|---|---|---|
| **260** | 5.757 ± 0.226 | -0.034 ± 0.225 | 0.153 ± 0.148 | 1.644 ± 0.211 | 0.515 ± 0.02 |
| **265** | 8.530 ± 0.098 | 0.779 ± 0.099 | 1.129 ± 0.074 | 2.505 ± 0.084 | 0.518 ± 0.02 |
| **267.5** | 13.444 ± 0.119 | 0.298 ± 0.125 | 0.652 ± 0.064 | 3.083 ± 0.107 | 0.531 ± 0.02 |

*The negative value here corresponds to very low or zero flux within the error bars.

## 7.2. SEPP-SABRE kinetics of IrHH hydrogens as a function of pressure

**Sample:** 4 mM of [Ir-$d_{22}$], 40 mM of $^{13}$C-pyr, and 20 mM of DMSO in 600 µL of methanol-$d_4$.

**Experiment:** In all the experiments below, we used a 9.4 T NMR system, at 265 K, and a SEPP-SABRE sequence for $^1$H hyperpolarization of the selected proton. For each pressure, 4 sets of data were measured. Set 1 - hyperpolarization of H$^a$-**3** at -29.1 ppm using SEPP-SABRE and following exchanging with H$^b$-**3** at -27.2 ppm, Ha-2 at -14.98 ppm, H$^b$-**2** at -24.08 ppm, and free $H_2$ at 4.6 ppm (**Figure S5-A, E**). Set 2 - hyperpolarization of H$^b$-**3** using SEPP-SABRE and following exchanging with H$^a$-**3**, H$^a$-**2**, H$^b$-**2**, and $H_2$ (**Figure S5-B, F**). Set 3 - hyperpolarization of H$^a$-**2** using SEPP-SABRE and following exchanging with H$^b$-**2**, H$^a$-**3**, H$^b$-**3**, and $H_2$ (**Figure S5-C, G**). Set 4 - hyperpolarization of H$^b$-**2** using SEPP-SABRE and following exchanging with H$^a$-**2**, H$^a$-**3**, H$^b$-**3**, and $H_2$ (**Figure S5-D, H**). $R_{H2}$ was measured for different temperatures (**Table S15**).

### 7.2.1. Complex 3 & 2

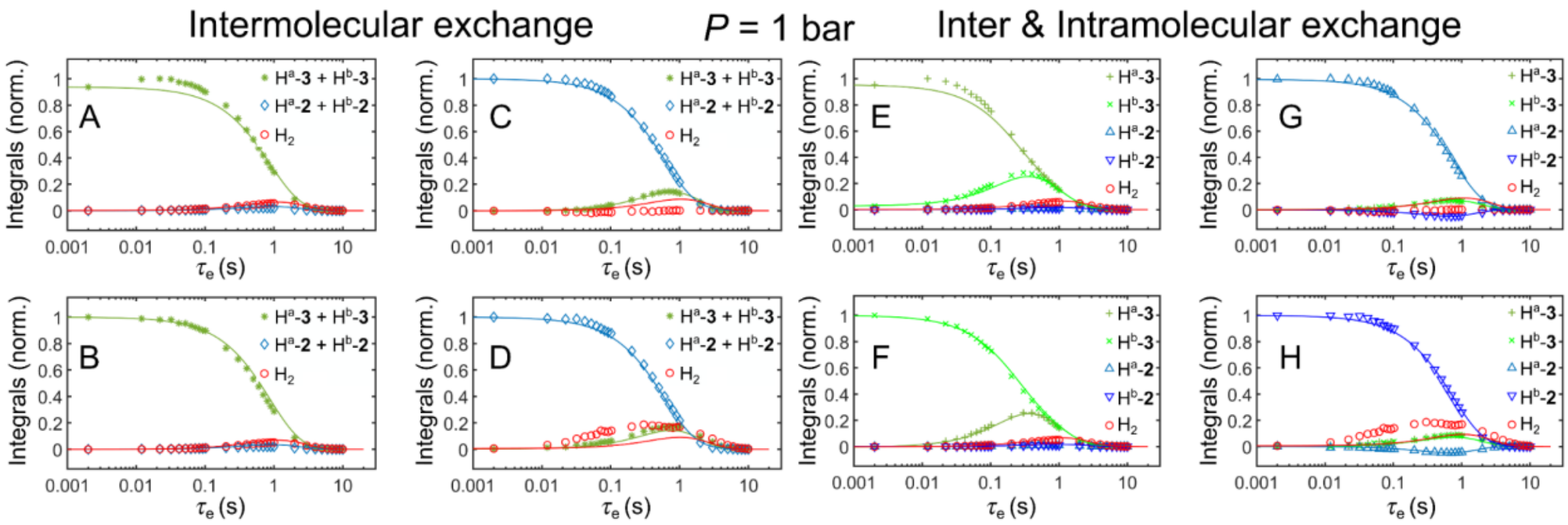

**Figure S5.1 bar: Normalized integrals of 3 and 2 SEPP-SABRE kinetics at 1 bar.** $^1$H NMR integrals of the hydride at -29.2 ppm (H$^a$-**3**, green plus), at -27.2 ppm (H$^b$-**3**, green cross), and 4.6 ppm ($H_2$, red circle) at 265 K. The summed integrals of **3**'s hydrogens (H$^a$-**3** + H$^b$-**3**, green star), and of **2**'s hydrogens (H$^a$-**2** + H$^b$-**2**, blue diamond) are used for the intermolecular exchange. The curves are fitted using a global fit of the model (**3**↔**2**)↔$H_2$ (A, B, C, D, Intermolecular exchange), and (E, F, G, H, Inter and Intramolecular exchange). Estimated exchange rate constants, dissociations, associations, and relaxation-exchanges are given in **Table S5.**

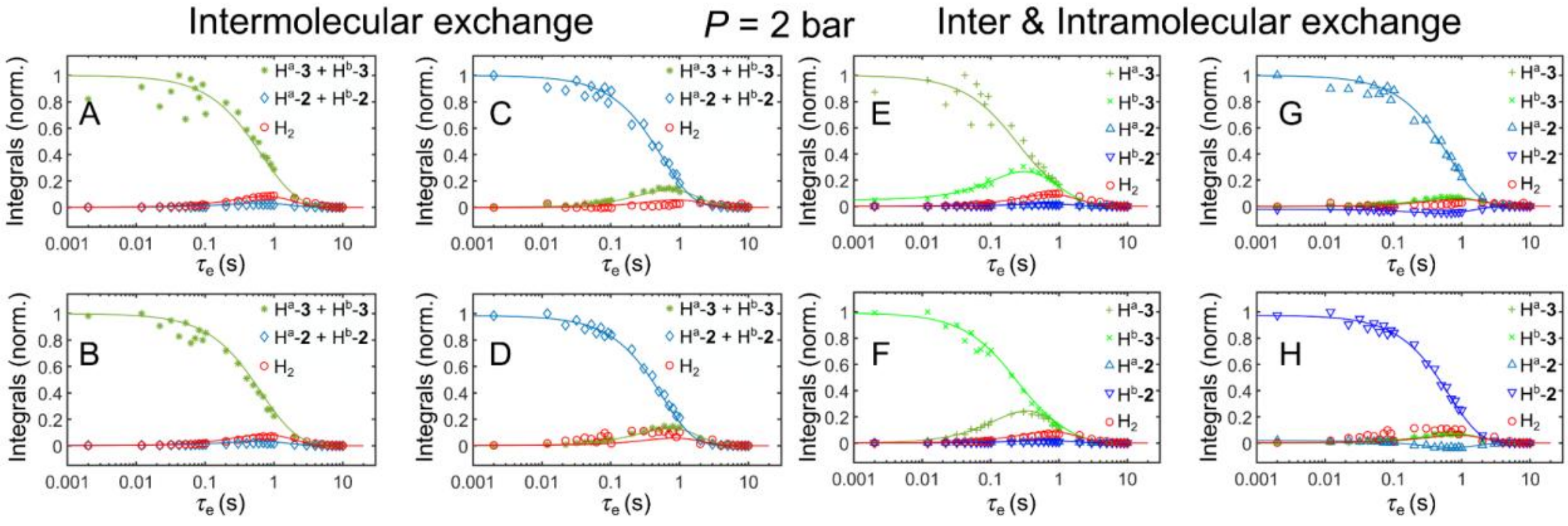

**Figure S5.2 bar: Normalized integrals of 3 and 2 SEPP-SABRE kinetics at 2 bar.** $^1$H NMR integrals of the hydride at -29.2 ppm (H$^a$-**3**, green plus), at -27.2 ppm (H$^b$-**3**, green cross), and 4.6 ppm ($H_2$, red circle) at 265 K. The summed integrals of **3**'s hydrogens (H$^a$-**3** + H$^b$-**3**, green star), and of **2**'s hydrogens (H$^a$-**2** + H$^b$-**2**, blue diamond) are used for the intermolecular exchange. The curves are fitted using a global fit of the model (**3**↔**2**)↔$H_2$ (A, B, C, D, Intermolecular exchange), and (E, F, G, H, Inter and Intramolecular exchange). Estimated exchange rate constants, dissociations, associations, and relaxation-exchanges are given in **Table S5.**

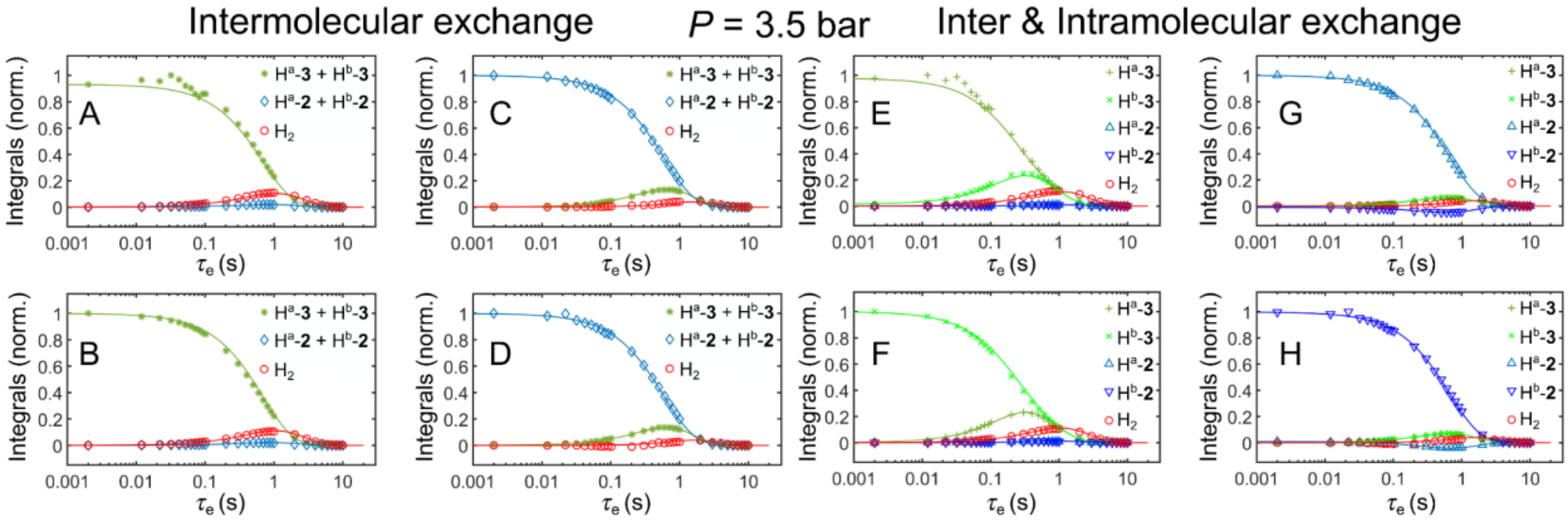

**Figure S5.3.5 bar: Normalized integrals of 3 and 2 SEPP-SABRE kinetics at 3.5 bar.** $^{1}$H NMR integrals of the hydride at -29.2 ppm ($H^a$-3, green plus), at -27.2 ppm ($H^b$-3, green cross), and 4.6 ppm ($H_2$, red circle) at 265 K. The summed integrals of **3**'s hydrogens ($H^a$-**3** + $H^b$-**3**, green star), and of **2**'s hydrogens ($H^a$-**2** + $H^b$-**2**, blue diamond) are used for the intermolecular exchange. The curves are fitted using a global fit of the model (**3**↔**2**)↔$H_2$ (A, B, C, D, Intermolecular exchange), and (E, F, G, H, Inter and Intramolecular exchange). Estimated exchange rate constants, dissociations, associations, and relaxation-exchanges are given in **Table S5.**

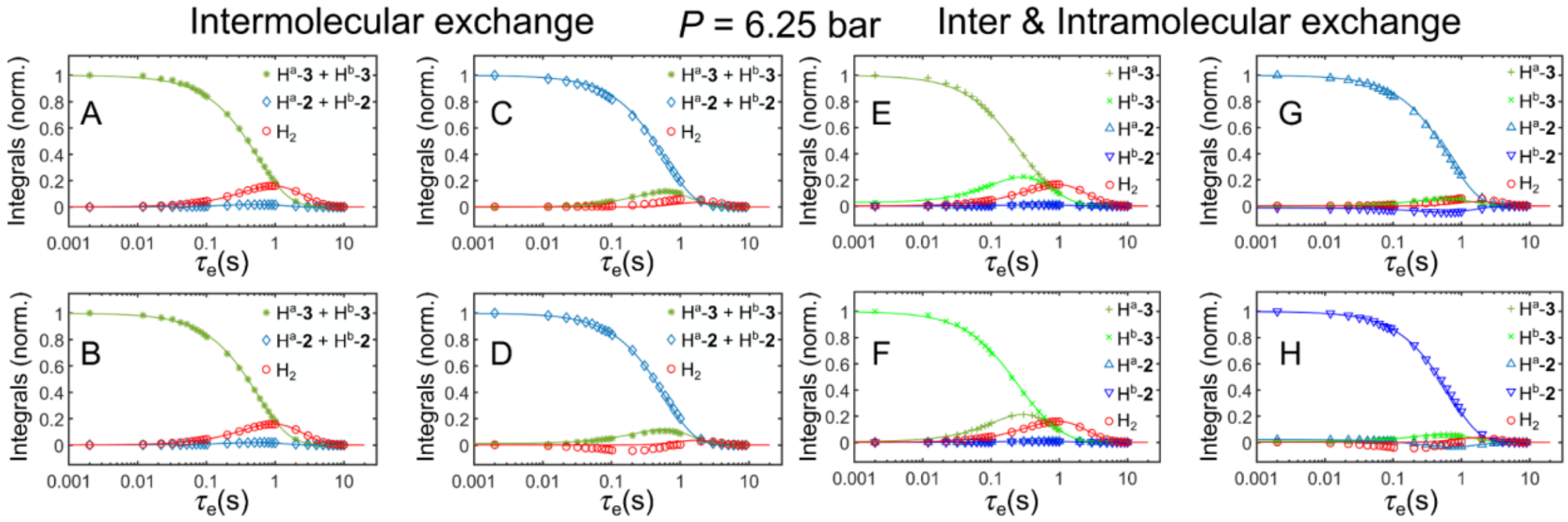

**Figure S5.6.25 bar: Normalized integrals of 3 and 2 SEPP-SABRE kinetics at 6.25 bar.** $^{1}$H NMR integrals of the hydride at -29.2 ppm ($H^a$-**3**, green plus), at -27.2 ppm ($H^b$-**3**, green cross), and 4.6 ppm ($H_2$, red circle) at 265 K. The summed integrals of **3**'s hydrogens ($H^a$-**3** + $H^b$-**3**, green star), and of **2**'s hydrogens ($H^a$-**2** + $H^b$-**2**, blue diamond) are used for the intermolecular exchange. The curves are fitted using a global fit of the model (**3**↔**2**)↔$H_2$ (A, B, C, D, Intermolecular exchange), and (E, F, G, H, Inter and Intramolecular exchange). Estimated exchange rate constants, dissociations, associations, and relaxation-exchanges are given in **Table S5.**

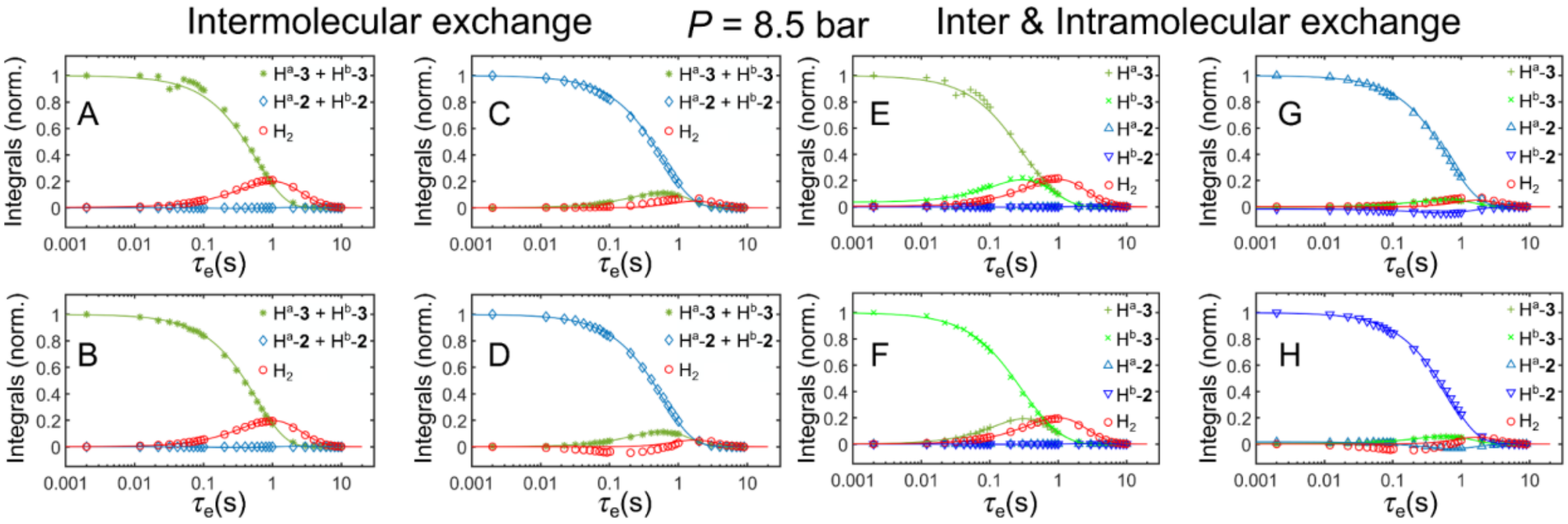

**Figure S5.8.5 bar: Normalized integrals of 3 and 2 SEPP-SABRE kinetics at 8.5 bar.** $^{1}$H NMR integrals of the hydride at -29.2 ppm ($H^a$-**3**, green plus), at -27.2 ppm ($H^b$-**3**, green cross), and 4.6 ppm ($H_2$, red circle) at 265 K. The summed integrals of **3**'s hydrogens ($H^a$-**3** + $H^b$-**3**, green star), and of **2**'s hydrogens ($H^a$-**2** + $H^b$-**2**, blue diamond) are used for the intermolecular exchange. The curves are fitted using a global fit of the model (**3**↔**2**)↔$H_2$ (A, B, C, D, Intermolecular exchange), and (E, F, G, H, Inter and Intramolecular exchange). Estimated exchange rate constants, dissociations, associations, and relaxation-exchanges are given in **Table S5.**

**Table S5**. Fitting parameters of **3** and **2** SEPP-SABRE kinetics for each pressure using 2 steps fitting process of model (**3**↔**2**)↔$H_2$. Parameters are defined in section 2.

| Press (bar) | $k_{3-H_2}$ (s$^{-1}$) | $k_{2-H_2}$ (s$^{-1}$) | $k_{3\to 2}$ (s$^{-1}$) | $k_{2\to 3}$ (s$^{-1}$) | $k_{3-\mathrm{ex}}^{\mathrm{HH}}$ (s$^{-1}$) | $k_{2-\mathrm{ex}}^{\mathrm{HH}}$ (s$^{-1}$) | $k_{3\mathrm{a}}'^{\ \mathrm{H}}$ (s$^{-1}$) | $k_{2\mathrm{a}}'^{\ \mathrm{H}}$ (s$^{-1}$) | $R_3$ (s$^{-1}$) | $R_2$ (s$^{-1}$) | $R_{\mathrm{H2}}$ (s$^{-1}$) |
|---|---|---|---|---|---|---|---|---|---|---|---|
| **1** | 0.175 ± 0.026 | 0.234 ± 0.030 | 0.134 ± 0.031 | 0.497 ± 0.035 | 2.109 ± 0.067 | -0.188 ± 0.020 | 0.218 ± 0.232 | 0.105 ± 0.247 | 0.942 ± 0.025 | 0.790 ± 0.030 | 0.518 ± 0.02 |
| **2** | 0.319 ± 0.055 | 0.181 ± 0.042 | 0.147 ± 0.053 | 0.509 ± 0.046 | 2.230 ± 0.078 | -0.184 ± 0.024 | 0.997 ± 0.466 | -0.053 ± 0.403 | 0.969 ± 0.028 | 1.000 ± 0.037 | 0.518 ± 0.02 |
| **3.5** | 0.317 ± 0.008 | 0.014 ± 0.006 | 0.119 ± 0.010 | 0.569 ± 0.008 | 2.088 ± 0.024 | -0.191 ± 0.008 | 0.202 ± 0.061 | -0.055 ± 0.054 | 1.063 ± 0.010 | 1.094 ± 0.012 | 0.518 ± 0.02 |
| **6.25** | 0.493 ± 0.087 | -0.032 ± 0.060 | 0.095 ± 0.107 | 0.539 ± 0.085 | 2.079 ± 0.025 | -0.177 ± 0.008 | 0.139 ± 0.448 | -0.032 ± 0.363 | 1.161 ± 0.011 | 1.177 ± 0.012 | 0.518 ± 0.02 |
| **8.5** | 0.578 ± 0.082 | -0.010 ± 0.059 | -0.032 ± 0.101 | 0.532 ± 0.084 | 1.710 ± 0.029 | -0.163 ± 0.010 | -0.049 ± 0.333 | 0.060 ± 0.269 | 1.139 ± 0.014 | 1.178 ± 0.016 | 0.518 ± 0.02 |

### 7.2.2. Complex 1

**Sample:** 4 mM of [Ir-*d*22], 40 mM of $^{13}$C-pyr, and 20 mM of DMSO in 600 µL of methanol-*d*4.

**Experiment:** In all the experiments below, we used a 9.4 T NMR system, at 265 K, and a SEPP-SABRE sequence for $^{1}$H hyperpolarization of the selected proton. For each pressure, 2 sets of data were measured. Set 1 - hyperpolarization of H$^{a}$-**1** at -15.56 ppm using SEPP-SABRE and following exchanging with H$^{b}$-**1** at -21.59 ppm, and free $H_2$ at 4.6 ppm (**Figure S5.1-top**). Set 2 - hyperpolarization of H$^{b}$-**1** using SEPP-SABRE and following exchange with H$^{a}$-**1**, and $H_2$ (**Figure S5.1-bottom**). $R_{\mathrm{H2}}$ was measured for different temperatures (**Table S15**).

**Simulations:**

To streamline the analysis, we employed a global fitting approach for model IrHH↔$H_2$ using MATLAB scripts to fit all parameters simultaneously using 2 sets of kinetics. Across both kinetics and for each temperature, we share 5 parameters: the dissociation rate constant $k_{1-\mathrm{H}_2}$, exchange constant $k_{1-\mathrm{ex}}^{\mathrm{HH}}$, association rate constant $k_{1\mathrm{a}}'^{\ \mathrm{H}}$, natural $H_2$ relaxation exchange $R_{\mathrm{H}_2}$ and relaxation exchange $R_{\mathbf{IrHH}}$. The reported errors are the results of such fits using MATLAB's nonlinear regression function “nlinfit”. The fitting script, together with integrals, is available in Zenodo DOI: https://doi.org/10.5281/zenodo.18449765.

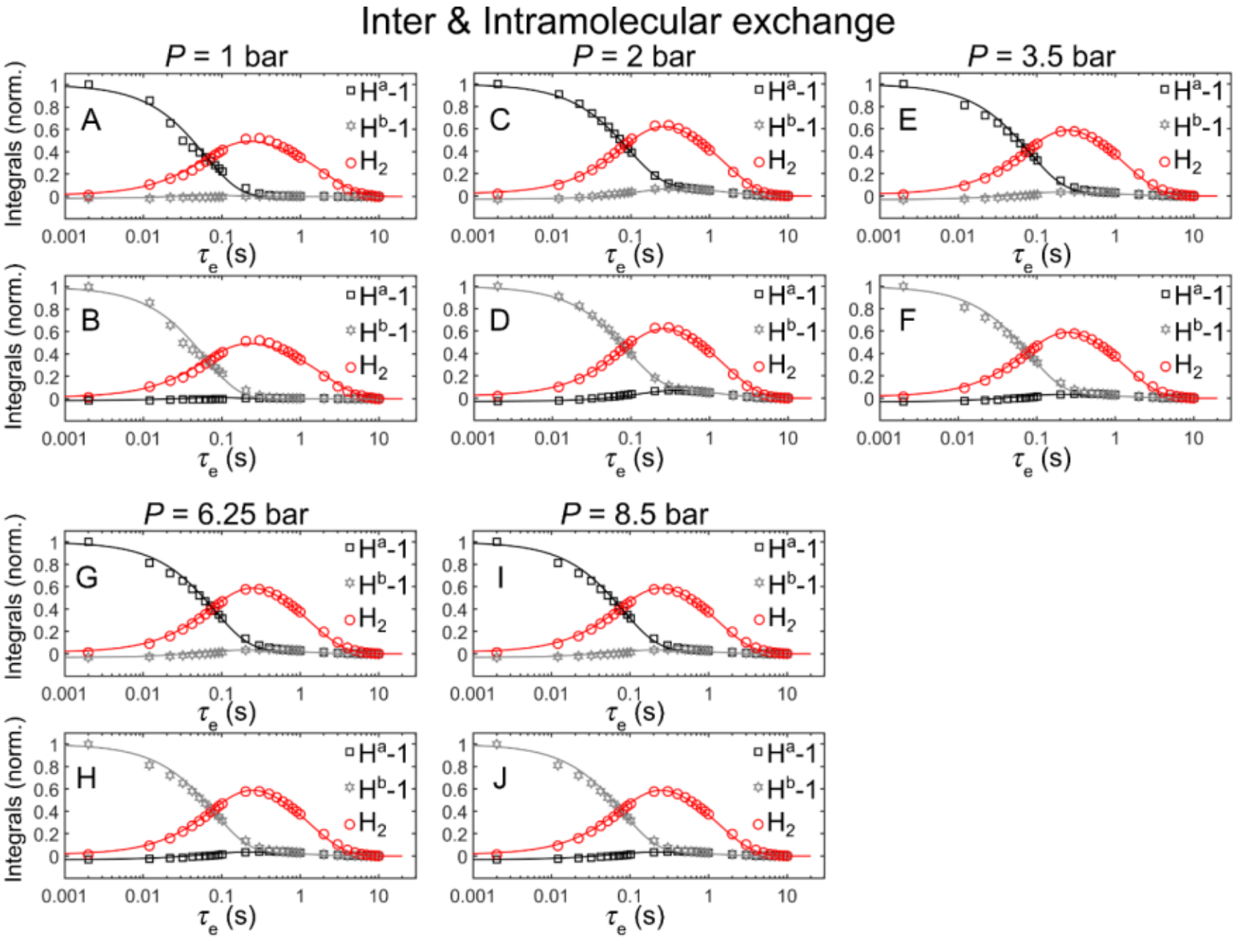

**Figure S5.1: Normalized integrals of SEPP-SABRE kinetics**. $^{1}$H NMR integrals of the hydride at -14.98 ppm (H$^{a}$-**1**, black square), at -21.59 ppm (H$^{b}$-**1**, grey star), and 4.6 ppm ($H_2$, red circle), at the pressure of 1 bar (A, B),

2 bar (C, D), 3.5 bar (E, F), 6.25 bar (G, H), and 8.5 bar (I, J). The curves are fitted using the model IrHH↔$H_2$ (Inter and Intramolecular exchange). Estimated exchange rate constants, dissociations, associations, and relaxation-exchanges are given in **Table S5.1**.

**Table S5.1**. Fitting parameters of **1** SEPP-SABRE kinetics for each pressure using model IrHH↔$H_2$. Parameters are defined in section 2.

| **Back pressure**<br>**Bubbling pressure** | $k_{1-H_2}$ ($s^{-1}$) | $k_{1-\text{ex}}^{\text{HH}}$ ($s^{-1}$) | $k'_{1\text{a}}{}^{\text{H}}$ ($s^{-1}$) | $R_{\text{IrHH}}$ ($s^{-1}$) | $R_{\text{H2}}$ ($s^{-1}$) |
|---|---|---|---|---|---|
| **5 PSI**<br>**1 bar** | 8.578 ± 0.147 | 0.680 ± 0.218 | -0.066 ± 0.041 | 7.089 ± 0.202 | 0.518 ± 0.02 |
| **20 PSI**<br>**2 bar** | 8.634 ± 0.053 | 0.378 ± 0.044 | 2.198 ± 0.045 | 0.979 ± 0.021 | 0.518 ± 0.02 |
| **40 PSI**<br>**3.5 bar** | 8.579 ± 0.101 | 0.790 ± 0.102 | 1.119 ± 0.077 | 2.518 ± 0.087 | 0.518 ± 0.02 |
| **75 PSI**<br>**6.25 bar** | 9.216 ± 0.090 | 0.347 ± 0.103 | 0.305 ± 0.047 | 3.217 ± 0.094 | 0.518 ± 0.02 |
| **100 PSI**<br>**8.5 bar** | 9.317 ± 0.096 | 0.353 ± 0.104 | 0.339 ± 0.052 | 2.848 ± 0.096 | 0.518 ± 0.02 |

## 7.3. SEPP-SABRE kinetics of IrHH hydrogens as a function of [DMSO]

**Sample:** 4 mM of [Ir-$d_{22}$], 40 mM of $^{13}$C-pyr, and a variable amount of DMSO between 20 and 300 mM in 600 µL of methanol-$d_4$.

**Experiment:** In all the experiments below, we used a 9.4 T NMR system, p$H_2$ at 8.5 bar pressure, 265 K, and a SEPP-SABRE sequence for $^1$H hyperpolarization of the selected proton. For each concentration, 4 sets of data were measured. Set 1 - hyperpolarization of $H^a$-**3** at -29.1 ppm using SEPP-SABRE and following exchanging with $H^b$-**3** at -27.2 ppm, $H^a$-**2** at -14.98 ppm, $H^b$-**2** at -24.08 ppm, and free $H_2$ at 4.6 ppm (**Figure S6-A, E**). Set 2 - hyperpolarization of $H^b$-**3** using SEPP-SABRE and following exchanging with $H^a$-**3**, $H^a$-**2**, $H^b$-**2**, and $H_2$ (**Figure S6-B, F**). Set 3 - hyperpolarization of $H^a$-**2** using SEPP-SABRE and following exchanging with $H^b$-**2**, $H^a$-**3**, $H^b$-**3**, and $H_2$ (**Figure S6-C, G**). Set 4 - hyperpolarization of $H^b$-2 using SEPP-SABRE and following exchanging with $H^a$-**2**, $H^a$-**3**, $H^b$-**3**, and $H_2$ (**Figure S6-D, H**). $R_{H2}$ was measured for different temperatures (**Table S15**).

### 7.3.1. Complex 3 & 2

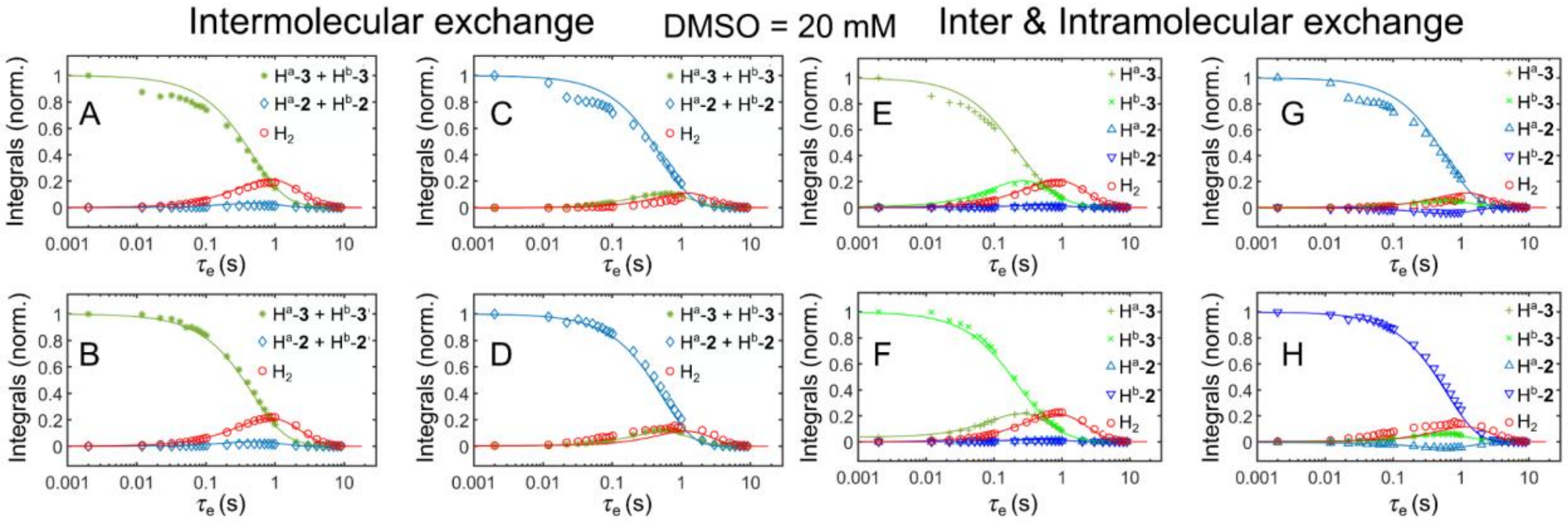

**Figure S6.20 mM: Normalized integrals of 3 and 2 SEPP-SABRE kinetics for 20 mM.** $^1$H NMR integrals of the hydride at -29.2 ppm (Ha-**3**, green plus), at -27.2 ppm ($H^b$-**3**, green cross), and 4.6 ppm ($H_2$, red circle) at 265 K. The summed integrals of **3**'s hydrogens ($H^a$-**3** + $H^b$-**3**, green star), and of **2**'s hydrogens ($H^a$-**2** + $H^b$-**2**, blue diamond) are used for the intermolecular exchange. The curves are fitted using a global fit of the model (**3**↔**2**)↔$H_2$ (A, B, C, D, Intermolecular exchange), and (E, F, G, H, Inter and Intramolecular exchange). Estimated exchange rate constants, dissociations, associations, and relaxation-exchanges are given in **Table S6.**

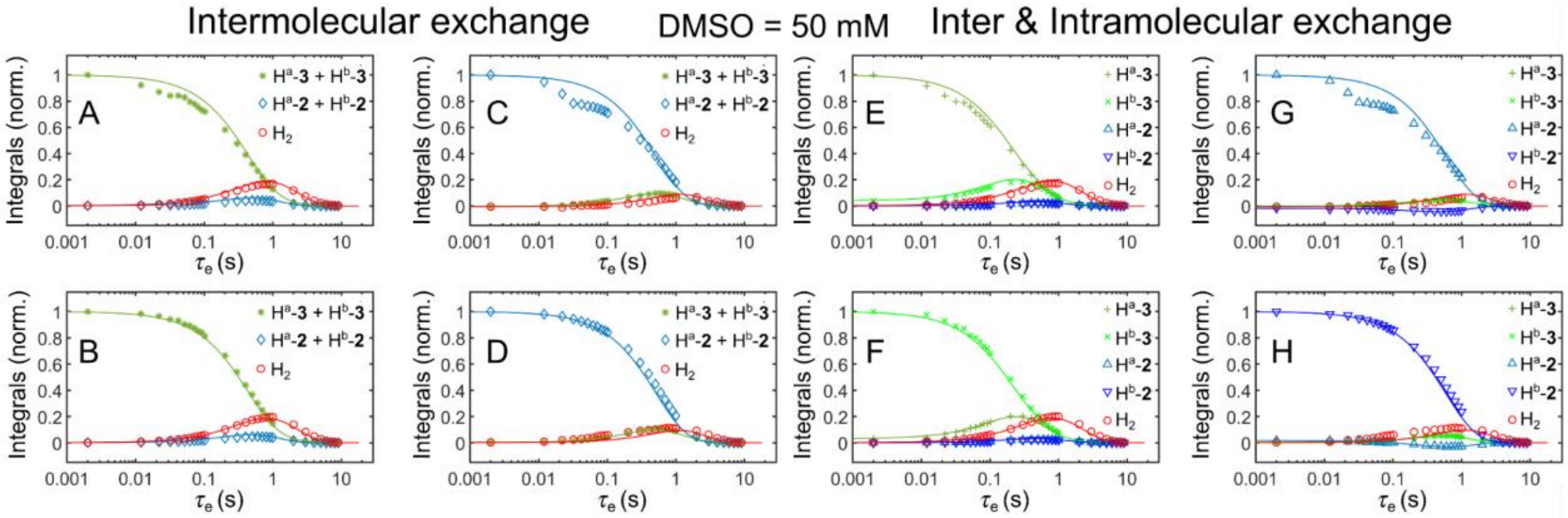

**Figure S6.50 mM: Normalized integrals of 3 and 2 SEPP-SABRE kinetics for 50 mM.** $^1$H NMR integrals of the hydride at -29.2 ppm ($H^a$-**3**, green plus), at -27.2 ppm ($H^b$-**3**, green cross), and 4.6 ppm ($H_2$, red circle) at 265 K. The summed integrals of **3**'s hydrogens ($H^a$-**3** + $H^b$-**3**, green star), and of **2**'s hydrogens ($H^a$-**2** + $H^b$-**2**, blue diamond) are used for the intermolecular exchange. The curves are fitted using a global fit of the model (**3**↔**2**)↔$H_2$ (A, B, C, D, Intermolecular exchange), and (E, F, G, H, Inter and Intramolecular exchange). Estimated exchange rate constants, dissociations, associations, and relaxation-exchanges are given in **Table S6.**

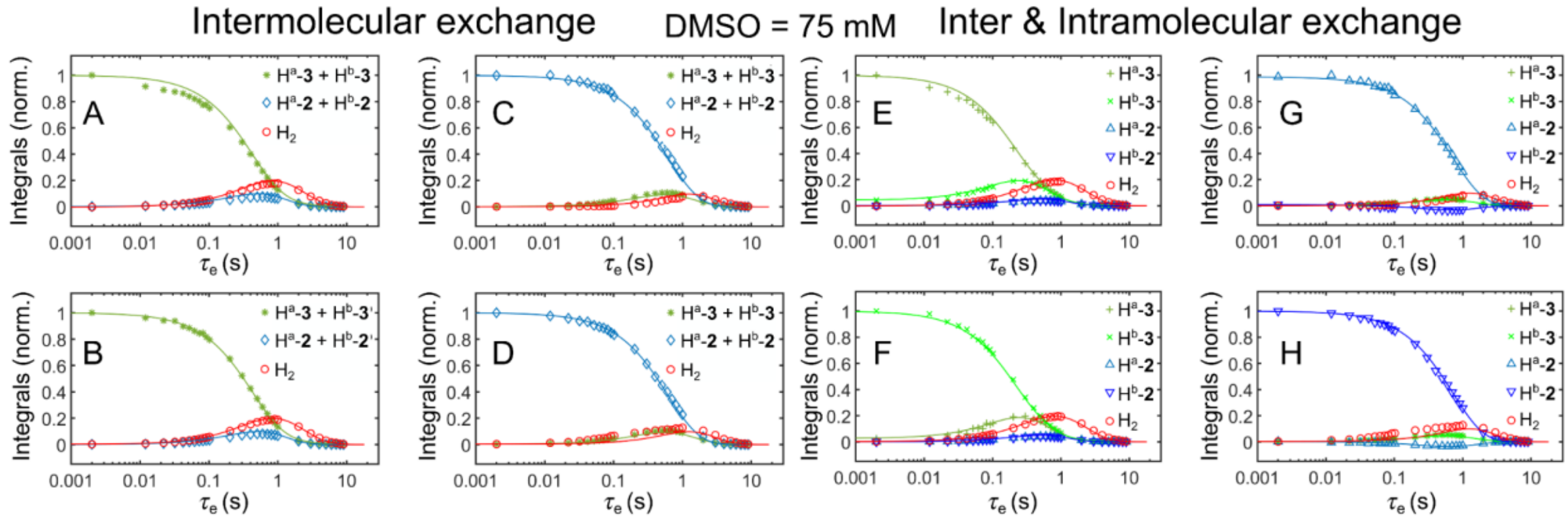

**Figure S6.75 mM: Normalized integrals of 3 and 2 SEPP-SABRE kinetics for 75 mM.** $^{1}$H NMR integrals of the hydride at -29.2 ppm ($H^a$-**3**, green plus), at -27.2 ppm ($H^b$-**3**, green cross), and 4.6 ppm ($H_2$, red circle) at 265 K. The summed integrals of **3**'s hydrogens ($H^a$-**3** + $H^b$-**3**, green star), and of **2**'s hydrogens ($H^a$-**2** + $H^b$-**2**, blue diamond) are used for the intermolecular exchange. The curves are fitted using a global fit of the model (**3**↔**2**)↔$H_2$ (A, B, C, D, Intermolecular exchange), and (E, F, G, H, Inter and Intramolecular exchange). Estimated exchange rate constants, dissociations, associations, and relaxation-exchanges are given in **Table S6.**

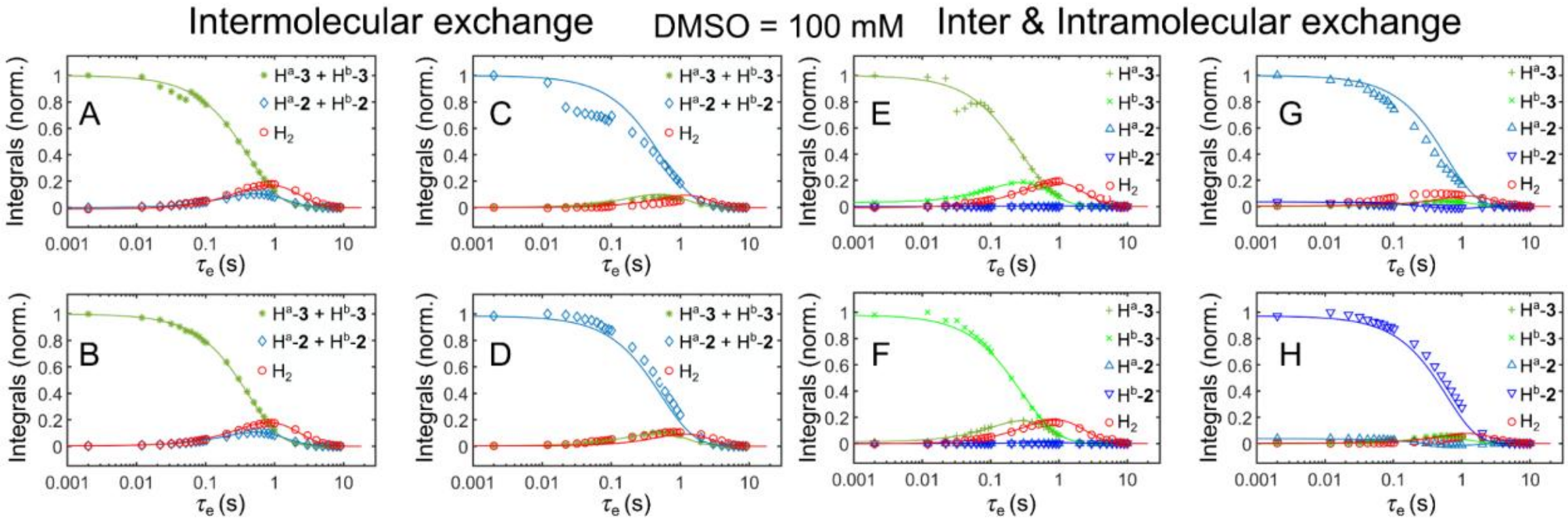

**Figure S6.100 mM: Normalized integrals of 3 and 2 SEPP-SABRE kinetics for 100 mM.** $^{1}$H NMR integrals of the hydride at -29.2 ppm ($H^a$-**3**, green plus), at -27.2 ppm ($H^b$-**3**, green cross), and 4.6 ppm ($H_2$, red circle) at 265 K. The summed integrals of **3**'s hydrogens ($H^a$-**3** + $H^b$-**3**, green star), and of **2**'s hydrogens ($H^a$-**2** + $H^b$-**2**, blue diamond) are used for the intermolecular exchange. The curves are fitted using a global fit of the model (**3**↔**2**)↔$H_2$ (A, B, C, D, Intermolecular exchange), and (E, F, G, H, Inter and Intramolecular exchange). Estimated exchange rate constants, dissociations, associations, and relaxation-exchanges are given in **Table S6.**

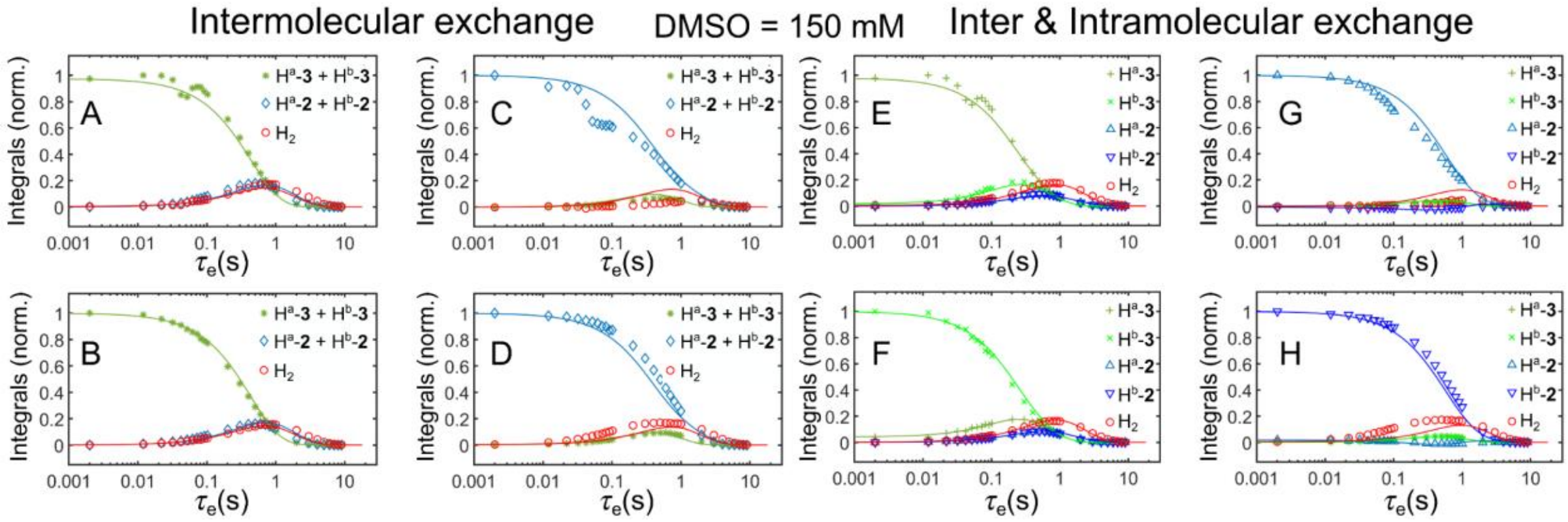

**Figure S6.150 mM: Normalized integrals of 3 and 2 SEPP-SABRE kinetics for 150 mM.** $^{1}$H NMR integrals of the hydride at -29.2 ppm ($H^a$-**3**, green plus), at -27.2 ppm ($H^b$-**3**, green cross), and 4.6 ppm ($H_2$, red circle) at 265 K. The summed integrals of **3**'s hydrogens ($H^a$-**3** + $H^b$-**3**, green star), and of **2**'s hydrogens ($H^a$-**2** + $H^b$-**2**, blue diamond) are used for the intermolecular exchange. The curves are fitted using a global fit of the model (**3**↔**2**)↔$H_2$ (A, B, C, D, Intermolecular exchange), and (E, F, G, H, Inter and Intramolecular exchange). Estimated exchange rate constants, dissociations, associations, and relaxation-exchanges are given in **Table S6.**

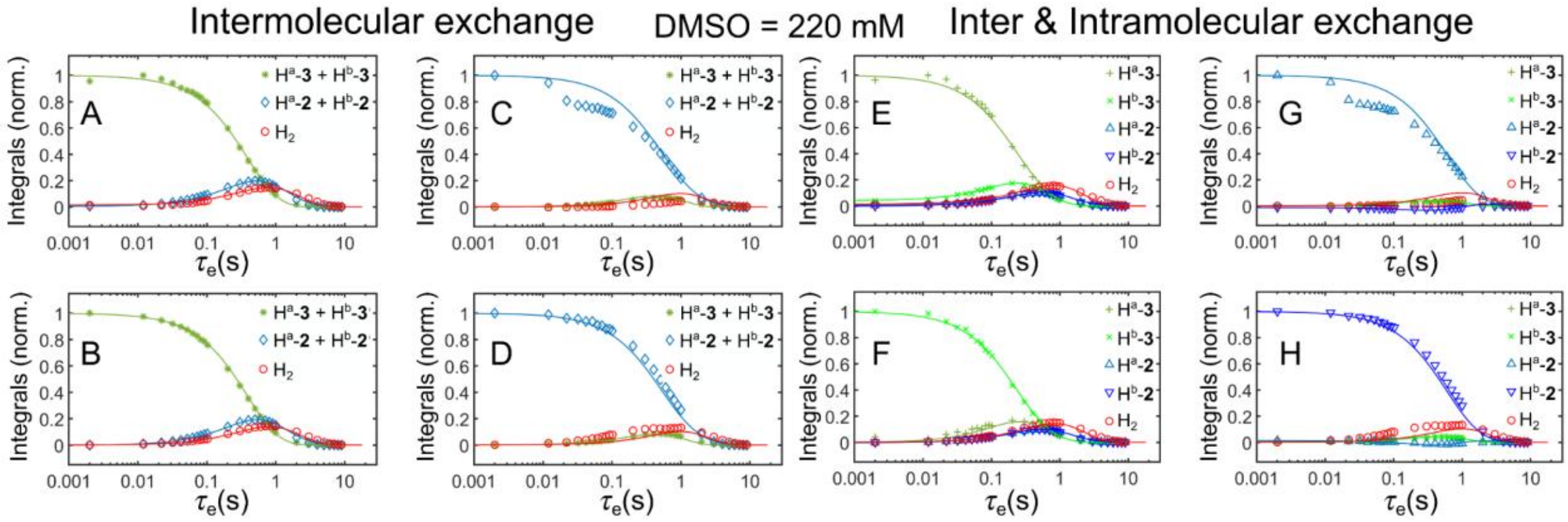

**Figure S6.220 mM: Normalized integrals of 3 and 2 SEPP-SABRE kinetics for 220 mM.** $^1$H NMR integrals of the hydride at -29.2 ppm ($H^a$-**3**, green plus), at -27.2 ppm ($H^b$-**3**, green cross), and 4.6 ppm ($H_2$, red circle) at 265 K. The summed integrals of **3**'s hydrogens ($H^a$-**3** + $H^b$-**3**, green star), and of **2**'s hydrogens ($H^a$-**2** + $H^b$-**2**, blue diamond) are used for the intermolecular exchange. The curves are fitted using a global fit of the model (**3**↔**2**)↔$H_2$ (A, B, C, D, Intermolecular exchange), and (E, F, G, H, Inter and Intramolecular exchange). Estimated exchange rate constants, dissociations, associations, and relaxation-exchanges are given in **Table S6.**

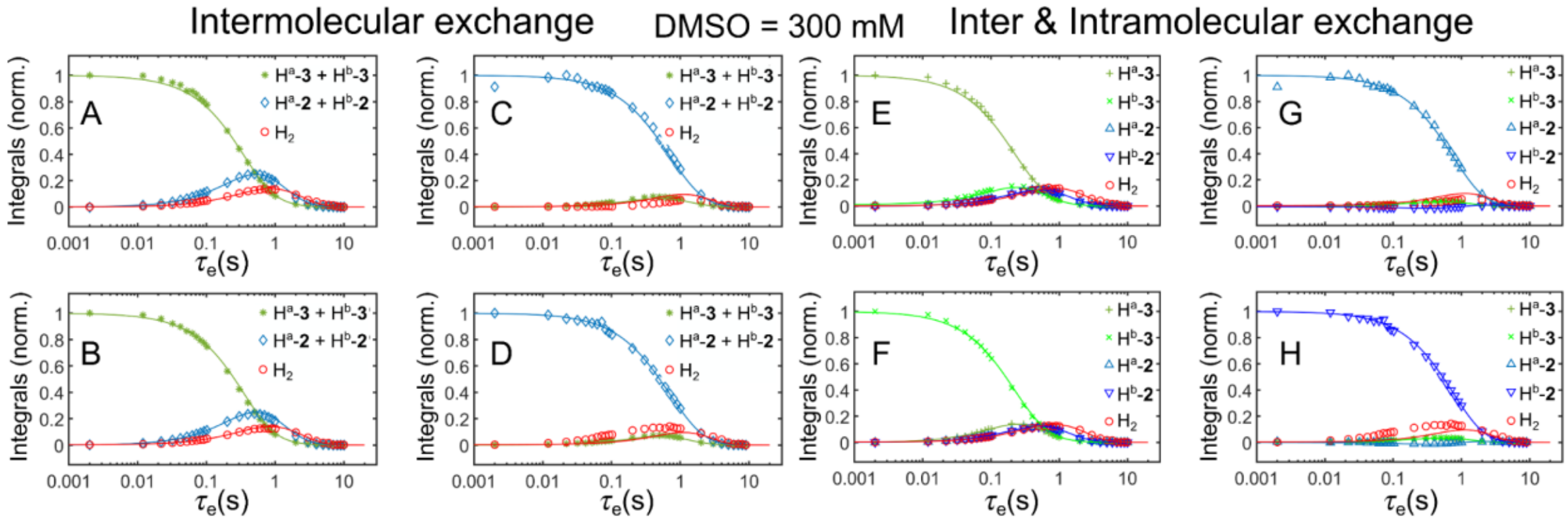

**Figure S6.300 mM: Normalized integrals of 3 and 2 SEPP-SABRE kinetics for 300 mM.** $^1$H NMR integrals of the hydride at -29.2 ppm ($H^a$-**3**, green plus), at -27.2 ppm ($H^b$-**3**, green cross), and 4.6 ppm ($H_2$, red circle) at 265 K. The summed integrals of **3**'s hydrogens ($H^a$-**3** + $H^b$-**3**, green star), and of **2**'s hydrogens ($H^a$-**2** + $H^b$-**2**, blue diamond) are used for the intermolecular exchange. The curves are fitted using a global fit of the model (**3**↔**2**)↔$H_2$ (A, B, C, D, Intermolecular exchange), and (E, F, G, H, Inter and Intramolecular exchange). Estimated exchange rate constants, dissociations, associations, and relaxation-exchanges are given in **Table S6.**

**Table S6**. Fitting parameters of **3** and **2** SEPP-SABRE kinetics for each concentration using a 2-step fitting process of model (**3**↔**2**)↔$H_2$. Parameters are defined in section 2.

| **[DMSO] (mM)** | $k_{3-H}$ (s$^{-1}$) | $k_{2-H_2}$ (s$^{-1}$) | $k_{3\to2}$ (s$^{-1}$) | $k_{2\to3}$ (s$^{-1}$) | $k_{3-\mathrm{ex}}^{\mathrm{HH}}$ (s$^{-1}$) | $k_{2-\mathrm{ex}}^{\mathrm{HH}}$ (s$^{-1}$) | $k'_{3\mathrm{a}}{}^{\mathrm{H}}$ (s$^{-1}$) | $k'_{2\mathrm{a}}{}^{\mathrm{H}}$ (s$^{-1}$) | $R_3$ (s$^{-1}$) | $R_2$ (s$^{-1}$) | $R_{\mathrm{H2}}$ (s$^{-1}$) |
|---|---|---|---|---|---|---|---|---|---|---|---|
| **20** | 0.726 ± 0.04 | 0.220 ± 0.028 | 0.185 ± 0.047 | 0.540 ± 0.039 | 2.248 ± 0.080 | -0.209 ± 0.023 | -0.162 ± 0.12 | -0.046 ± 0.114 | 1.168 ± 0.036 | 1.056 ± 0.036 | 0.518 ± 0.02 |
| **50** | 0.678 ± 0.04 | 0.186 ± 0.029 | 0.344 ± 0.051 | 0.527 ± 0.041 | 2.115 ± 0.081 | -0.189 ± 0.024 | 0.248 ± 0.163 | -0.023 ± 0.137 | 1.287 ± 0.040 | 1.161 ± 0.037 | 0.518 ± 0.02 |
| **75** | 0.701 ± 0.02 | 0.167 ± 0.014 | 0.525 ± 0.026 | 0.463 ± 0.021 | 1.949 ± 0.039 | -0.170 ± 0.010 | 0.280 ± 0.081 | -0.077 ± 0.065 | 1.154 ± 0.020 | 0.951 ± 0.015 | 0.518 ± 0.02 |
| **100** | 0.634 ± 0.05 | 0.172 ± 0.036 | 0.552 ± 0.066 | 0.559 ± 0.054 | 1.837 ± 0.103 | -0.203 ± 0.033 | 0.021 ± 0.15 | 0.190 ± 0.190 | 1.103 ± 0.055 | 1.153 ± 0.050 | 0.518 ± 0.02 |
| **150** | 0.540 ± 0.04 | 0.345 ± 0.034 | 0.799 ± 0.056 | 0.409 ± 0.047 | 1.631 ± 0.077 | -0.178 ± 0.025 | -0.298 ± 0.181 | 0.276 ± 0.106 | 0.938 ± 0.045 | 1.044 ± 0.036 | 0.518 ± 0.02 |
| **220** | 0.513 ± 0.05 | 0.262 ± 0.034 | 1.025 ± 0.062 | 0.534 ± 0.049 | 1.738 ± 0.085 | -0.192 ± 0.025 | -0.433 ± 0.224 | 0.758 ± 0.219 | 0.963 ± 0.051 | 1.067 ± 0.037 | 0.518 ± 0.02 |
| **300** | 0.498 ± 0.02 | 0.213 ± 0.016 | 1.350 ± 0.031 | 0.362 ± 0.025 | 1.586 ± 0.043 | -0.103 ± 0.011 | 0.039 ± 0.127 | 0.211 ± 0.112 | 1.007 ± 0.029 | 0.894 ± 0.015 | 0.518 ± 0.02 |

*The negative values of $k_{2-\mathrm{ex}}^{\mathrm{HH}}$ are reflective of the NOE contribution dominating under limited/zero exchange. Other negative values correspond to very low or zero flux within the error bars.

### 7.3.2. Complex 1

**Sample:** 4 mM of [Ir-*d*22], 40 mM of $^{13}$C-pyr, and a variable amount of DMSO between 20 and 300 mM in 600 µL of methanol-*d*4.

**Experiment:** In all the experiments below, we used a 9.4 T NMR system, at 265 K, and a SEPP-SABRE sequence for $^1$H hyperpolarization of the selected proton. For each pressure, 2 sets of data were measured. Set 1 - Hyperpolarization of H$^a$-**1** at -15.56 ppm using SEPP-SABRE, followed by exchange with H$^b$-**1** at -21.59 ppm, and free $H_2$ at 4.6 ppm (**Figure S6.1-top**). Set 2 - Hyperpolarization of H$^b$-**1** using SEPP-SABRE and subsequent exchange with H$^a$-**1** and $H_2$ (**Figure S6.1, Bottom**). $R_{H2}$ was measured for different temperatures (**Table S15**).

**Simulations:** To streamline the analysis, we employed a global fitting approach for model IrHH↔H2 using MATLAB scripts to fit all parameters simultaneously using 2 sets of kinetics. Across both kinetics and for each temperature, we share 5 parameters: the dissociation rate constant $k_{1-\mathrm{H}_2}$, exchange constant $k_{1-\mathrm{ex}}^{\mathrm{HH}}$, association rate constant $k_{1\mathrm{a}}'^{\mathrm{H}}$, natural $H_2$ relaxation exchange $R_{\mathrm{H}_2}$ and relaxation exchange $R_{\mathrm{IrHH}}$. The reported errors are the results of such fits using MATLAB's nonlinear regression function "nlinfit". The fitting script, together with integrals, are available in Zenodo DOI: https://doi.org/10.5281/zenodo.18449765.

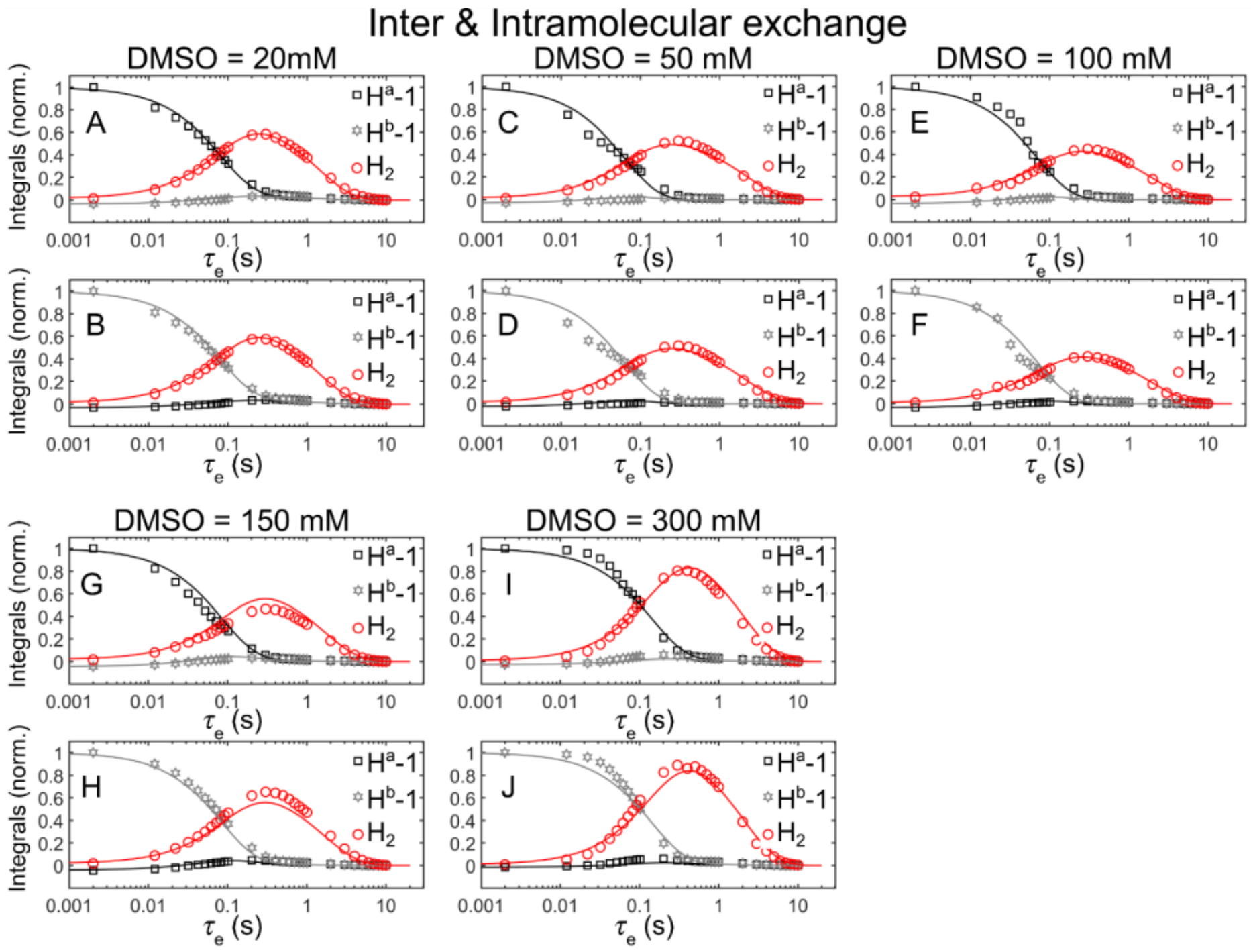

**Figure S6.1: Normalized integrals of SEPP-SABRE kinetics**. $^1$H NMR integrals of the hydride at -14.98 ppm (H$^a$-**1**, black square), at -21.59 ppm (H$^b$-**1**, grey star), and 4.6 ppm ($H_2$, red circle), of 20 mM (A, B), 50 mM (C, D), 100 mM (E, F), 150 mM (G, H), and 300 mM (I, J). The curves are fitted using the model IrHH↔H2 (Inter and Intramolecular exchange). Estimated exchange rate constants, dissociations, associations, and relaxation-exchanges are given in **Table S6.1**.

**Table S6.1**. Fitting parameters of **1** SEPP-SABRE kinetics for each DMSO concentration using model IrHH↔H2. Parameters are defined in section 2

| **[DMSO] (mM)** | $k_{1-H_2}$ (s$^{-1}$) | $k_{1-\mathrm{ex}}^{\mathrm{HH}}$ (s$^{-1}$) | $k_{1\mathrm{a}}'^{\mathrm{H}}$ (s$^{-1}$) | $R_{\mathrm{IrHH}}$ (s$^{-1}$) | $R_{\mathrm{H2}}$ (s$^{-1}$) |
|---|---|---|---|---|---|
| **20** | 8.530 ± 0.098 | 0.779 ± 0.099 | 1.129 ± 0.074 | 2.505 ± 0.084 | 0.518 ± 0.02 |

| 50 | 7.982 ± 0.201 | 1.431 ± 0.312 | -0.129 ± 0.055 | 6.580 ± 0.273 | 0.518 ± 0.02 |
|---|---|---|---|---|---|
| 100 | 6.193 ± 0.174 | 1.354 ± 0.279 | -0.053 ± 0.060 | 6.778 ± 0.269 | 0.518 ± 0.02 |
| 150 | 6.787 ± 0.230 | 1.623 ± 0.296 | 0.271 ± 0.133 | 3.033 ± 0.255 | 0.518 ± 0.02 |
| 300 | 6.834 ± 0.171 | 0.471 ± 0.160 | 0.150 ± 0.087 | 0.477 ± 0.074 | 0.518 ± 0.02 |

## 7.4. SEPP-SABRE kinetics of IrHH hydrogens as a function of [pyruvate]

**Sample:** 4 mM of [Ir-$d_{22}$], 20 mM of DMSO, and a variable amount of $^{13}$C-pyr between 40 and 174 mM in 600 µL of methanol-$d_4$.

**Experiment:** In all the experiments below, we used a 9.4 T NMR system, p$H_2$ at 8.5 bar pressure, 265 K, and a SEPP-SABRE sequence for $^1$H hyperpolarization of the selected proton. For each concentration of pyruvate, 4 sets of data were measured. Set 1 - hyperpolarization of $H^a$-3 at -29.1 ppm using SEPP-SABRE and following exchanging with $H^b$-**3** at -27.2 ppm, $H^a$-**2** at -14.98 ppm, $H^b$-**2** at -24.08 ppm, and free $H_2$ at 4.6 ppm (**Figure S7-A, E**). Set 2 - hyperpolarization of $H^b$-**3** using SEPP-SABRE and following exchanging with $H^a$-**3**, $H^a$-**2**, $H^b$-**2**, and $H_2$ (**Figure S7-B, F**). Set 3 - hyperpolarization of $H^a$-**2** using SEPP-SABRE and following exchanging with $H^b$-**2**, $H^a$-**3**, $H^b$-**3**, and $H_2$ (**Figure S7-C, G**). Set 4 - hyperpolarization of $H^b$-**2** using SEPP-SABRE and following exchanging with $H^a$-**2**, $H^a$-**3**, $H^b$-**3**, and $H_2$ (**Figure S7-D, H**). $R_{H2}$ was measured for different temperatures (**Table S15**).

### 7.4.1. Complex 3 & 2

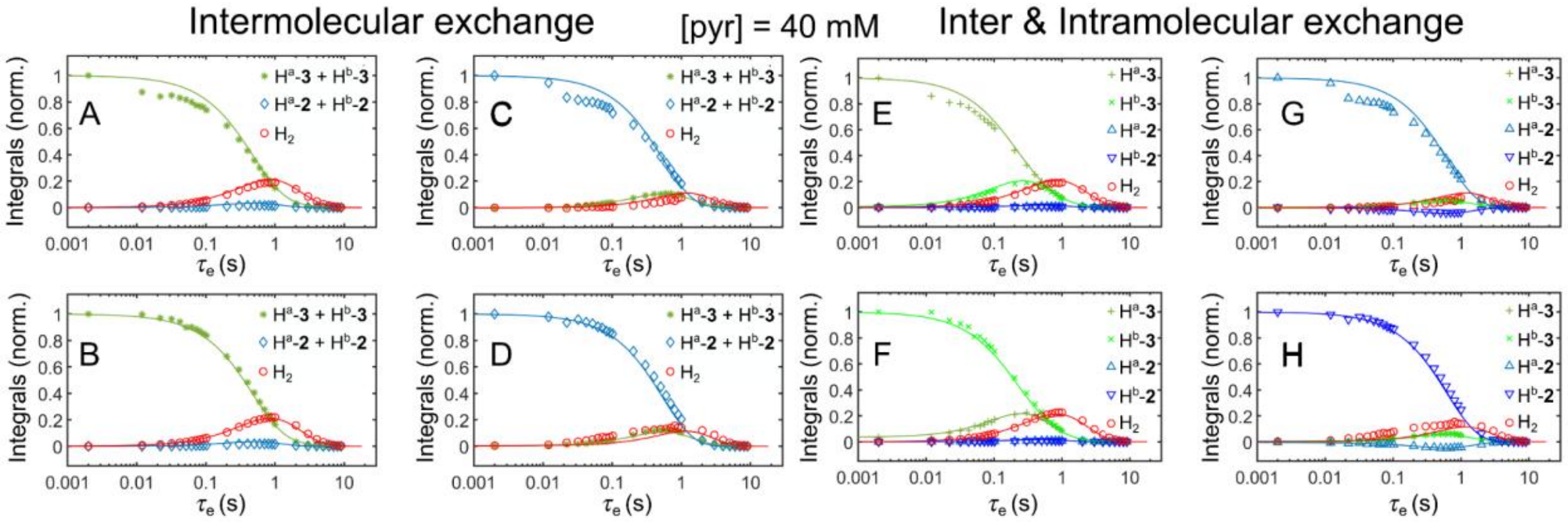

**Figure S7.40 mM: Normalized integrals of 3 and 2 SEPP-SABRE kinetics for 40 mM.** $^1$H NMR integrals of the hydride at -29.2 ppm ($H^a$-**3**, green plus), at -27.2 ppm ($H^b$-**3**, green cross), and 4.6 ppm ($H_2$, red circle) at 265 K. The summed integrals of **3**'s hydrogens ($H^a$-**3** + $H^b$-**3**, green star) and of **2**'s hydrogens ($H^a$-**2** + $H^b$-**2**, blue diamond) are used for the intermolecular exchange. The curves are fitted using a global fit of the model (**3**↔**2**)↔$H_2$ (A, B, C, D, Intermolecular exchange), and (E, F, G, H, Inter and Intramolecular exchange). Estimated exchange rate constants, dissociations, associations, and relaxation-exchanges are given in **Table S7.**

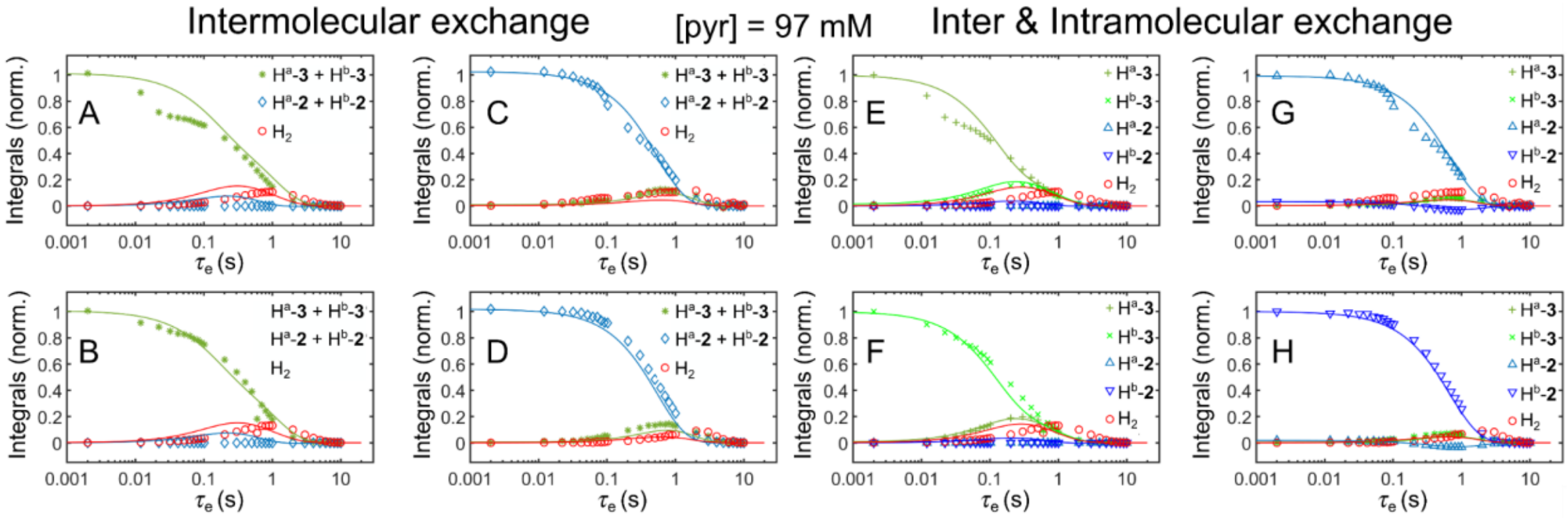

**Figure S7.100 mM: Normalized integrals of 3 and 2 SEPP-SABRE kinetics for 100 mM.** $^{1}$H NMR integrals of the hydride at -29.2 ppm ($H^a$-**3**, green plus), at -27.2 ppm ($H^b$-**3**, green cross), and 4.6 ppm ($H_2$, red circle) at 265 K. The summed integrals of **3**'s hydrogens ($H^a$-**3** + $H^b$-**3**, green star), and of **2**'s hydrogens ($H^a$-**2** + $H^b$-**2**, blue diamond) are used for the intermolecular exchange. The curves are fitted using a global fit of the model (**3**↔**2**)↔$H_2$ (A, B, C, D, Intermolecular exchange), and (E, F, G, H, Inter and Intramolecular exchange). Estimated exchange rate constants, dissociations, associations, and relaxation-exchanges are given in **Table S7.**

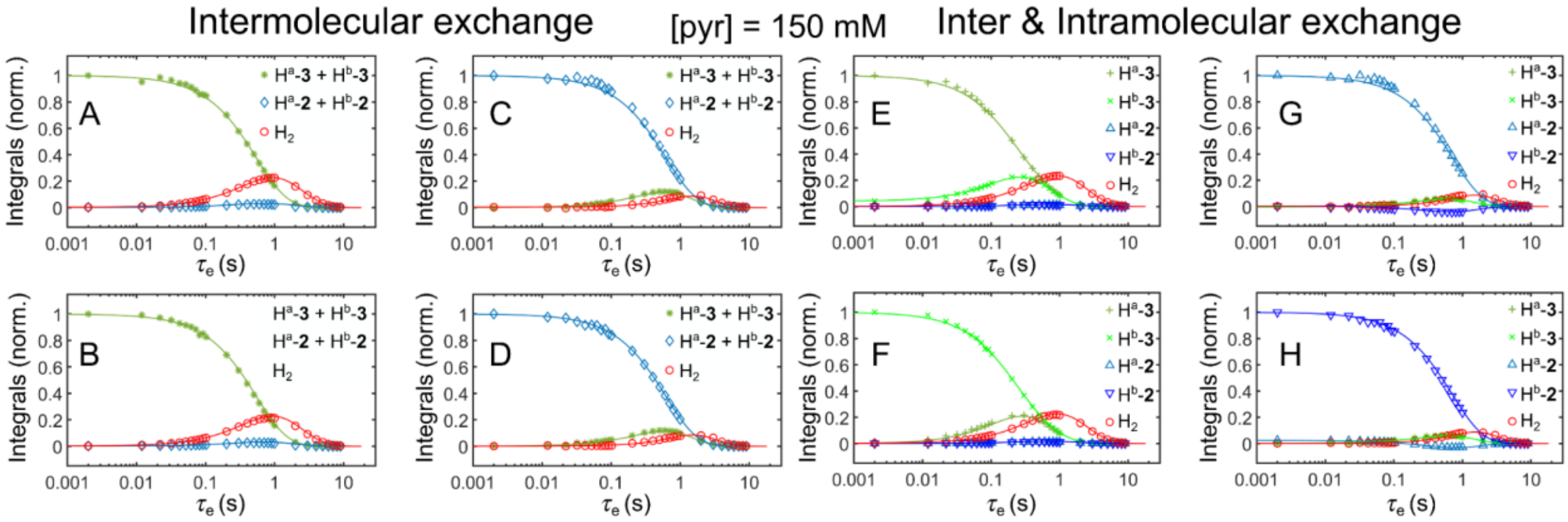

**Figure S7.150 mM: Normalized integrals of 3 and 2 SEPP-SABRE kinetics for 150 mM.** $^{1}$H NMR integrals of the hydride at -29.2 ppm ($H^a$-**3**, green plus), at -27.2 ppm ($H^b$-**3**, green cross), and 4.6 ppm ($H_2$, red circle) at 265 K. The summed integrals of **3**'s hydrogens ($H^a$-**3** + $H^b$-**3**, green star), and of **2**'s hydrogens ($H^a$-**2** + $H^b$-**2**, blue diamond) are used for the intermolecular exchange. The curves are fitted using a global fit of the model (**3**↔**2**)↔$H_2$ (A, B, C, D, Intermolecular exchange), and (E, F, G, H, Inter and Intramolecular exchange). Estimated exchange rate constants, dissociations, associations, and relaxation-exchanges are given in **Table S7.**

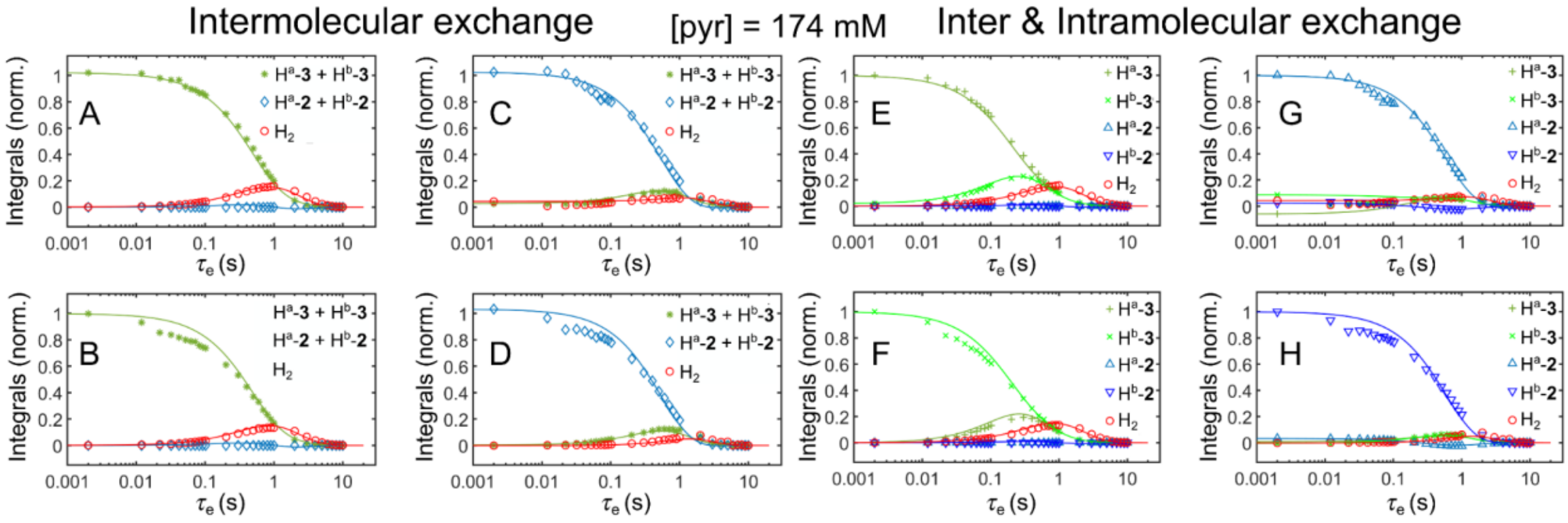

**Figure S7.174 mM: Normalized integrals of 3 and 2 SEPP-SABRE kinetics for 174 mM.** $^{1}$H NMR integrals of the hydride at -29.2 ppm ($H^a$-**3**, green plus), at -27.2 ppm ($H^b$-**3**, green cross), and 4.6 ppm ($H_2$, red circle) at 265 K. The summed integrals of **3**'s hydrogens ($H^a$-**3** + $H^b$-**3**, green star), and of **2**'s hydrogens ($H^a$-**2** + $H^b$-**2**, blue diamond) are used for the intermolecular exchange. The curves are fitted using a global fit of the model (**3**↔**2**)↔$H_2$ (A, B, C, D, Intermolecular exchange), and (E, F, G, H, Inter and Intramolecular exchange). Estimated exchange rate constants, dissociations, associations, and relaxation-exchanges are given in **Table S7.**

**Table S7**. Fitting parameters of **3** and **2** SEPP-SABRE kinetics for each concentration using 2 steps fitting process of model (**3**↔**2**)↔$H_2$. Parameters are defined in section 2.

| [pyr] (mM) | $k_{3-H}$ ($s^{-1}$) | $k_{2-H_2}$ ($s^{-1}$) | $k_{3\to2}$ ($s^{-1}$) | $k_{2\to3}$ ($s^{-1}$) | $k_{3ex}^{HH}$ ($s^{-1}$) | $k_{2ex}^{HH}$ ($s^{-1}$) | ${k'_{3a}}^{H}$ ($s^{-1}$) | ${k'_{2a}}^{H}$ ($s^{-1}$) | $R_3$ ($s^{-1}$) | $R_2$ ($s^{-1}$) | $R_{H2}$ ($s^{-1}$) |
|---|---|---|---|---|---|---|---|---|---|---|---|
| **40** | 0.726 ± 0.04 | 0.220 ± 0.028 | 0.185 ± 0.047 | 0.540 ± 0.039 | 2.248 ± 0.080 | -0.209 ± 0.023 | -0.162 ± 0.12 | -0.046 ± 0.114 | 1.168 ± 0.036 | 1.056 ± 0.036 | 0.518 ± 0.02 |
| **100** | 0.683 ± 0.02 | 0.088 ± 0.011 | 0.075 ± 0.019 | 0.553 ± 0.015 | 1.950 ± 0.036 | -0.176 ± 0.011 | 0.014 ± 0.053 | 0.074 ± 0.043 | 0.997 ± 0.017 | 1.022 ± 0.017 | 0.518 ± 0.02 |
| **150** | 0.701 ± 0.01 | 0.071 ± 0.007 | 0.122 ± 0.012 | 0.527 ± 0.009 | 2.109 ± 0.024 | -0.165 ± 0.007 | 0.015 ± 0.034 | 0.073 ± 0.027 | 1.010 ± 0.010 | 0.965 ± 0.010 | 0.518 ± 0.02 |
| **174** | 0.515 ± 0.03 | 0.046 ± 0.023 | 0.132 ± 0.041 | 0.534 ± 0.033 | 2.470 ± 0.079 | -0.141 ± 0.024 | 0.579 ± 0.179 | -0.331 ± 0.142 | 1.334 ± 0.034 | 1.275 ± 0.036 | 0.518 ± 0.02 |

### 7.4.2. Complex 1

**Sample:** 4 mM of [Ir-*d*22], 20 mM of DMSO, and a variable amount of $^{13}C$-pyr between 40 and 174 mM in 600 µL of methanol-*d*4.

**Experiment:** In all the experiments below, we used a 9.4 T NMR system, $pH_2$ at 8.5 bar pressure, 265 K, and a SEPP-SABRE sequence for $^1H$ hyperpolarization of the selected proton. For each temperature, 2 sets of data were measured. Set 1 - hyperpolarization of Ha-1 at -15.56 ppm using SEPP-SABRE and following exchanging with $H^b$-1 at -21.59 ppm, and free $H_2$ at 4.6 ppm (**Figure S7.1-top**). Set 2 - hyperpolarization of $H^b$-1 using SEPP-SABRE and following the exchange with $H^a$-1, and $H_2$ (**Figure S7.1-bottom**). $R_{H2}$ was measured for different temperatures (**Table S15**).

**Simulations:**

To streamline the analysis, we employed a global fitting approach for model IrHH↔$H_2$ using MATLAB scripts to fit all parameters simultaneously using 2 sets of kinetics. Across both kinetics and for each temperature, we share 5 parameters: the dissociation rate constant $k_{1-H_2}$, exchange constant $k_{1-ex}^{HH}$, association rate constant ${k'_{1a}}^{H}$, natural $H_2$ relaxation exchange $R_{H_2}$ and relaxation exchange $R_{IrHH}$. The reported errors are the results of such fits using MATLAB's nonlinear regression function “nlinfit”. The fitting script, together with integrals, is available in Zenodo DOI: https://doi.org/10.5281/zenodo.18449765.

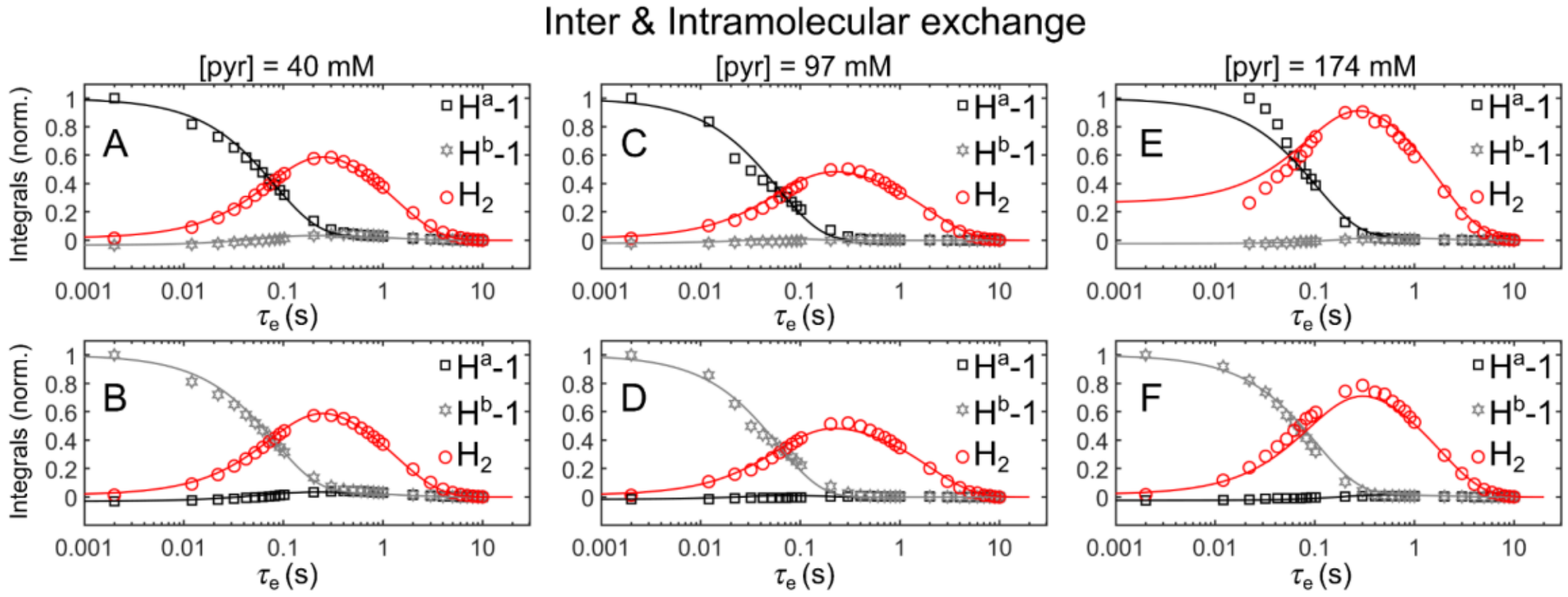

**Figure S7.1: Normalized integrals of 1 SEPP-SABRE kinetics.** $^1H$ NMR integrals of the hydride at -14.98 ppm ($H^a$-**1**, black square), at -21.59 ppm ($H^b$-**1**, grey star), and 4.6 ppm ($H_2$, red circle), of pyruvate concentration of 40 mM (A, B), 97 mM (C, D), and 174 mM (E, F). The curves are fitted using the model IrHH↔$H_2$ (Inter and Intramolecular exchange). Estimated exchange rate constants, dissociations, associations, and relaxation-exchanges are given in **Table S7.1**.

**Table S7.1.** Fitting parameters of **1** SEPP-SABRE kinetics for each pyruvate concentration using model IrHH↔$H_2$. Parameters are defined in section 2

| **pyr (mM)** | $k_{1-H_2}$ (s-1) | $k_{1-\mathrm{ex}}^{\mathrm{HH}}$ (s-1) | $k'_{1\mathrm{a}}{}^{\mathrm{H}}$ (s-1) | $R_{\mathrm{IrHH}}$ (s-1) | $R_{\mathrm{H2}}$ (s-1) |
|---|---|---|---|---|---|
| **40** | 8.530 ± 0.098 | 0.779 ± 0.099 | 1.129 ± 0.074 | 2.505 ± 0.084 | 0.518 ± 0.02 |
| **97** | 8.556 ± 0.165 | 0.735 ± 0.249 | -0.071 ± 0.044 | 7.451 ± 0.231 | 0.518 ± 0.02 |
| **174** | 8.509 ± 0.268 | 0.457 ± 0.156 | -0.152 ± 0.234 | 1.153 ± 0.184 | 0.518 0.02 |

## 7.5 Thermal kinetics of IrHH exchange of 3 as a function of temperature

**Sample:** 4 mM of [Ir], 40 mM of $^{13}$C-pyr, and 20 mM of DMSO in 600 µL of methanol-$d_4$ at 6 bar.

**Experiments**:

To monitor the hydride ligand exchange pathways, selective $^{1}$H exchange spectroscopy (SEXSY) refs (76KesslerJMR,77BauerJMR) was used (**Figure S8**, **SI**). Due to a poor signal-to-noise ratio, each measurement required four hours of data acquisition (selnogpzs.2 experiment, 256 scans, 2-second repetition time, 16 kinetic time points; see also **Section 3 in SI**). This unsatisfactory situation stemmed from the relatively weak thermally derived $^{1}$H NMR signals of IrHH hydrides at the typical ~1 mM iridium complex concentration used. Although magnetization transfer and species interchange were evident in the resulting NMR spectra, the inherent limitations in the accuracy of this technique, despite its ability to encode effects up to and including the rate-limiting step, rendered it insufficient for producing complete insights.

**Results:**

This section aims to show thermal exchange kinetics measured by varying temperatures from 247 K to 264 K. Selnogpzs.2 experiment was used, which is a 1D NOESY pulse sequence using a selective refocusing shaped pulse (see above references and also **Section 3 in SI**). The curves are fitted using the model IrHH↔$H_2$ exchange model described in **Section 1.1 of the SI**. **Figure S8** shows the experimental points and the fitting results, whereas **Table S8** presents the obtained reaction kinetic rates.

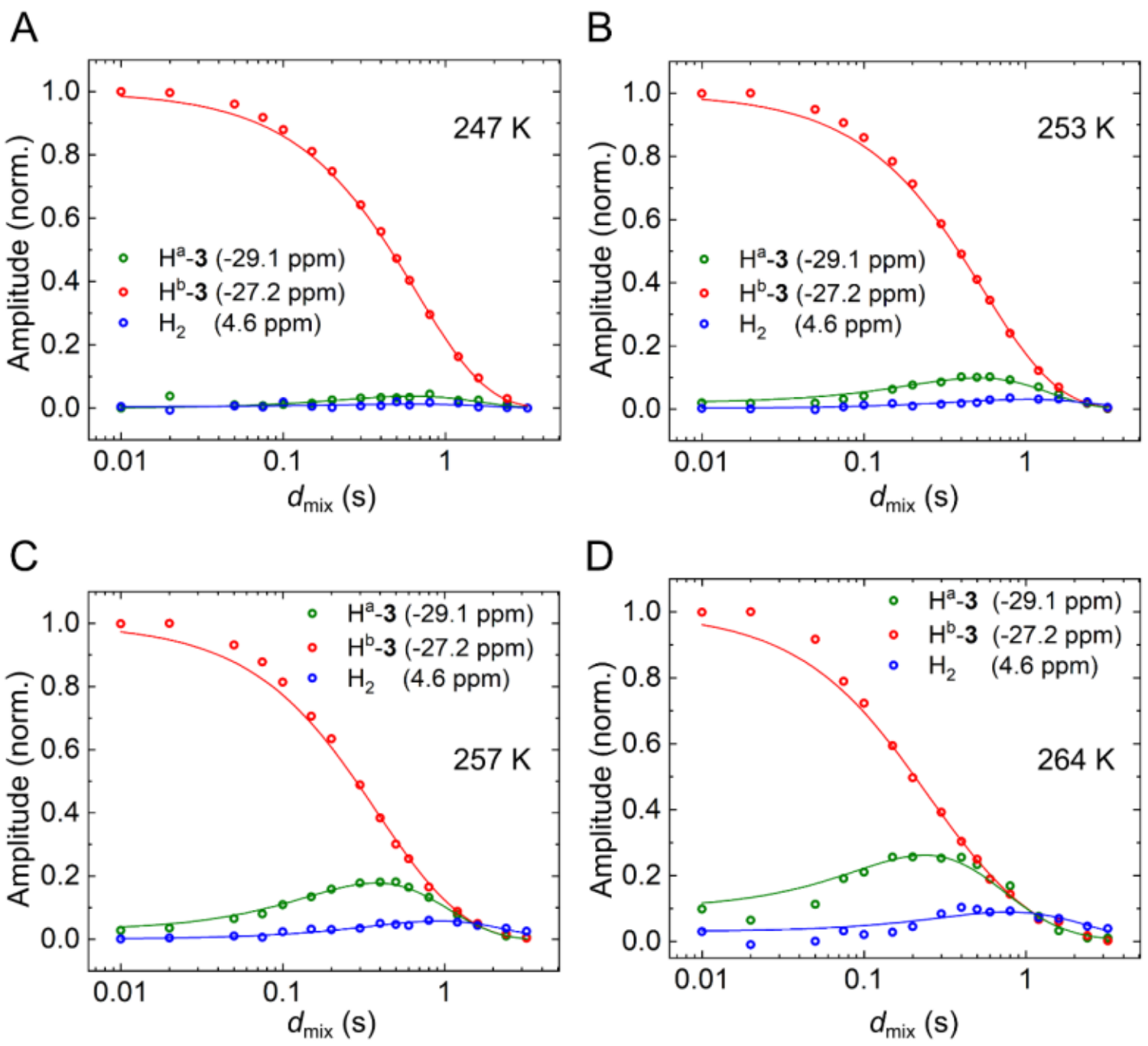

**Figure S8: Normalized integrals of selnogpzs.2 thermal kinetics.** $^{1}$H NMR integrals of the hydride at -29.2 ppm (Ha, green), at -27.2 ppm (Hb, red), and 4.6 ppm ($H_2$, blue) at 247 K (A), 253 K (B), 257 K (C), and 264 K (D). The curves are fitted using the model IrHH↔$H_2$. Estimated exchange rate constants, dissociations, associations, and relaxation-exchanges are given in Table S8.

**Table S8:** Fitting parameters selnogpzs.2 thermal kinetics for each temperature.

| Temperature (K) | $k_{3\text{-H2}}$ (s$^{-1}$) | $k_{3-\mathrm{ex}}^{\mathrm{HH}}$ (s$^{-1}$) | $k'_{3\mathrm{a}}{}^{\mathrm{H}}$ (s$^{-1}$) | $R_{\mathrm{H2}}$ (s$^{-1}$) | $R_{\mathrm{Hab\text{-}3}}$ (s$^{-1}$) |
|---|---|---|---|---|---|
| **247** | 0.04 ± 0.03 | 0.168 ± 0.02 | -1.201 ± 1.927 | 1.29 ± 1.20 | 1.305 ± 0.03 |
| **253** | 0.08 ± 0.01 | 0.45 ± 0.01 | -0.152 ± 0.452 | 0.77 ± 0.5 | 1.31 ± 0.03 |
| **257** | 0.166 ± 0.02 | 1.175 ± 0.04 | -0.267 ± 0.365 | 1.00 ± 0.4 | 1.316 ± 0.02 |
| **264** | 0.20 ± 0.04 | 2.30 ± 0.15 | 0.659 ± 0.558 | 0.01 ± 0.54 | 1.606 ± 0.09 |

# 8. Pyruvate exchange

One hypothesis is that the dissociation of the pyruvate from the SABRE complex can associate again in different positions, influencing a change of both hydride' chemical shifts at -29.1 ppm and -27.2 ppm. Therefore, we measured pyruvate's exchange rates to understand the mechanism of exchange. We used the SEPP-SPINEPTplus-SABRE pulses sequence, which consists of three parts:

## 8.1. SEPP-SPINEPTplus – SOT to hyperpolarize pyruvate

First part frequency-selective excitation of parahydrogen-derived polarization SOT sequence (SEPP) SOT converts the PASADENA two-spin order into the magnetization of one of them. In the second part, insensitive nuclei enhanced by polarization transfer (INEPT) with selective pulses (SP) with an additional 180$^{o}$ pulse on both channels (SPINEPTplus) to transfer polarization from the proton to the targeted nuclei to have a net magnetization on the bound $^{13}$C-pyruvate (**Figure S9 A**).

The first part, $\tau_1$ and $\tau_2$ delays of SEPP were optimized previously by Assaf et al., (43AssafJPCL). As for the second part SPINEPTplus, we performed temperature dependence experiments from 260 K to 275.5 K as a function of $\tau_3$. Polarization is maximal when the timing is properly set for $\tau_3$ (**Figure S9 B**). The main reason for polarization loss is the increase in the total duration of the pulse sequence, and with high temperatures, the dissociation rate increases.

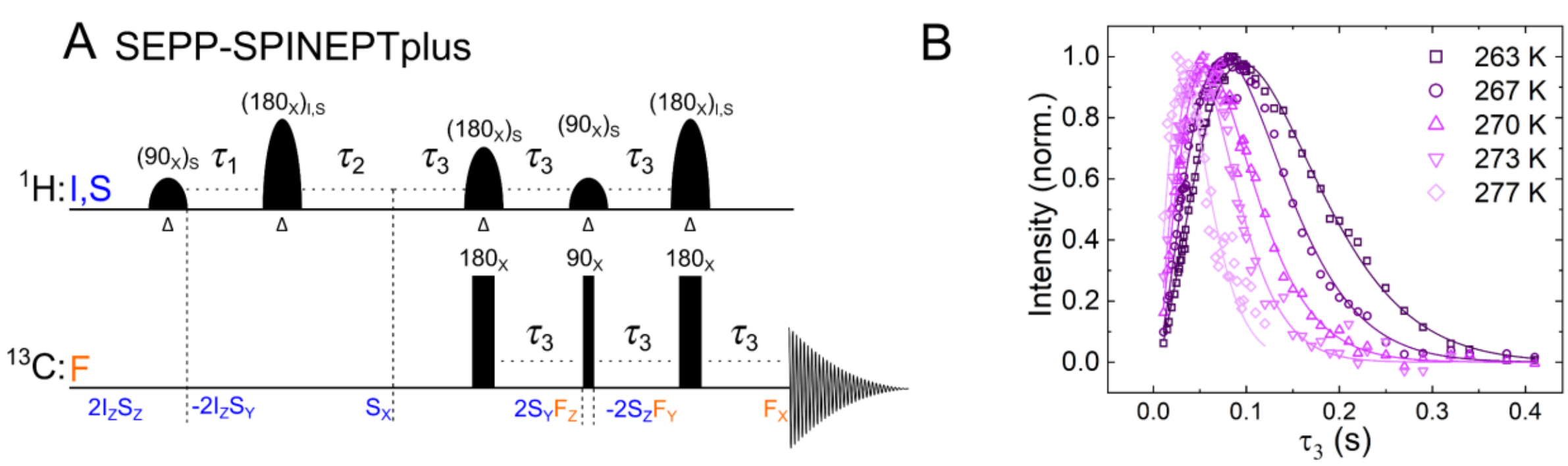

**Figure S9:** Scheme of SEPP-SPINEPT+ pulse sequence (A) and dependence of the SEPP-SPINEPT+ enhanced NMR signal of 2-$^{13}$C-ePyr (-29.1 ppm $^{1}$H of **3** was selectively excited with SEPP-SPINEPTplus) on the inter-pulse delay $\tau_3$ at different temperature (B): 263 K (squares), 267 K (circles); 270 K (up-triangles), 273 K (down-triangles), 277 K (diamond). Lines represent the global fitting by a sinus decay function $y0+A\times\sin(2\pi J\tau_3)^2 \times \exp(-4R\tau_3)$. The *J*-coupling was treated as a shared parameter across all datasets, while the relaxation rates *R* were fitted independently for each temperature. This yielded a *J*-coupling value of 0.61 ± 0.15 Hz. The extracted relaxation rates *R* exhibit a clear temperature dependence, with values of 5.0 ± 0.16 $s^{-1}$, 6.43 ± 0.14 $s^{-1}$, 8.6 ± 0.13 $s^{-1}$, 10.56 ± 0.13 $s^{-1}$, and 15.5 ± 0.2 $s^{-1}$, respectively. Previously, this *J* value was estimated to be 0.9 ± 0.1 Assaf et al. (43AssafJPCL).

## 8.2. SEPP-SPINEPTplus-SABRE – SOT to measure pyruvate exchange

After SEPP-SPINEPTplus, we have a net magnetization on the $^{13}$C-ePyr. Then, the magnetization of all spins is rotated, and a dephasing gradient is applied (**Figure S10**). In the ideal case, this would destroy the magnetization of all nuclear spins, leaving only z-magnetization of the excited carbon. Then, after mixing time, $\tau_e$, and subsequent excitation, all exchanging species can be observed (**Figure S11,12,13,14**).

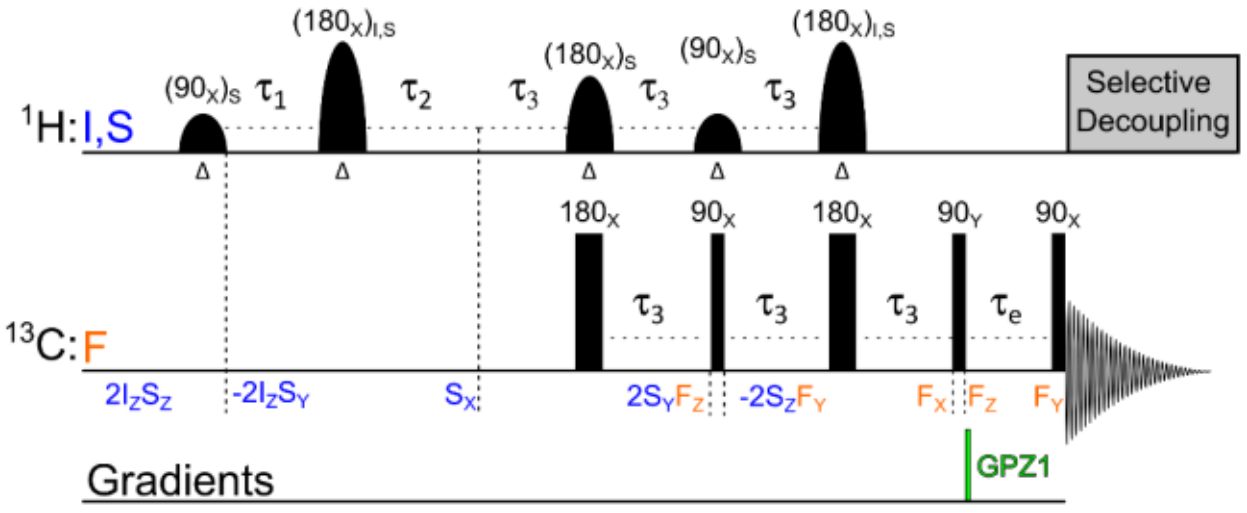

**Figure S10: Scheme of SEPP-SPINEPTplus-SABRE pulse sequence.** During acquisition, a selective decoupling $\{^{1}H\}$ at 2.36 ppm was added to decouple the methyl group's interaction. GPZ1 is a gradient pulse of 2 ms with SMSQ10.100 shape at 31%.

These exchange rates were measured as a function of temperature, bubbling pressure, and concentrations of DMSO and pyruvate. During these measurements, the dissociation of the bound pyruvate doesn't take a direct path to the free form; however, an intermediate exchange happens before total dissociation. We observed an exchange between $^{13}C^{2}$-pyr **3** at 206.59 ppm and $^{13}C^{2}$-pyr **2** at 196.86 ppm, followed by dissociation to $^{13}C^{2}$-free at 202.5 ppm. We fitted the data using the carbon exchange model (**$^{13}C^{2}$-3 ↔ $^{13}C^{2}$-2)↔ $^{13}C^{2}$-free.**

## 8.3. Pyruvate exchange kinetics of $^{13}C^2$ of 3 and 2 in the function of temperature with 20 mM of DMSO

**Sample:** 4 mM of [Ir-*d22*], 40 mM of $^{13}$C-pyr, and 20 mM of DMSO in 600 µL of methanol-*d4*.

**Experiment:** In all the experiments below, we used a 9.4 T NMR system, $pH_2$ pressure of 8.5 bar, and a SEPP-SPINEPTplus-SABRE sequence for $^{13}$C hyperpolarization of the selected carbon. We experimented twice: first by hyperpolarizing the $^{13}C^2$-pyr of **3**, and observing its transition to **2** and then to free pyruvate **(Figure S11-top)**. In the second experiment, we hyperpolarized the $^{13}C^2$-pyr of **2** and tracked its conversion to **3** before dissociation **(Figure S11-bottom)**. The temperature was varied from 260K to 270.9 K.

**Simulations:** These experiments provided two sets of data, which we analyzed using a global fit with the model **($^{13}C^2$-3 ↔$^{13}C^2$-2)↔$^{13}C^2$-free**, as explained in Section 2, to extract the relevant kinetic parameters. $R_{\mathrm{pyr}}$ was measured for the free carbon.

**Results:**

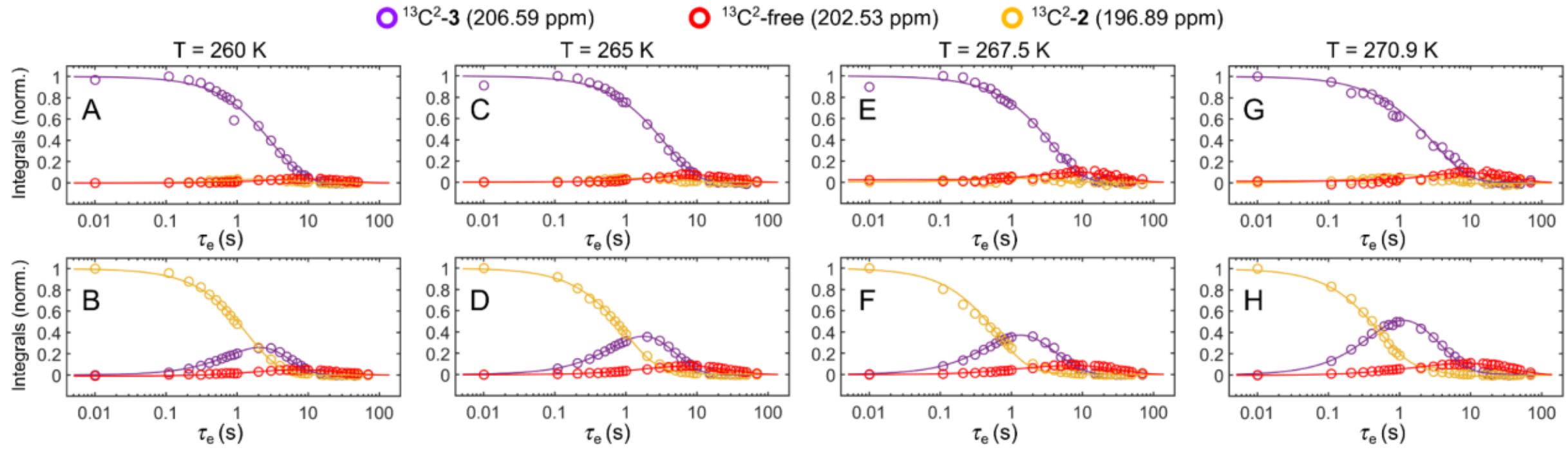

**Figure S11: Normalized integrals of SEPP-SPINEPTplus-SABRE kinetics.** $^{13}$C NMR integrals of the equatorial $^{13}C^2$-epyr **3** at 206.59 ppm (purple), free $^{13}C^2$-free at 202.53 ppm (red), and equatorial $^{13}C^2$-ePyr **2** at 196.89 ppm (orange) at the nominal temperatures of 260 K (A,B), 265 K (C,D), 267.5 K (E,F), and 270.9 K (G,H). The curves are fitted using model **($^{13}C^2$-3 ↔$^{13}C^2$-2)↔$^{13}C^2$-free**. Estimated exchange rate constants, dissociations, associations, and relaxation-exchanges are given in Table S9.

**Table S9**. Fitting parameters of SEPP-SPINEPTplus-SABRE kinetics of both experiments for each temperature using model ($^{13}C^2$-**3** ↔$^{13}C^2$-**2**)↔$^{13}C^2$-free. Parameters are defined in section 2.

| Temp (K) | Fig | $k_{3-pyr'}$ (s$^{-1}$) | $k_{2-pyr'}$ (s$^{-1}$) | $k_{3-2}$ (s$^{-1}$) | $k_{2-3}$ (s$^{-1}$) | $k_{\mathrm{a}}'^{\mathrm{P[3]}}$ (s$^{-1}$) | $k_{\mathrm{a}}'^{\mathrm{P[2]}}$ (s$^{-1}$) | $R_3$ (s$^{-1}$) | $R_2$ (s$^{-1}$) | $R_{\mathrm{pyr}}$ (s$^{-1}$) |
|---|---|---|---|---|---|---|---|---|---|---|
| **260** | A&B | 0.014 ± 0.002 | 0.034 ± 0.004 | 0.049 ± 0.007 | 0.350 ± 0.009 | 0.019 ± 0.042 | -0.014 ± 0.043 | 0.270 ± 0.006 | 0.312 ± 0.011 | 0.034 ± 0.004 |
| **265** | C&D | 0.016 ± 0.001 | 0.042 ± 0.004 | 0.121 ± 0.006 | 0.630 ± 0.007 | 0.025 ± 0.026 | -0.019 ± 0.027 | 0.208 ± 0.004 | 0.467 ± 0.009 | 0.033 ± 0.003 |
| **267.5** | E&F | 0.034 ± 0.003 | 0.087 ± 0.009 | 0.089 ± 0.013 | 0.849 ± 0.018 | -0.002 ± 0.031 | 0.015 ± 0.030 | 0.210 ± 0.008 | 0.611 ± 0.022 | 0.029 ± 0.002 |
| **270.9** | G&H | 0.042 ± 0.004 | 0.119 ± 0.012 | 0.121 ± 0.008 | 1.251 ± 0.021 | 0.008 ± 0.032 | 0.041 ± 0.039 | 0.198 ± 0.010 | 0.470 ± 0.049 | 0.029 ± 0.003 |

## 8.4. Pyruvate exchange kinetics of $^{13}C^2$ of 3 and 2 in the function of temperature with 100mM of DMSO

**Sample:** 4 mM of [Ir-*d22*], 40 mM of $^{13}$C-pyr, and 100 mM of DMSO in 600 µL of methanol-*d4*.

**Experiment:** In all the experiments below, we used a 9.4 T NMR system, $pH_2$ pressure of 8.5 bar, and a SEPP-SPINEPTplus-SABRE sequence for $^{13}$C hyperpolarization of the selected carbon. We experimented twice: first by hyperpolarizing the $^{13}C^2$-pyr of **3**, and observing its transition to **2** and then to free pyruvate **(Figure S12-top)**. In the second experiment, we hyperpolarized the $^{13}C^{2}$-pyr of **2** and monitored its conversion to **3** prior to dissociation **(Figure S12-bottom)**. The temperature was varied from 260K to 270.9 K.

**Simulations:** These experiments provided two sets of data, which we analyzed using a global fit with the model **($^{13}C^{2}$-3 ↔$^{13}C^{2}$-2)↔$^{13}C^{2}$-free**, as explained in Section 2, to extract the relevant kinetic parameters. $R_{\text{pyr}}$ was measured for the free carbon.

**Results:** Increasing of temperature increases dissociation and exchange rates.

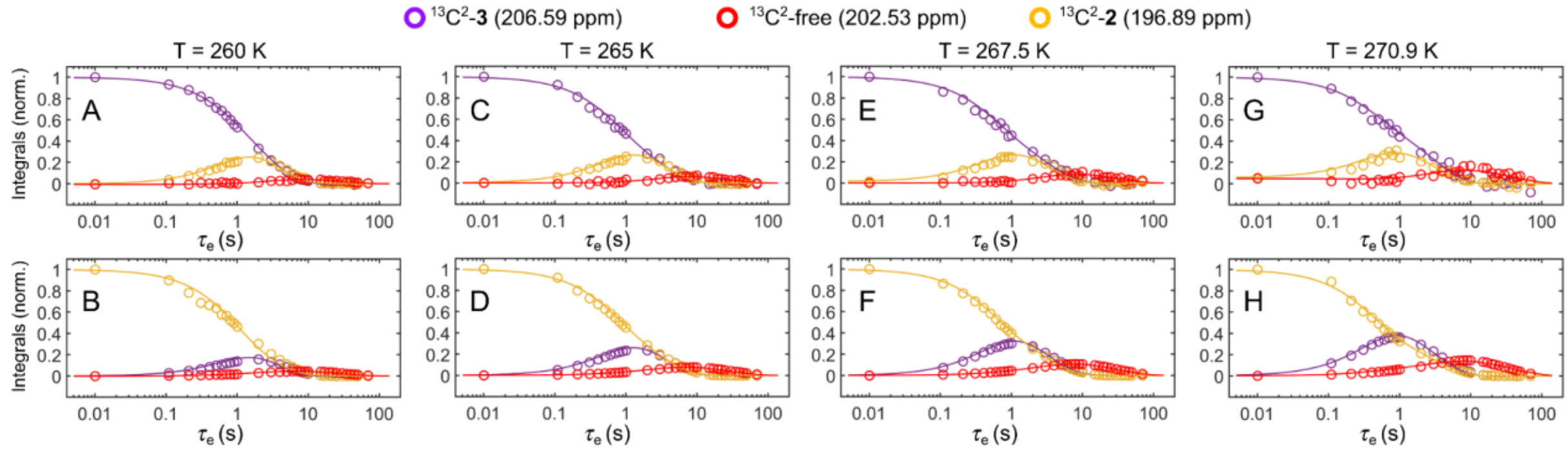

**Figure S12: Normalized integrals of SEPP-SPINEPTplus-SABRE kinetics.** $^{13}C$ NMR integrals of the equatorial $^{13}C^{2}$-epyr **3** at 206.59 ppm (purple), free $^{13}C^{2}$-free at 202.53 ppm (red), and equatorial $^{13}C^{2}$-epyr **2** at 196.89 ppm (orange) at the nominal temperatures of 260 K (A, B), 265 K (C, D), 267.5 K (E, F), and 270.9 K (G, H). The curves are fitted using model **($^{13}C^{2}$-3 ↔$^{13}C^{2}$-2)↔$^{13}C^{2}$-free**. Estimated exchange rate constants, dissociations, associations, and relaxation-exchanges are given in Table S10.

**Table S10**. Fitting parameters of SEPP-SPINEPTplus-SABRE kinetics of both experiments for each temperature using model ($^{13}C^{2}$-**3** ↔$^{13}C^{2}$-**2**)↔$^{13}C^{2}$-free. Parameters are defined in section 2.

| **Temp (K)** | **Fig** | $k_{3-\text{pyr}'}$ (s$^{-1}$) | $k_{2-\text{pyr}'}$ (s$^{-1}$) | $k_{3\to2}$ (s$^{-1}$) | $k_{2\to3}$ (s$^{-1}$) | ${k'_{3a}}^{\text{P}}$ (s$^{-1}$) | ${k'_{2a}}^{\text{P}}$ (s$^{-1}$) | $R_3$ (s$^{-1}$) | $R_2$ (s$^{-1}$) | $R_{\text{pyr}}$ (s$^{-1}$) |
|---|---|---|---|---|---|---|---|---|---|---|
| **260** | A&B | 0.010 ± 0.004 | 0.031 ± 0.005 | 0.482 ± 0.013 | 0.338 ± 0.012 | -0.106 ± 0.069 | 0.113 ± 0.070 | 0.172 ± 0.013 | 0.468 ± 0.014 | 0.034 ± 0.004 |
| **265** | C&D | 0.010 ± 0.005 | 0.048 ± 0.005 | 0.618 ± 0.014 | 0.614 ± 0.014 | -0.020 ± 0.052 | 0.025 ± 0.053 | 0.305 ± 0.013 | 0.284 ± 0.013 | 0.033 ± 0.003 |
| **267.5** | E&F | 0.013 ± 0.006 | 0.071 ± 0.006 | 0.723 ± 0.018 | 0.890 ± 0.019 | 0.020 ± 0.048 | -0.021 ± 0.05 | 0.339 ± 0.015 | 0.221 ± 0.017 | 0.029 ± 0.002 |
| **270.9** | G&H | 0.031 ± 0.009 | 0.089 ± 0.012 | 0.931 ± 0.036 | 1.319 ± 0.040 | -0.022 ± 0.065 | 0.017 ± 0.071 | 0.362 ± 0.025 | 0.173 ± 0.032 | 0.029 ± 0.003 |

## 8.5. Pyruvate exchange kinetics of $^{13}C^2$ of 3 and 2 in the function of pressure with 20 mM of DMSO

**Sample:** 4 mM of [Ir-$d_{22}$], 40 mM of $^{13}$C-pyr, and 20 mM of DMSO in 600 µL of methanol-$d_4$.

**Experiment:** In all the experiments below, we used a 9.4 T NMR system at 265 K, and a SEPP-SPINEPTplus-SABRE sequence for $^{13}$C hyperpolarization of the selected carbon. We experimented twice: first by hyperpolarizing the $^{13}C^2$-pyr of **3**, and observing its transition to **2** and then to free pyruvate **(Figure S13-top)**. In the second experiment, we hyperpolarized the $^{13}C^2$-pyr of **2** and tracked its conversion to **3** before dissociation **(Figure S13-bottom)**. The pressure was varied from 1 bar to 8.5 bar.

**Simulations:** These experiments provided two sets of data, which we analyzed using a global fit with the model **($^{13}C^2$-3 ↔$^{13}C^2$-2)↔$^{13}C^2$-free**, as explained in Section 2, to extract the relevant kinetic parameters. $R_{\mathrm{pyr}}$ was measured for the free carbon.

**Results:**

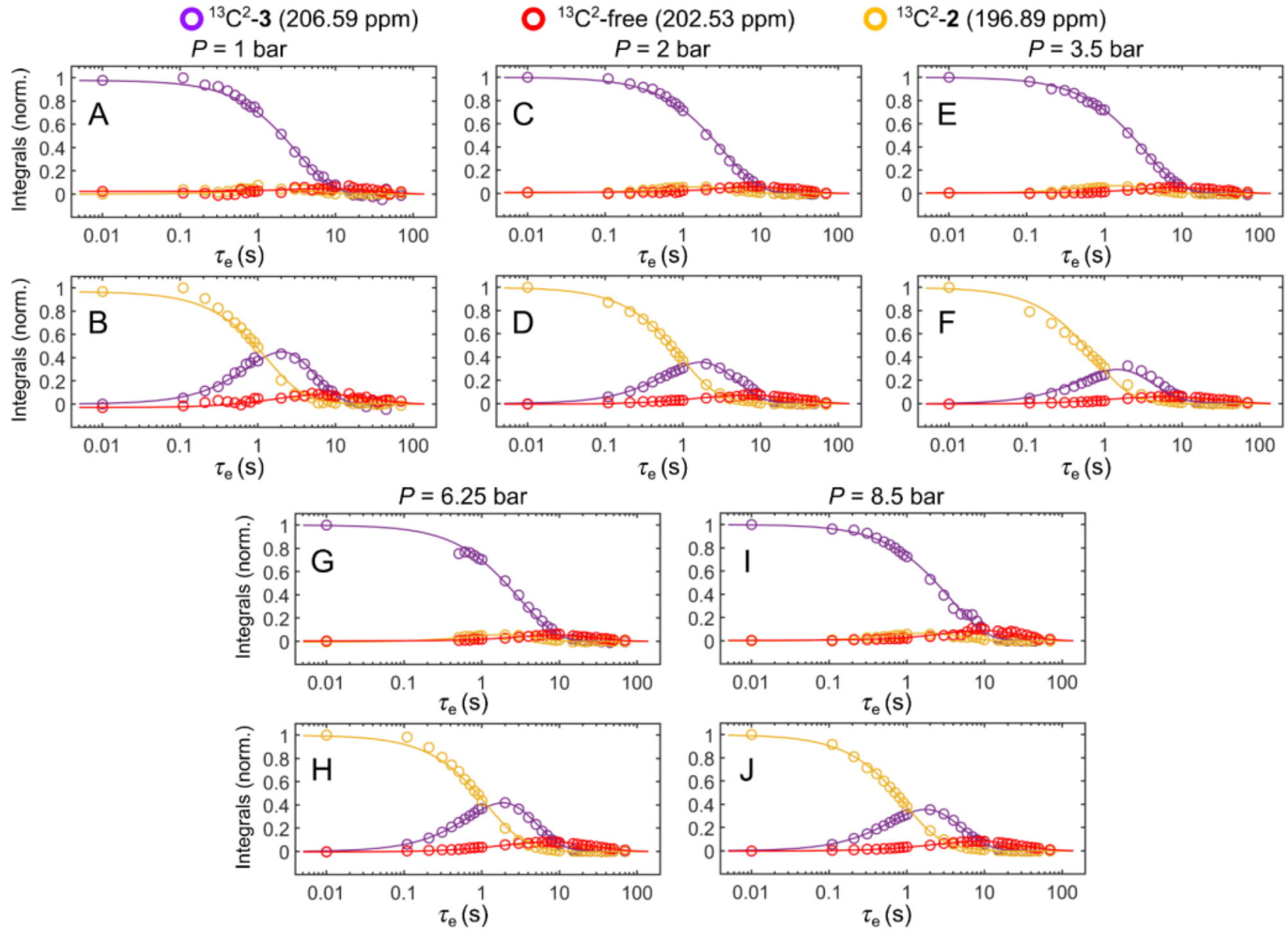

**Figure S13: Normalized integrals of SEPP-SPINEPTplus-SABRE kinetics.** $^{13}$C NMR integrals of the equatorial $^{13}C^2$-epyr **3** at 206.59 ppm (purple), free $^{13}C^2$-free at 202.53 ppm (red), and equatorial $^{13}C^2$-epyr **2** at 196.89 ppm (orange) at 1 bar (A, B), 2 bar (C, D), 3.5 bar (E, F), 6.25 bar (G, H), and 8.5 bar (I, J) at 265 K. The curves are fitted using model **($^{13}C^2$-3↔$^{13}C^2$-2)↔$^{13}C^2$-free**. Estimated exchange rate constants, dissociations, associations, and relaxation-exchanges are given in Table S11.

**Table S11**. Fitting parameters of SEPP-SPINEPTplus-SABRE kinetics of both experiments for each pressure using model ($^{13}C^2$-**3** ↔$^{13}C^2$-**2**)↔$^{13}C^2$-free. Parameters are defined in section 2.

| Back pressure Bubbling pressure | Fig | $k_{3-\mathrm{pyr}}$ (s$^{-1}$) | $k_{2-\mathrm{pyr}}$ (s$^{-1}$) | $k_{3\rightarrow 2}$ (s$^{-1}$) | $k_{2\rightarrow 3}$ (s$^{-1}$) | $k'^{\mathrm{P}}_{3a}$ (s$^{-1}$) | $k'^{\mathrm{P}}_{2a}$ (s$^{-1}$) | $R_3$ (s$^{-1}$) | $R_2$ (s$^{-1}$) | $R_{\mathrm{pyr}}$ (s$^{-1}$) |
|---|---|---|---|---|---|---|---|---|---|---|
| **5 PSI 1 bar** | A,B | 0.008 ± 0.0023 | 0.074 ± 0.007 | 0.050 ± 0.0011 | 0.642 ± 0.015 | -0.047 ± 0.050 | 0.050 ± 0.056 | 0.287 ± 0.007 | 0.120 ± 0.016 | 0.033 ± 0.003 |
| **20 PSI 2 bar** | C,D | 0.014 ± 0.002 | 0.059 ± 0.004 | 0.098 ± 0.006 | 0.613 ± 0.008 | 0.041 ± 0.027 | -0.042 ± 0.028 | 0.243 ± 0.004 | 0.329 ± 0.010 | 0.033 ± 0.003 |
| **40 PSI 3.5 bar** | E,F | 0.011 ± 0.003 | 0.064 ± 0.007 | 0.141 ± 0.012 | 0.602 ± 0.015 | 0.094 ± 0.045 | -0.100 ± 0.046 | 0.212 ± 0.009 | 0.617 ± 0.020 | 0.033 ± 0.003 |
| **75 PSI 6.25 bar** | G,H | 0.017 ± 0.002 | 0.049 ± 0.004 | 0.092 ± 0.007 | 0.647 ± 0.009 | 0.059 ± 0.032 | -0.058 ± 0.033 | 0.275 ± 0.005 | 0.126 ± 0.010 | 0.033 ± 0.003 |
| **100 PSI 8.5 bar** | I,J | 0.016 ± 0.001 | 0.042 ± 0.004 | 0.121 ± 0.006 | 0.630 ± 0.007 | 0.025 ± 0.026 | -0.019 ± 0.027 | 0.208 ± 0.004 | 0.467 ± 0.009 | 0.033 ± 0.003 |

## 8.6. Pyruvate exchange kinetics of $^{13}C^2$ of 3 and 2 in the function of pressure with 100 mM of DMSO

**Sample:** 4 mM of [Ir-$d_{22}$], 40 mM of $^{13}$C-pyr, and 100 mM of DMSO in 600 µL of methanol-$d_4$

**Experiment:** In all the experiments below, we used a 9.4 T NMR system at 265 K, and a SEPP-SPINEPTplus-SABRE sequence for $^{13}$C hyperpolarization of the selected carbon. We experimented twice: first by hyperpolarizing the $^{13}C^2$-pyr of **3**, and observing its transition to **2** and then to free pyruvate **(Figure S14-top)**. In the second experiment, we hyperpolarized the $^{13}C^2$-pyr of **2** and tracked its conversion to **3** before dissociation **(Figure S14-bottom)**. The pressure was varied from 1 bar to 8.5 bar.

**Simulations:** These experiments provided two sets of data, which we analyzed using a global fit with the model **($^{13}C^2$-3 ↔$^{13}C^2$-2)↔$^{13}C^2$-free**, as explained in Section 2, to extract the relevant kinetic parameters. $R_{\mathrm{pyr}}$ was measured for the free carbon.

**Results:**

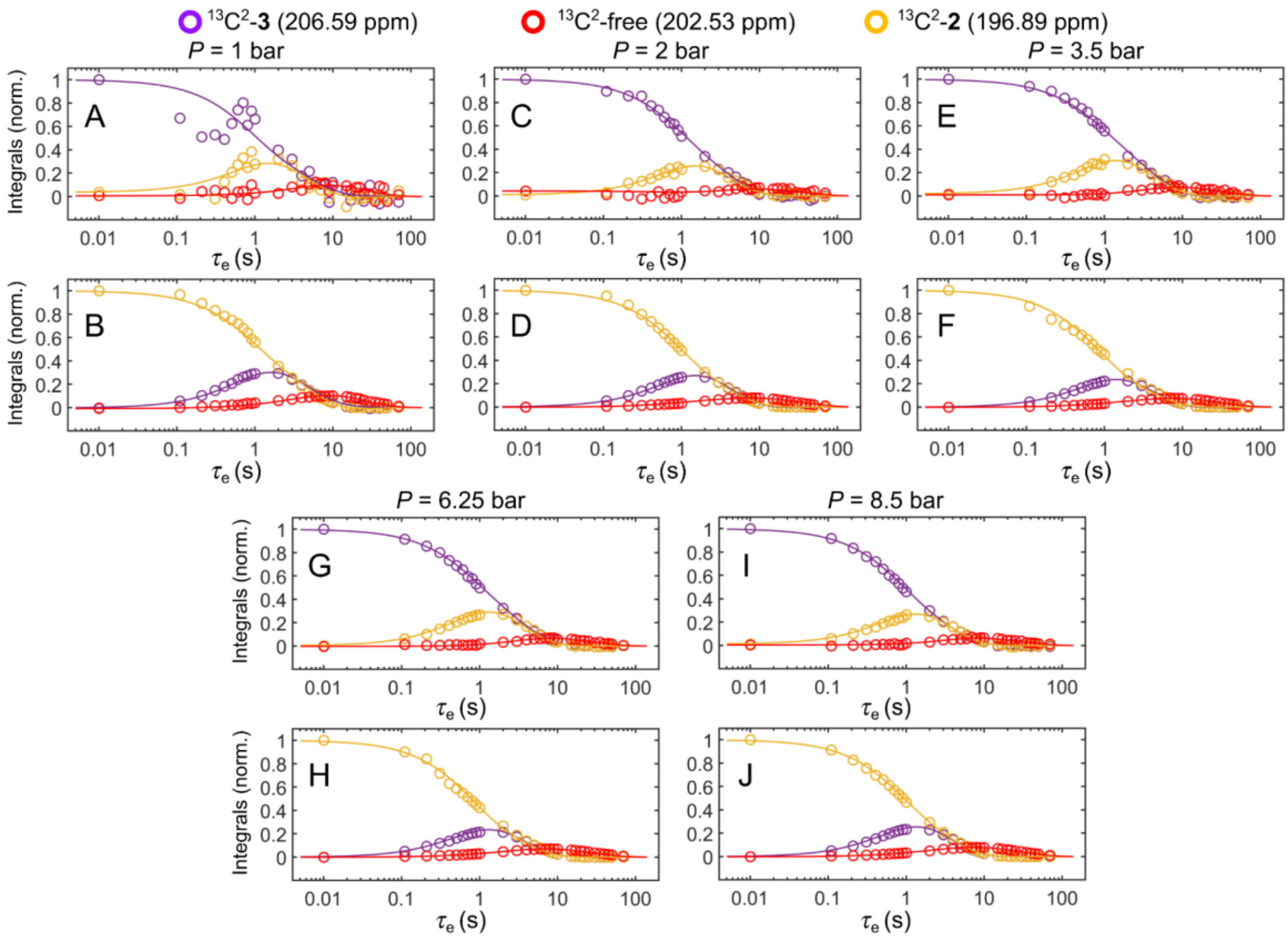

**Figure S14: Normalized integrals of SEPP-SPINEPTplus-SABRE kinetics.** $^{13}$C NMR integrals of the equatorial $^{13}C^2$-epyr **3** at 206.59 ppm (purple), free $^{13}C^2$-fpyr at 202.53 ppm (red), and equatorial $^{13}C^2$-epyr **2** at 196.89 ppm (orange) at 1 bar (A, B), 2 bar (C, D), 3.5 bar (E, F), 6.25 bar (G, H), and 8.5 bar (I, J) at 265 K. The curves are fitted using model **($^{13}C^2$-3↔$^{13}C^2$-2)↔$^{13}C^2$-free**. Estimated exchange rate constants, dissociations, associations, and relaxation-exchanges are given in Table S12.

**Table S12.** Fitting parameters of SEPP-SPINEPTplus-SABRE kinetics of both experiments for each pressure using model ($^{13}C^2$-**3** ↔$^{13}C^2$-**2**)↔$^{13}C^2$-free. Parameters are defined in section 2.

| Back pressure<br>Bubbling pressure | Fig | $k_{3-\mathrm{pyr}}$ (s$^{-1}$) | $k_{2-\mathrm{pyr}}$ (s$^{-1}$) | $k_{3\to2}$ (s$^{-1}$) | $k_{2\to3}$ (s$^{-1}$) | $k'^{\mathrm{P}}_{3a}$ (s$^{-1}$) | $k'^{\mathrm{P}}_{2a}$ (s$^{-1}$) | $R_3$ (s$^{-1}$) | $R_2$ (s$^{-1}$) | $R_{\mathrm{pyr}}$ (s$^{-1}$) |
|---|---|---|---|---|---|---|---|---|---|---|
| **5 PSI**<br>**1 bar** | A,B | 0.023 ± 0.020 | 0.045 ± 0.019 | 0.522 ± 0.050 | 0.576 ± 0.051 | 0.177 ± 0.135 | -0.169 ± 0.137 | 0.278 ± 0.043 | 0.117 ± 0.044 | 0.033 ± 0.003 |
| **20 PSI**<br>**2 bar** | C,D | 0.013 ± 0.004 | 0.041 ± 0.004 | 0.505 ± 0.010 | 0.531 ± 0.011 | -0.048 ± 0.036 | 0.050 ± 0.038 | 0.215 ± 0.009 | 0.271 ± 0.010 | 0.033 ± 0.003 |
| **40 PSI**<br>**3.5 bar** | E,F | 0.010 ± 0.004 | 0.049 ± 0.005 | 0.650 ± 0.007 | 0.493 ± 0.007 | -0.057 ± 0.034 | 0.085 ± 0.046 | 0.058 ± 0.010 | 0.454 ± 0.012 | 0.033 ± 0.003 |
| **75 PSI**<br>**6.25 bar** | G,H | 0.012 ± 0.003 | 0.047 ± 0.004 | 0.663 ± 0.009 | 0.537 ± 0.009 | 0.004± 0.031 | 0.001 ± 0.032 | 0.130 ± 0.008 | 0.442 ± 0.010 | 0.033 ± 0.003 |
| **100 PSI**<br>**8.5 bar** | I,J | 0.006 ± 0.003 | 0.051 ± 0.003 | 0.587 ± 0.007 | 0.567 ± 0.007 | -0.004 ± 0.025 | 0.012 ± 0.026 | 0.297 ± 0.007 | 0.268 ± 0.007 | 0.033 ± 0.003 |

## 8.7. Pyruvate exchange kinetics of $^{13}C^2$ of 3 and 2 in the function of DMSO

**Sample:** 4 mM of [Ir-*d*$_{22}$], 40 mM of $^{13}$C-pyr, and a variable amount of DMSO between 20 and 300 mM in 600 µL of methanol-*d*$_4$.

**Experiment:** In all the experiments below, we used a 9.4 T NMR system at 265 K, p$H_2$ pressure of 8.5 bar, and a SEPP-SPINEPTplus-SABRE sequence for $^{13}$C hyperpolarization of the selected carbon. We experimented twice: first by hyperpolarizing the $^{13}C^2$-pyr of **3**, and observing its transition to **2** and then to free pyruvate **(Figure S15-top)**. In the second experiment, we hyperpolarized the $^{13}C^2$-pyr of **2** and tracked its conversion to **3** before dissociation **(Figure S15-bottom)**.

**Simulations:** These experiments provided two sets of data, which we analyzed using a global fit with the model **($^{13}C^2$-3 ↔$^{13}C^2$-2)↔$^{13}C^2$-free**, as explained in Section 2, to extract the relevant kinetic parameters. $R_{\mathrm{pyr}}$ was measured for the free carbon.

**Results:**

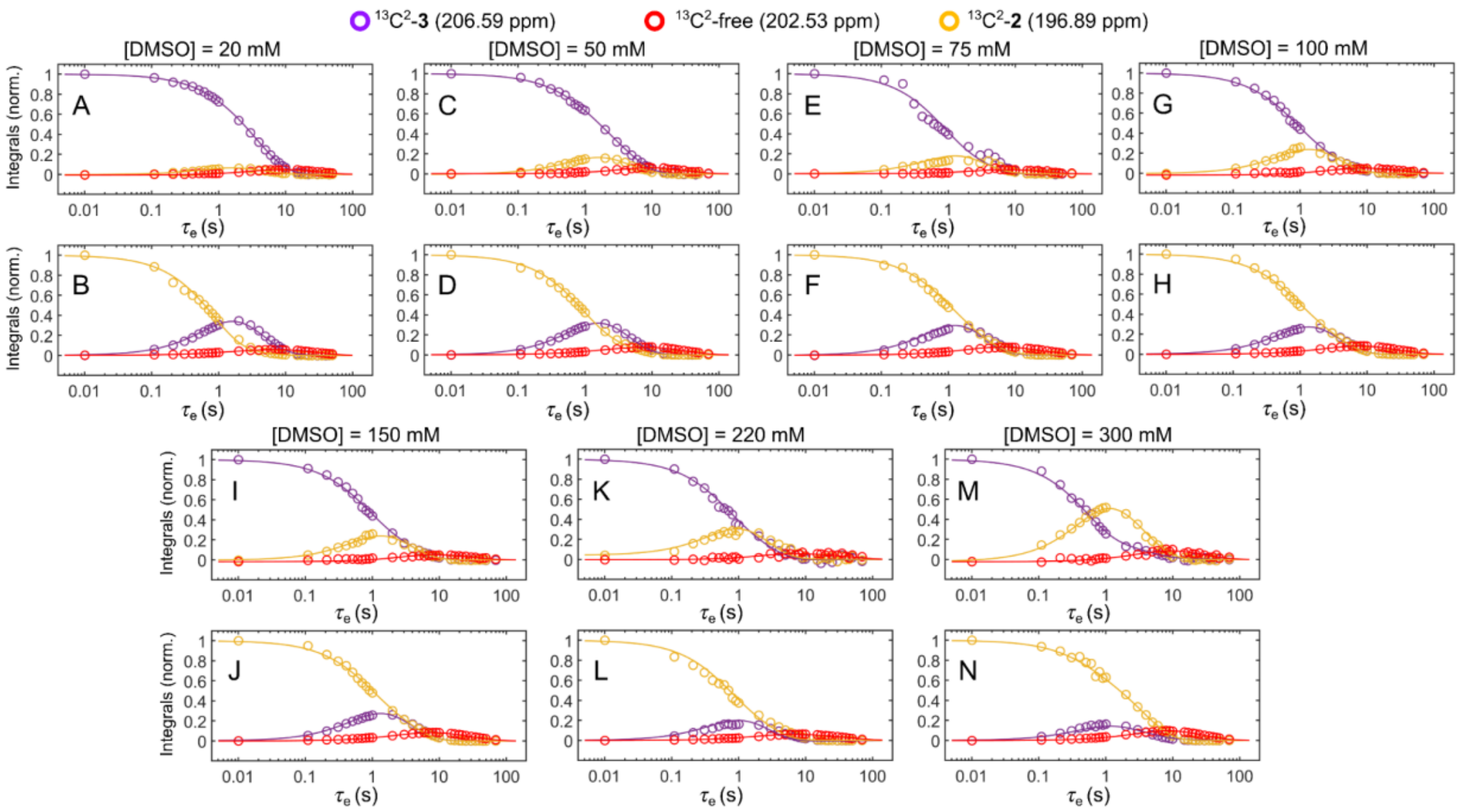

**Figure S15: Normalized integrals of SEPP-SPINEPTplus-SABRE kinetics.** $^{13}$C NMR integrals of the equatorial $^{13}C^2$-epyr **3** at 206.59 ppm (purple), free $^{13}C^2$-fpyr at 202.53 ppm (red), and equatorial $^{13}C^2$-epyr **2** at 196.89 ppm (orange) for 20 mM of [DMSO] (A, B), 50 mM (C, D), 75mM (E, F), 100 mM (G, H), 150 mM (I, J), 220 mM (K, L),

and 300 mM (M, N) at 265 K. The curves are fitted using model **($^{13}C^2$-3↔$^{13}C^2$-2)↔$^{13}C^2$-free**. Estimated exchange rate constants, dissociations, associations, and relaxation-exchanges are given in Table S13.

**Table S13**. Fitting parameters of SEPP-SPINEPTplus-SABRE kinetics of both experiments for each concentration using model ($^{13}C^2$-**3** ↔$^{13}C^2$-**2**)↔$^{13}C^2$-free. Parameters are defined in section 2.

| DMSO (mM) | Fig | $k_{3-\mathrm{pyr}}$, ($s^{-1}$) | $k_{2-\mathrm{pyr}}$, ($s^{-1}$) | $k_{3\to2}$ ($s^{-1}$) | $k_{2\to3}$ ($s^{-1}$) | $k'^{\mathrm{P}}_{3a}$ ($s^{-1}$) | $k'^{\mathrm{P}}_{2a}$ ($s^{-1}$) | $R_3$ ($s^{-1}$) | $R_2$ ($s^{-1}$) | $R_{\mathrm{pyr}}$ ($s^{-1}$) |
|---|---|---|---|---|---|---|---|---|---|---|
| **20** | A,B | 0.016 ± 0.001 | 0.042 ± 0.004 | 0.121 ± 0.006 | 0.630 ± 0.007 | 0.025 ± 0.026 | -0.019 ± 0.027 | 0.208 ± 0.004 | 0.467 ± 0.009 | 0.033 ± 0.003 |
| **50** | C,D | 0.012 ± 0.001 | 0.049 ± 0.002 | 0.311 ± 0.005 | 0.596 ± 0.006 | 0.012 ± 0.018 | -0.012 ± 0.019 | 0.213 ± 0.004 | 0.314 ± 0.006 | 0.033 ± 0.003 |
| **75** | E,F | 0.012 ± 0.009 | 0.048 ± 0.009 | 0.414 ± 0.023 | 0.650 ± 0.026 | -0.049 ± 0.082 | 0.06 ± 0.085 | 0.593 ± 0.025 | 0.070 ± 0.024 | 0.033 ± 0.003 |
| **100** | G,H | 0.022 ± 0.003 | 0.040 ± 0.004 | 0.533 ± 0.009 | 0.607 ± 0.010 | 0.024 ± 0.034 | -0.02 ± 0.034 | 0.383 ± 0.009 | 0.188 ± 0.009 | 0.033 ± 0.003 |
| **150** | I,J | 0.020 ± 0.010 | 0.040 ± 0.009 | 0.849 ± 0.031 | 0.595 ± 0.028 | -0.155 ± 0.114 | 0.144 ± 0.114 | 0.338 ± 0.030 | 0.489 ± 0.026 | 0.033 ± 0.003 |
| **220** | K,L | 0.012 ± 0.013 | 0.045 ± 0.009 | 1.308 ± 0.036 | 0.592 ± 0.028 | -0.317± 0.113 | 0.320 ± 0.114 | 0.132 ± 0.031 | 0.418 ± 0.020 | 0.033 ± 0.003 |
| **300** | M,N | 0.007 ± 0.009 | 0.048 ± 0.004 | 1.556 ± 0.022 | 0.483 ± 0.015 | 0.000 ± 0.049 | 0.000 ± 0.050 | 0.477 ± 0.021 | 0.209 ± 0.009 | 0.033 ± 0.003 |

## 8.8. Pyruvate exchange kinetics of $^{13}C^2$ of 3 and 2 in the function of pyruvate

**Sample:** 4 mM of [Ir-*d*$_{22}$], 20 mM of DMSO, and a variable amount of $^{13}$C-pyr between 40 and 174 mM in 600 µL of methanol-*d*$_4$.

**Experiment:** In all the experiments below, we used a 9.4 T NMR system at 265 K, $pH_2$ pressure of 8.5 bar, and a SEPP-SPINEPTplus-SABRE sequence for $^{13}$C hyperpolarization of the selected carbon. We experimented twice: first by hyperpolarizing the $^{13}C^2$-pyr of **3**, and observing its transition to **2** and then to free pyruvate **(Figure S16-top)**. In the second experiment, we hyperpolarized the $^{13}C^2$-pyr of **2** and tracked its conversion to **3** before dissociation **(Figure S16-bottom)**.

**Simulations:** These experiments provided two sets of data, which we analyzed using a global fit with the model **($^{13}C^2$-3 ↔$^{13}C^2$-2)↔$^{13}C^2$-free**, as explained in Section 2, to extract the relevant kinetic parameters. $R_{\mathrm{pyr}}$ was measured for the free carbon.

**Results:**

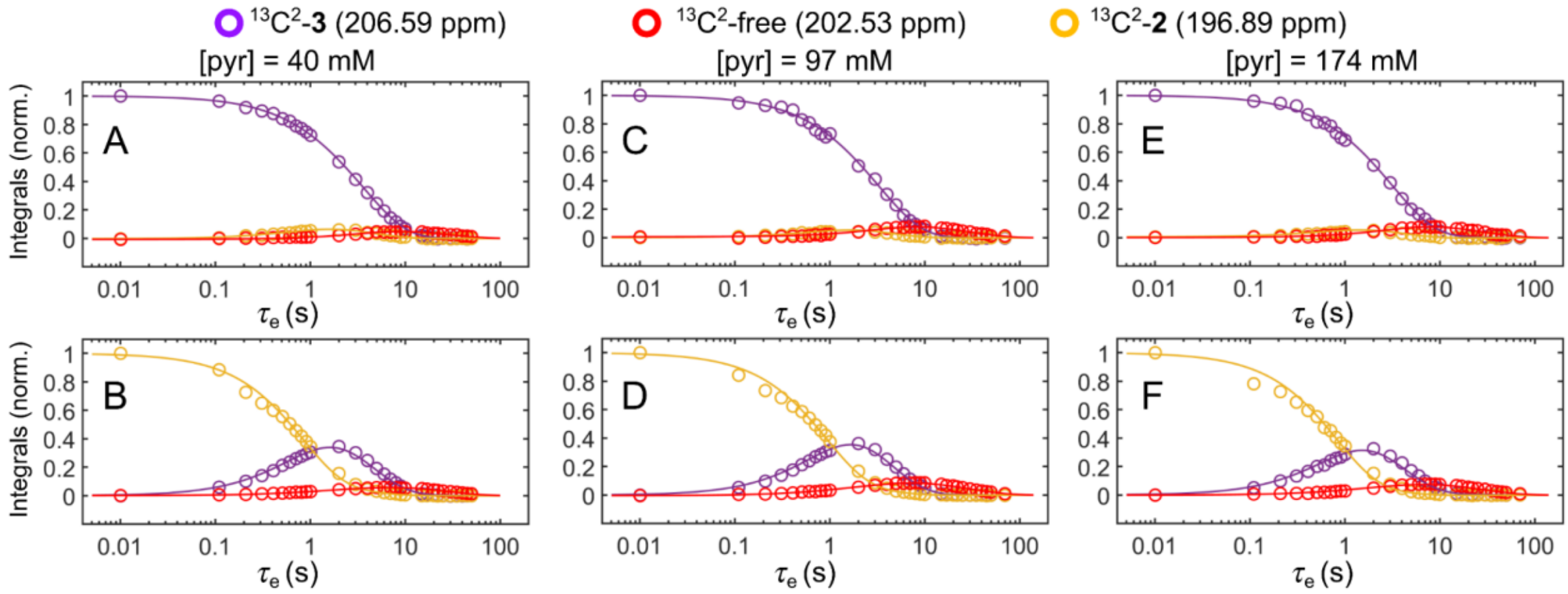

**Figure S16: Normalized integrals of SEPP-SPINEPTplus-SABRE kinetics.** $^{13}$C NMR integrals of the equatorial $^{13}C^2$-epyr **3** at 206.59 ppm (purple), free $^{13}C^2$-fpyr at 202.53 ppm (red), and equatorial $^{13}C^2$-epyr **2** at 196.89 ppm (orange) for 40 mM of pyruvate (A, B), 97 mM (C, D), and 174 mM (E, F). The curves are fitted using model **($^{13}C^2$-3↔$^{13}C^2$-2)↔$^{13}C^2$-free**. Estimated exchange rate constants, dissociations, associations, and relaxation-exchanges are given in Table S14.

**Table S14**. Fitting parameters of SEPP-SPINEPTplus-SABRE kinetics of both experiments for each concentration using model ($^{13}C^{2}$-**3** ↔ $^{13}C^{2}$-**2**)↔$^{13}C^{2}$-free. Parameters are defined in section 2.

| **Pyr (mM)** | **Fig** | $k_{3-\mathrm{pyr}}$, (s$^{-1}$) | $k_{2-\mathrm{pyr}}$, (s$^{-1}$) | $k_{3\to2}$ (s$^{-1}$) | $k_{2\to3}$ (s$^{-1}$) | $k'^{\mathrm{P}}_{3\mathrm{a}}$ (s$^{-1}$) | $k'^{\mathrm{P}}_{2\mathrm{a}}$ (s$^{-1}$) | $R_{\mathbf{3}}$ (s$^{-1}$) | $R_{\mathbf{2}}$ (s$^{-1}$) | $R_{\mathrm{pyr}}$ (s$^{-1}$) |
|---|---|---|---|---|---|---|---|---|---|---|
| **40** | A&B | 0.016 ± 0.001 | 0.042 ± 0.004 | 0.121 ± 0.006 | 0.630 ± 0.007 | 0.025 ± 0.026 | -0.019 ± 0.027 | 0.208 ± 0.004 | 0.467 ± 0.009 | 0.033 ± 0.003 |
| **97** | E&F | 0.024 ± 0.002 | 0.059 ± 0.004 | 0.095 ± 0.007 | 0.627 ± 0.009 | 0.014 ± 0.024 | -0.011 ± 0.025 | 0.240 ± 0.005 | 0.348 ± 0.011 | 0.033 ± 0.003 |
| **174** | G&H | 0.027 ± 0.003 | 0.054 ± 0.006 | 0.105 ± 0.009 | 0.608 ± 0.011 | 0.046 ± 0.029 | -0.041 ± 0.03 | 0.237 ± 0.007 | 0.505 ± 0.015 | 0.033 ± 0.003 |

## 9. Dissociation constant rates of $^{13}C^2$-3 & $^{13}C^2$-2

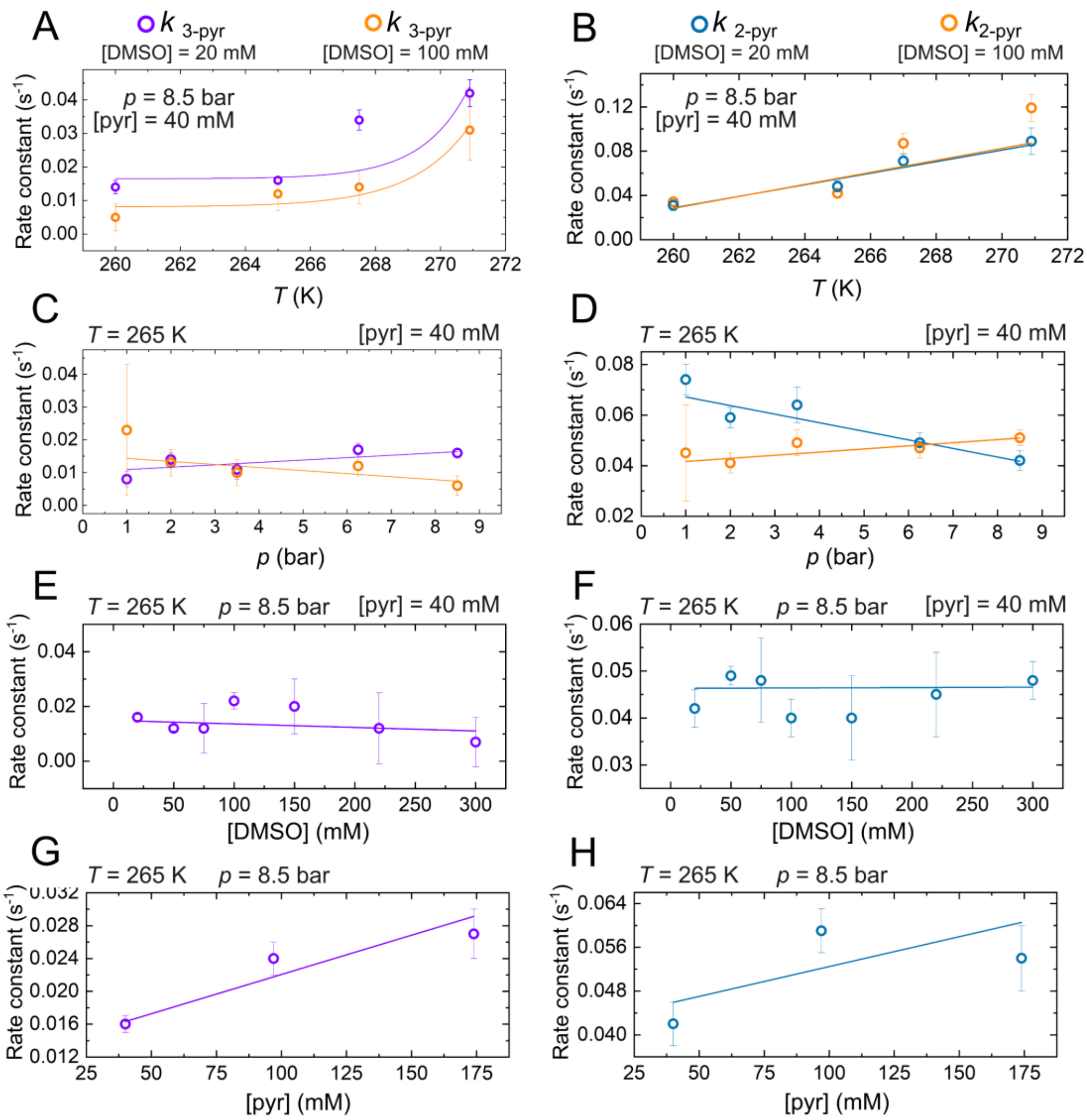

**Figure S17: Dissociation rate of $^{13}C^2$ of 3 and 2.** Dissociation constant rates $k_{3-\mathrm{pyr}}$, and $k_{2-\mathrm{pyr}}$, of tables S7-12 are plotted as a function of temperature (A, B), [DMSO] (C, D), $pH_2$ pressure (E, F), and [pyr] (G, H) by performing a global fit of the model IrHH↔$H_2$. $k_{3-\mathrm{pyr}}$ dissociation constant rate of $^{13}C^2$-**3** (purple) with 20 mM of DMSO in the sample and 100 mM of DMSO (orange). $k_{2-\mathrm{pyr}}$ dissociation constant rate of $^{13}C^2$-**2** (blue) with 20 mM of DMSO in the sample and 100 mM of DMSO (orange).

## 10. $T_1$ measurements for $H_2$ (4.6 ppm) and free-$^{13}C^2$ (202.5 ppm) as a function of temperature

**Sample:** 60 mM of $^{13}$C-pyr, in 400 μL of methanol-$d_4$.

**Experiment:** We performed an inversion recovery experiment to measure the T1 of the system's free hydrogen and free carbon, thereby reducing the number of fitting parameters in the models. The sample was bubbled with $H_2$ for 30 min at 8.5 bar to ensure that no more oxygen remained, which might affect the relaxation time. The sample was maintained at a pressure of 8.5 bar throughout the measurement.

**Results:**

**Table S15**: $T_1$ measurements of $H_2$ and free-$^{13}C^2$.

| Temperature (K) | $T_1$ of $H_2$ (s) | $T_1$ of Free-$^{13}C^2$ (s) |
|---|---|---|
| 260 | 1.94 ± 0.07 | 29.24 ± 1.28 |
| 265 | 1.93 ± 0.07 | 30.27 ± 1.35 |
| 267.5 | 1.88 ± 0.06 | 34.02 ± 1.26 |
| 270.9 | 1.90 ± 0.03 | 34.34 ± 1.74 |
| 275.5 | 1.82 ± 0.06 | 35.17 ± 1.28 |

# 11. Exchange rates of $H^{a,b}$-1

**Sample:** In this section, 4 sets of experiments were performed. In the case of [DMSO] and [pyr] experiments, the concentration of a key component was systematically varied.

1. **Temperature and pressure experiments (A, C):** 4 mM of [Ir-*d*22], 20 mM of DMSO, and 40 mM $^{13}$C-pyr in 600 µL of methanol-*d*3.
2. **[DMSO] experiments (B):** 4 mM of [Ir-*d*22], a variable amount of DMSO, and 40 mM $^{13}$C-pyr in 600 µL of methanol-*d*3.
3. **[pyr] experiments (D):** 4 mM of [Ir-*d*22], 20 mM of DMSO, and a variable amount of $^{13}$C-pyr in 600 µL of methanol-*d*3.

**Experiment:** We used a 9.4 T NMR system at 265 K, and a $pH_2$ pressure of 8.5 bar.

**Results:** we plotted the exchange rate $k_{1ex}^{\mathrm{HH}}$ and dissociation rate $k_{1-\mathrm{H2}}$ of tables 4.1, 5.1, 6.1 and 7.1.

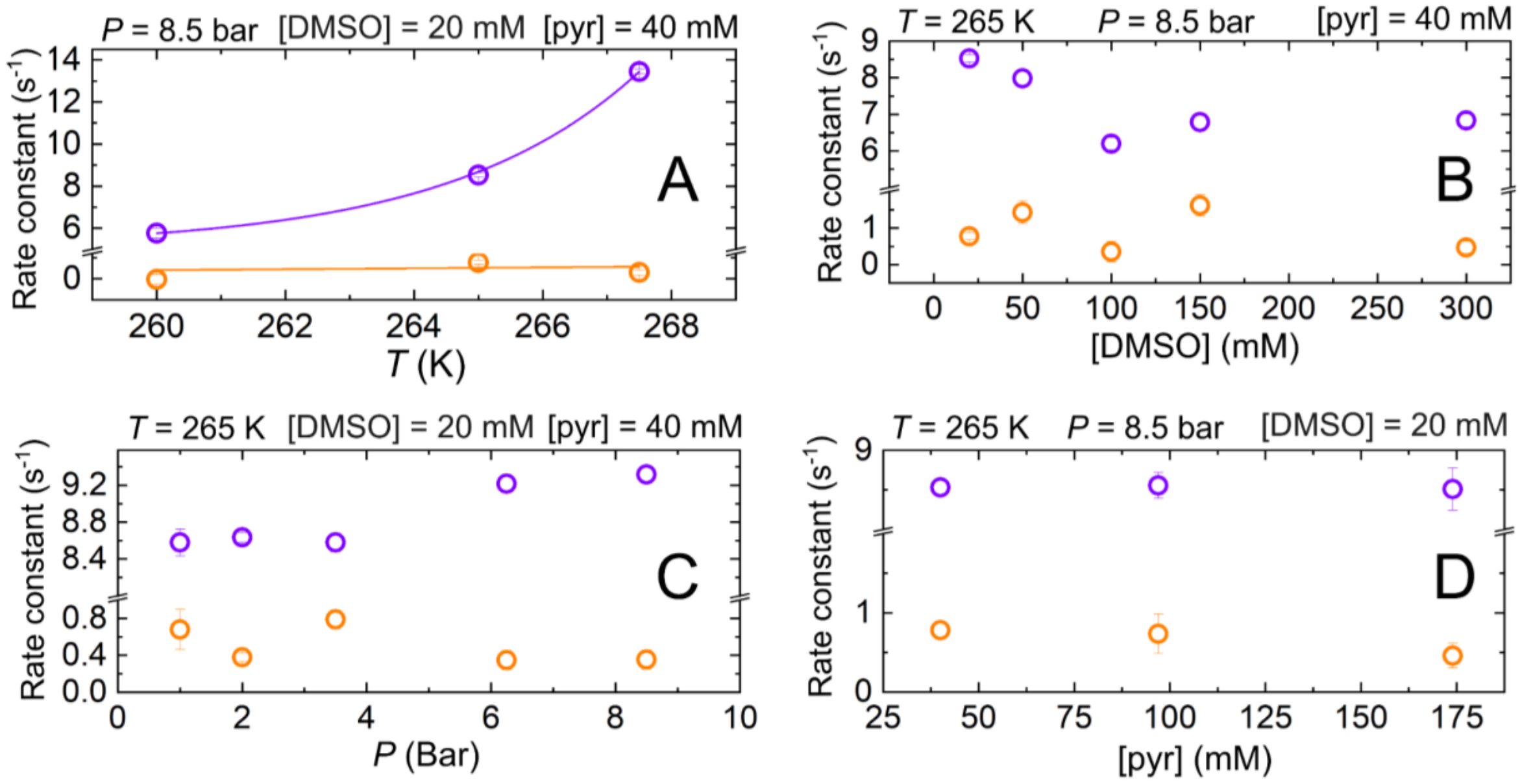

**Figure S18: IrHH exchange rates of $H^{a,b}$-1.** Extracted parameters as a function of temperature (A), [DMSO] (B), $pH_2$ pressure (C), and [pyr] (D) by performing a global fit of the model IrHH↔$H_2$. Both hydrogens can exchange within the complex, with the exchange rate $k_{1ex}^{\mathrm{HH}}$ (orange), dissociate from the complex with the exchange rate $k_{1-\mathrm{H2}}$ (purple).

# 12. Thermal polarized studies on complexes 1, 2 & 3 to determine their relative populations

**Sample:** In this section, three sets of experiments were performed. In each case, the concentration of a key component was systematically varied.

1. **Temperature experiments:** 4 mM of [Ir-$d_{22}$], 20 mM of DMSO, and 40 mM $^{13}$C-pyr in 600 µL of methanol-$d_3$.
2. **[DMSO] experiments:** 4 mM of [Ir-$d_{22}$], a variable amount of DMSO, and 40 mM $^{13}$C-pyr in 600 µL of methanol-$d_3$.
3. **[pyr] experiments:** 4 mM of [Ir-$d_{22}$], 20 mM of DMSO, and a variable amount of $^{13}$C-pyr in 600 µL of methanol-$d_3$.

**Experiment:** We used a 9.4 T NMR system at 265 K, a p$H_2$ pressure of 8.5 bar, and 100 scans of a $^1$H thermal spectrum using the "zgesgp" sequence to suppress the water signal.

**Table S16.1** Normalized integrals of IrHH resonances of complexes **1**, **2**, and **3** as a function of temperature with 20 mM of DMSO and 40 mM of pyruvate. All data were normalized, and relative to the total hydride signal at the maximum of all complexes.

| Temp (K) | **3** | **2** | **1** | **1 + 2 + 3** | All complexes |
|---|---|---|---|---|---|
| 260 | 0.37724 | 0.12694 | 0.21128 | 0.71546 | 1 |
| 265 | 0.39852 | 0.11821 | 0.18681 | 0.70353 | 0.94847 |
| 267.5 | 0.32135 | 0.05676 | 0.10184 | 0.47995 | 0.88735 |
| 270.9 | 0.37841 | 0.09491 | 0.12507 | 0.59838 | 0.84191 |
| 275.5 | 0.36409 | 0.07416 | 0.09432 | 0.53257 | 0.76438 |

**Table S16.2** Normalized integrals of IrHH resonances of complexes **1**, **2**, and **3** as a function of DMSO concentrations at 265 K. All data were normalized to the maximum of all complexes.

| [DMSO] (mM) | **3** | **2** | **1** | **1 + 2 + 3** | All complexes |
|---|---|---|---|---|---|
| 20 | 0.34179 | 0.09387 | 0.1612 | 0.59686 | 0.91268 |
| 70 | 0.1189 | 0.25328 | 0.30509 | 0.67727 | 0.9604 |
| 120 | 0.08108 | 0.26103 | 0.32552 | 0.66763 | 0.98339 |
| 170 | 0.02448 | 0.28346 | 0.3804 | 0.68834 | 1 |

**Table S16.3** Normalized integrals of IrHH lines of complexes **1**, **2**, and **3** as a function of pyruvate concentrations at 265 K. All data were normalized to the maximum of all complexes.

| [pyr] (mM) | **3** | **2** | **1** | **1 + 2 + 3** | All complexes |
|---|---|---|---|---|---|
| 30 | 0.388 | 0.09647 | 0.1414 | 0.62596 | 1 |
| 50 | 0.36278 | 0.09605 | 0.11784 | 0.57668 | 0.98863 |
| 64 | 0.33224 | 0.07449 | 0.09774 | 0.50447 | 0.93334 |
| 100 | 0.27734 | 0.08418 | 0.11759 | 0.47911 | 0.88515 |

**Table S16.4** Normalized integrals of IrHH lines of complexes **1**, **2**, and **3** as a function of temperature from 220 K to 259 K with 20 mM of DMSO and 40 mM of pyruvate. All data were normalized to the maximum of all complexes. All these experiments were done in Finland.

| Temp (K) | **3** | **2** | **1** | **1 + 2 + 3** | All complexes |
|---|---|---|---|---|---|

| 220 | 0.05608 | 0.43972 | 0.23668 | 0.73248 | 1 |
|---|---|---|---|---|---|
| 235 | 0.13281 | 0.25325 | 0.30696 | 0.69301 | 0.91001 |
| 247 | 0.24789 | 0.12268 | 0.19273 | 0.5633 | 0.69963 |
| 259 | 0.35108 | 0.10711 | 0.14667 | 0.60487 | 0.84764 |

All experiments in Table S16.1-3 were conducted in Kiel, Germany, and experiments in Table S16.4 were conducted in Oulu, Finland. The data collected by both groups show consistent and continuous trends: we measured in the temperature range of 260-275.5 K, while the other set of measurements was performed between 220-260 K. The overlapping pattern across both datasets supports the validity and continuity of our results. Specifically, the integrals of complex **3** increase as a function of temperature, while the integrals corresponding to complexes **1** and **2** decrease accordingly.

# 13. $^{1}H$-$^{1}H$-NOE

**Sample:** 4 mM of [Ir-$d_{22}$], 20 mM of DMSO, and 40 mM $^{13}$C-pyr in 600 µL of methanol-$d_4$.

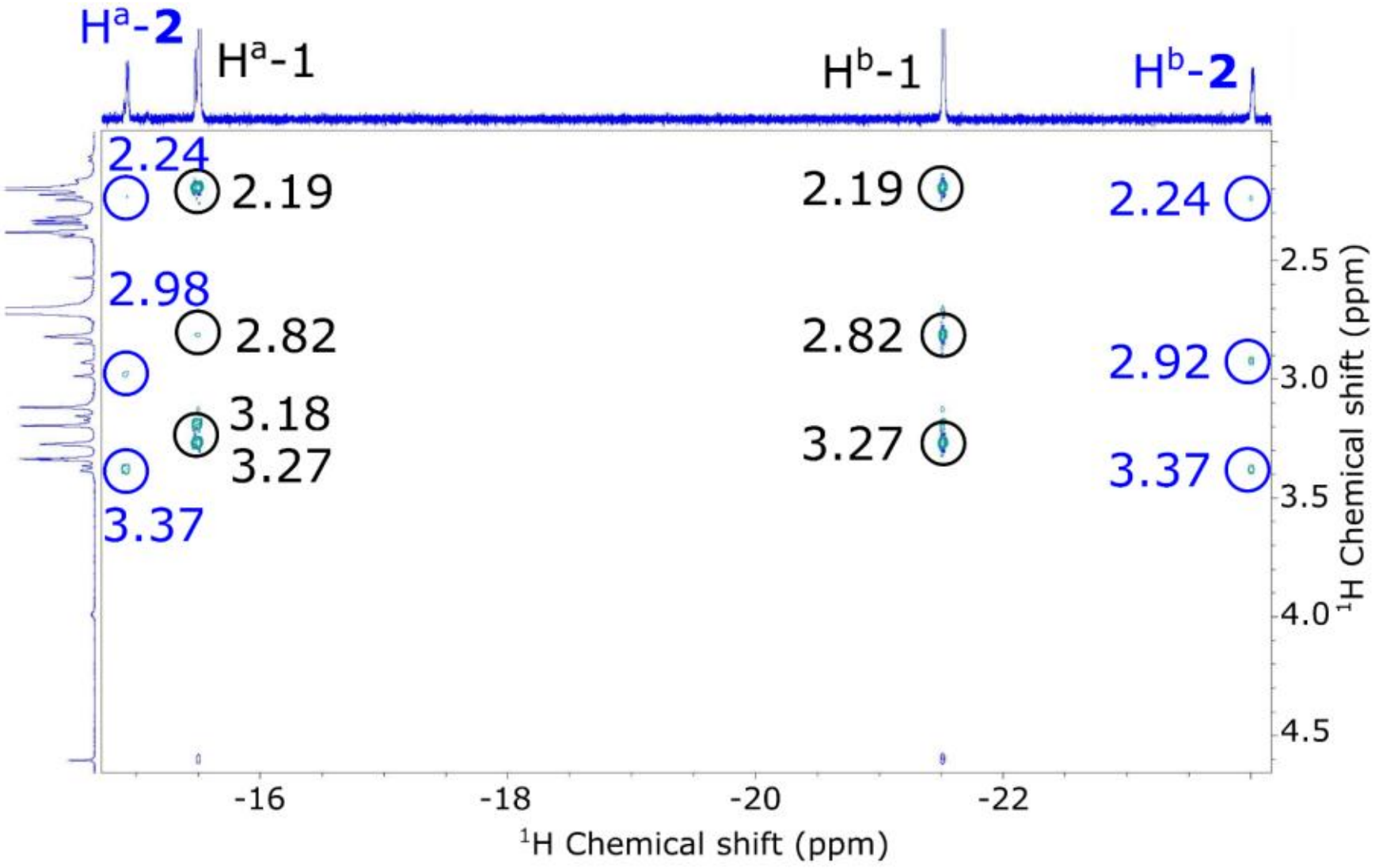

**Figure S19:** $^{1}H$-$^{1}H$ NOESY experiment. The experiment was conducted at 243 K, revealing interactions between the IrHH hydrides of **1** and **2** and the DMSO and pyruvate hydrogens.

# 14. $^{13}$C-DMSO-HMQC

**Sample:** [Ir] = 4.9 mM, [not labeled Pyr] = 51 mM, [$^{13}$C-DMSO]=20 mM, Meod-$d_3$ = 400 uL.

**Experiment:** $^1$H-$^{13}$C HMQC at 263K, at 8.5 bar of normal hydrogen. The $J$ interaction was set to 10 Hz, ns = 45 scans, aq = 0.81 s, d1 = 3 s. estimated $T_1$ = 1 s.

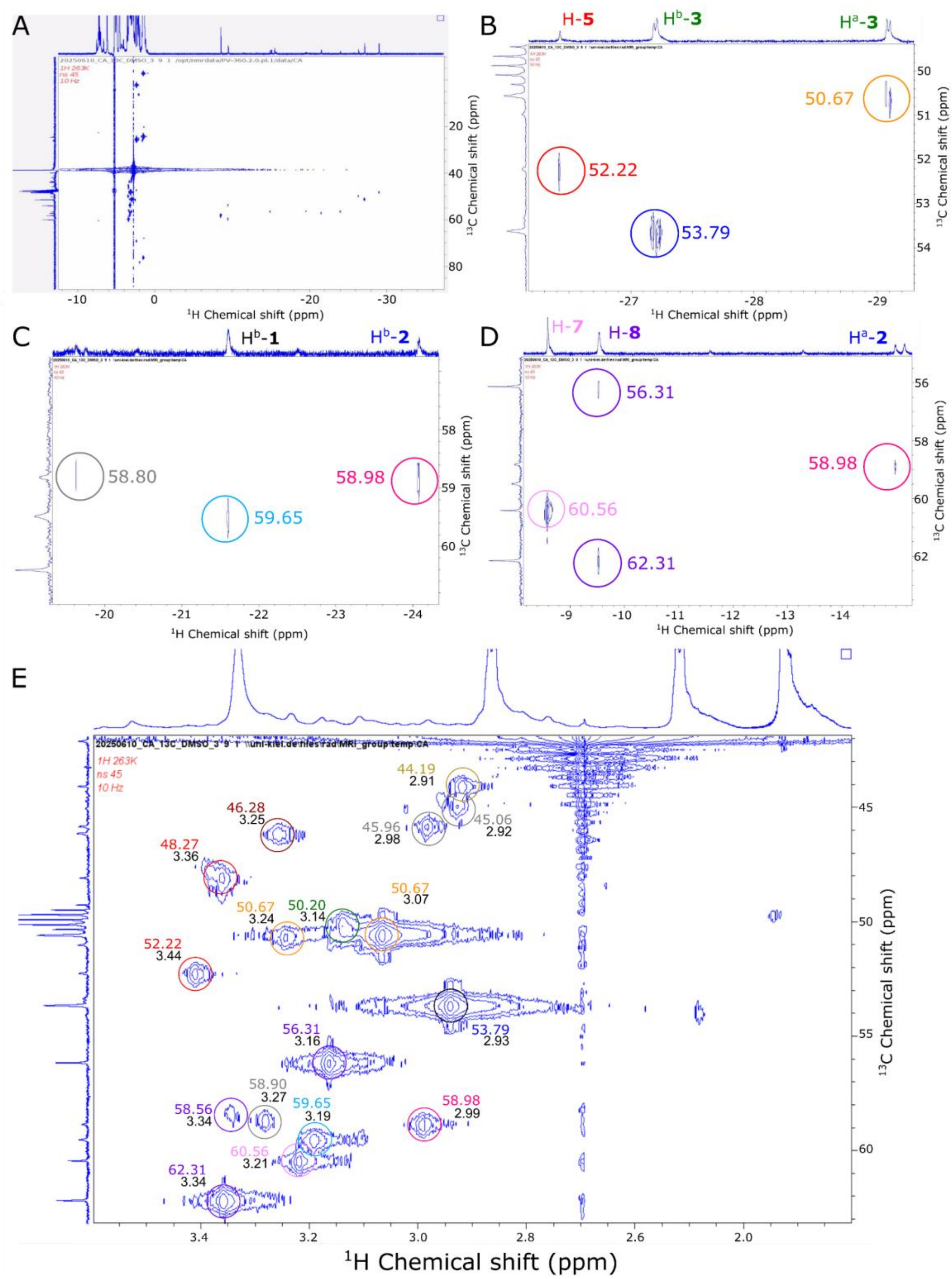

**Figure S20:** $^1$H-$^{13}$C HMQC spectra optimized for a 10 Hz. $^1$H-$^{13}$C $J$ coupling recorded in methanol-$d_3$ at 263 K.

# 15. SEPP-SPINEPT of $^{13}$C-DMSO

**Sample:** [Ir] = 4.9 mM, [not labeled Pyr] = 51 mM, [$^{13}$C-DMSO]=20 mM, Meod-$d_3$ = 400 uL.

**Experiment:** We applied the SEPP-SPINEPT pulse sequence to detect the $^{1}$H-$^{13}$C interactions between the hydride ligands and $^{13}$C-labeled DMSO in each complex. This experiment aimed to confirm the presence of two DMSO molecules in complex **2**.

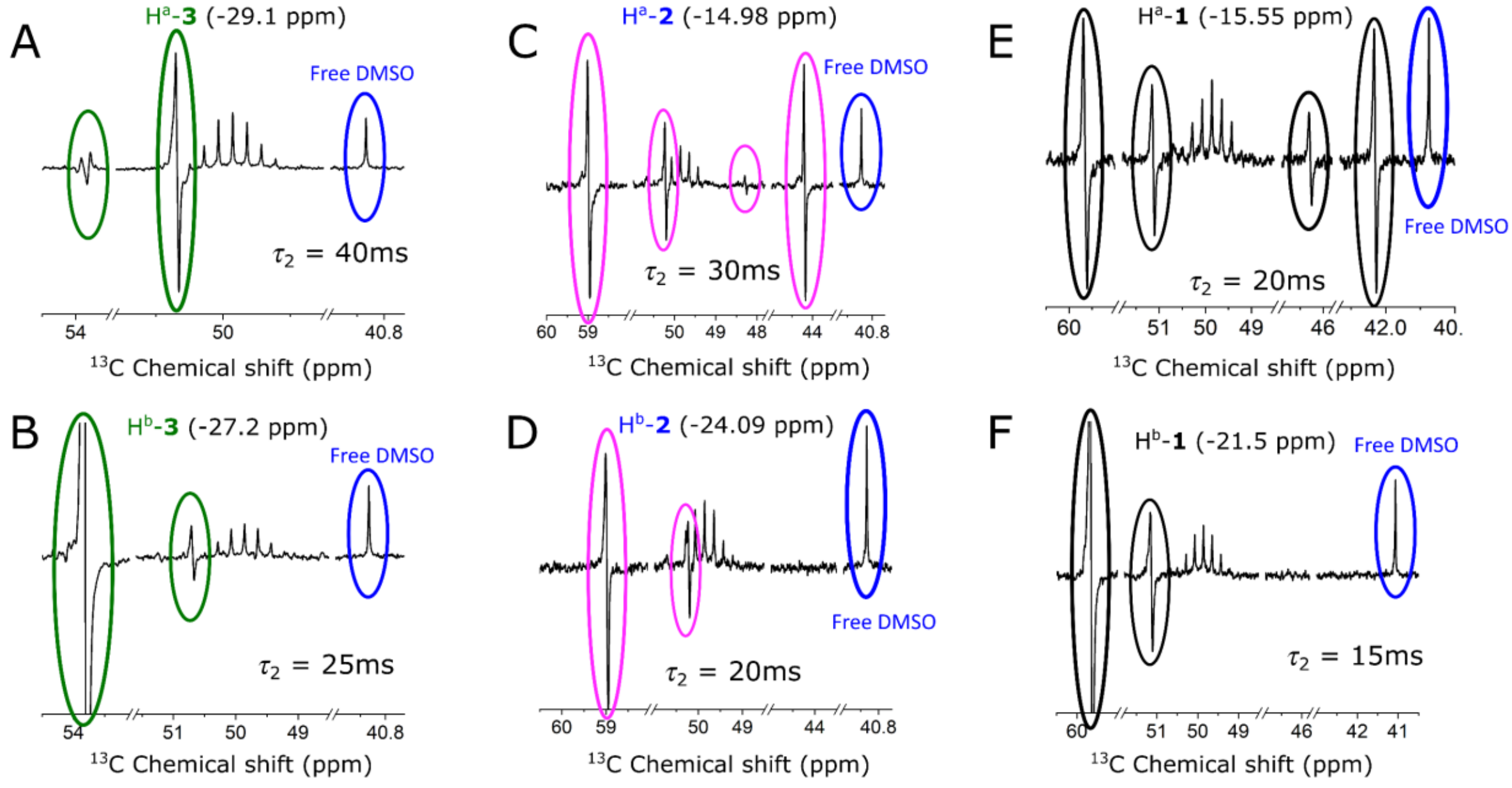

**Figure S21:** Phased NMR spectra at a maximum of polarization of the absolute intensity $^{13}$C NMR signals of bound DMSO after SEPP-SPINEPT at 267K. (A) This NMR spectrum shows polarization transferred from H$^{a}$-**3** (-29.1 ppm) to the bound $^{13}$C-DMSO of **3,** while (B) shows polarization transferred from H$^{b}$-**3** (-27.2 ppm) to the bound $^{13}$C-DMSO of **3.** (C) The spectrum shows polarization transferred from H$^{a}$-**2** (-14.98 ppm) to the bound $^{13}$C-DMSO of **2,** and (D) shows polarization transferred from H$^{b}$-**2** (-24.09 ppm) to the bound $^{13}$C-DMSO of **2.** (E) The spectrum shows polarization transferred from H$^{a}$-**1** (-15.55 ppm) to the bound $^{13}$C-DMSO of **1,** and (F) shows polarization transferred from H$^{b}$-**1** (-21.5 ppm) to the bound $^{13}$C-DMSO of **1.** The peak at 41 ppm corresponds to free DMSO in the sample. The multiplet (septet) at 49.86 ppm is assigned to the methanol.

**Table S17**: NMR resonances of **2**. The resonance labels correspond to those shown in **Figure S19-21.**

| Methanol-$d_4$ at 267K | | | | |
|---|---|---|---|---|
| **Interaction Number** | **$^{1}$H (ppm)** | **$^{13}$C (ppm)** | **$^{1}$H (ppm) DMSO** | **Type of interaction** |
| 1 | -14.98 | 58.98 | 2.99 | IrHH-DMSO 1 |
| 2 | -14.98 | 50.20 | 3.14 | IrHH-DMSO 2 |
| 3 | -14.98 | 48.27 | 3.37 | IrHH-DMSO 3 |
| 4 | -14.98 | 44.19 | 2.91 | IrHH-DMSO 4 |
| 5 | -24.09 | 58.98 | 2.99 | IrHH-DMSO 1 |
| 6 | -24.09 | 50.20 | 3.14 | IrHH-DMSO 2 |

# 16. Effect of intramolecular hydrogen exchange on SABRE

**Effect of intramolecular hydrogen exchange on SLIC-SABRE efficacy:** The SLIC-SABRE experiment is the most efficient for SABRE polarization of pyruvate at low field reported to date (70PravdivtsevJPCC,78SchmidtJPCL). Neglecting pyruvate exchange at low temperatures, and considering only hydrogen exchange (as previously reported in 70PravdivtsevJPCC) means $pH_2$ exchange should occur at an optimum rate (in this case, approximately 1-10 $s^{-1}$), for more rapid exchange, there is insufficient time to transfer polarization, while for too slow an exchange, relaxation destroys the spin order of $pH_2$. Intramolecular exchange of two hydride protons will reduce the polarization level (**Figure S22**). This is because for efficient polarization transfer, the scalar couplings between $pH_2$ and the target must be different, and these average under exchange, so polarization transfer slows down, and efficacy drops.

**The simulations were achieved using the following model.** To simulate the SABRE experiment, we used the formalism developed for the linear exchange model in (70PravdivtsevJPCC, 79KnechtRSCadv). Note the critical difference: we assume that only $H_2$ exchanges, while the substrate stays bound to the Ir-complex. This is a typical case for **3** at low temperatures. These assumptions can be expressed with the following model chemical reaction

$$\mathrm{IrHHPyr} \xrightarrow{\mathrm{H_2\ to\ pH_2\ exchange,}\ k_\mathrm{d}} \mathrm{IrH'H'Pyr}$$

$$\mathrm{IrH'HPyr} \xrightarrow{\mathrm{intramolecular\ H_2\ exchange,}\ k_\mathrm{ex}} \mathrm{IrHH'Pyr}$$

where $\mathrm{IrHHPyr}$ states for the active SABRE complex and prime symbol ( $'$ ) is used to distinguish hydrogens before and after exchange. For this chemical exchange, the following generalized Liouville-von Neuman equation has to be solved numerically:

$$\frac{d\hat{\rho}_{\mathrm{IrHHPyr}}}{dt} = \left[\hat{\hat{L}}_{\mathrm{IrHHPyr}} + k_{\mathrm{ex}}\left(\hat{\hat{Q}}^{\mathrm{HH}}_{\mathrm{H1}\leftrightarrow\mathrm{H2}} - \hat{\hat{1}}_{\mathrm{IrHHPyr}}\right) + k_{\mathrm{d}}\left(\hat{\hat{D}}^{\mathrm{H}_2^*}_{\mathrm{IrPyr}}\widehat{\widehat{Tr}}^{\mathrm{HH}}_{\mathrm{IrHHPyr}} - \hat{\hat{1}}_{\mathrm{IrHHPyr}}\right)\right]\hat{\rho}_{\mathrm{IrHHPyr}}$$

Here $\hat{\rho}_{\mathrm{IrHHPyr}}$ is the density matrix for the active Ir-complex with $H_2$ and pyruvate (Iyr), $\hat{\hat{L}}$ is the corresponding Liouville superoperator, $\hat{\hat{1}}$ is a unitary superoperator, $\hat{\hat{D}}^{\mathrm{H}_2^*}_{\mathrm{IrPyr}}$ is the superoperator of the direct product that adds $pH_2$ to IrPyr and $\widehat{\widehat{Tr}}^{\mathrm{HH}}_{\mathrm{IrHHPyr}}$ is a superoperator that removes HH from $\mathrm{IrHHPyr}$, $\hat{\hat{Q}}^{\mathrm{HH}}_{\mathrm{H1}\leftrightarrow\mathrm{H2}}$ is a super operator that exchanges two protons.

For example, for $k_d$ = 10 $s^{-1}$, predicted polarization is falling from 67.2% to 43.7% (or by 35%) when intramolecular exchange increases from 0 to 10 $s^{-1}$. For 3, the $k^{HH}{}_{3ex}$ ranged between 2 and 7 $s^{-1}$ in the studied temperature range.

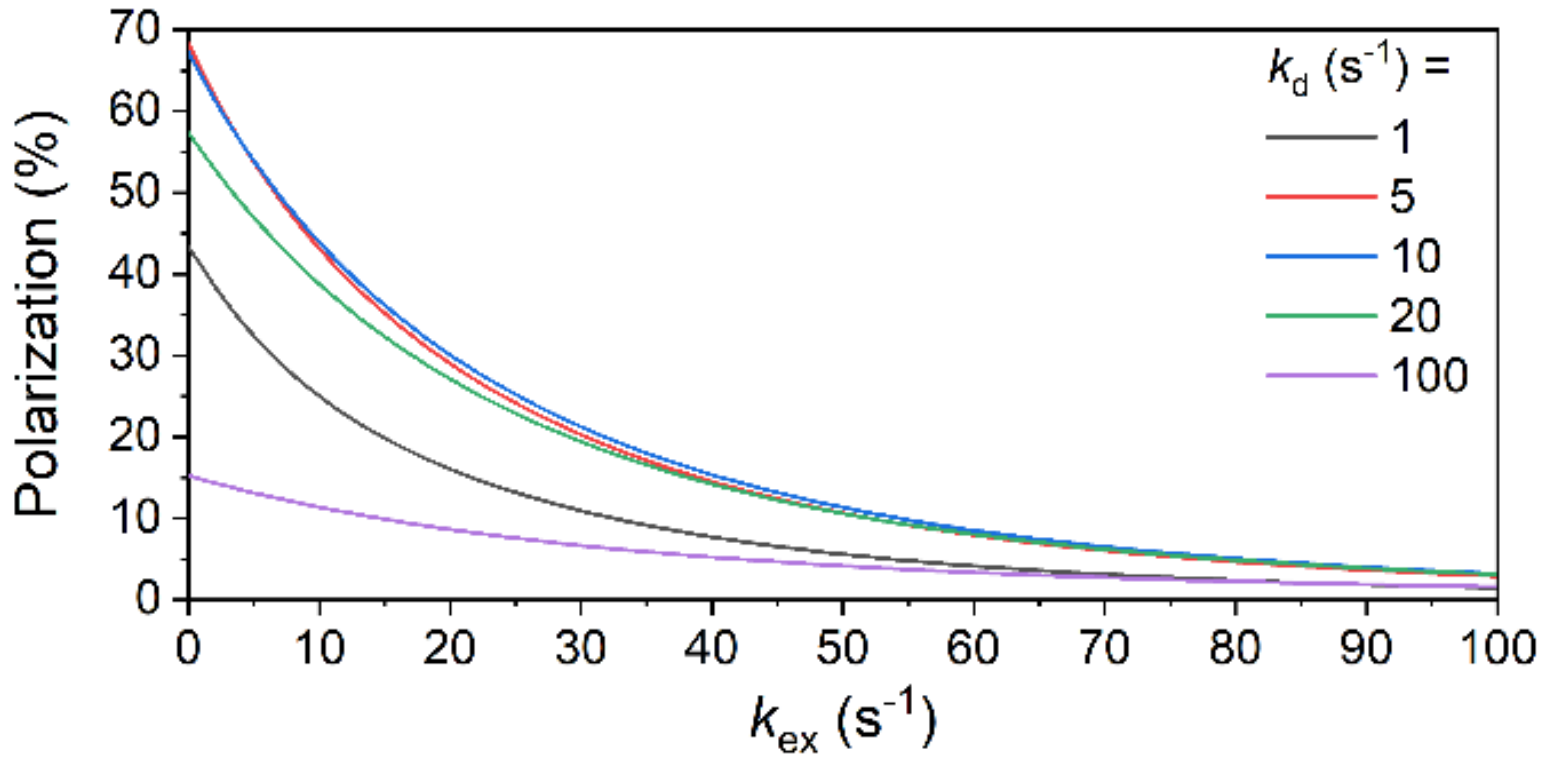

**Figure S22. SLIC-SABRE [1-$^{13}$C]pyruvate polarization as a function of intramolecular hydrogen exchange.** Intramolecular exchange is always detrimental to polarization transfer. *Simulation parameters*: three spin systems consisting of two protons and one $^{13}$C resembling 1-$^{13}$C of pyruvate. $T_1$ of $^1$H was 1 s, $T_1$ of $^{13}$C was 25 s, chemical shifts were -27.2 ppm and -

29.1 ppm for protons and 164 ppm for $^{13}C$, and $J$ coupling constants between protons and carbon were -0.55, 0.014 Hz, and between protons -10.48 Hz. Simulations were performed for a 50 μT magnetic field using the SLIC pulse with an amplitude of 1 μT along the X axis in the rotating frame of reference, resonant with $^{13}C$, for 60 s. Polarization along the X axis in the rotation frame of reference was generated by SLIC and is displayed here.

# 17. Other hyperpolarized complexes

**Sample:** 4 mM of [Ir-*d22*], 300 mM of DMSO, and 40 mM $^{13}$C-pyr in 600 µL of methanol-*d*$_4$.

**Experiment:** We used a 9.4 T NMR system at 265 K, a p$H_2$ pressure of 8.5 bar, and a SEPP-SABRE sequence for $^1$H hyperpolarization of the selected proton. Set 1 - hyperpolarization of H$^a$-**1** at -15.56 ppm using SEPP-SABRE, and following exchanging with H$^b$-**1** at -21.59 ppm, and free $H_2$ at 4.6 ppm (**Figure S16-left**). Set 2 - hyperpolarization of H$^b$-**1** using SEPP-SABRE and following exchange with H$^a$-**1**, and $H_2$ (**Figure S16-right**).

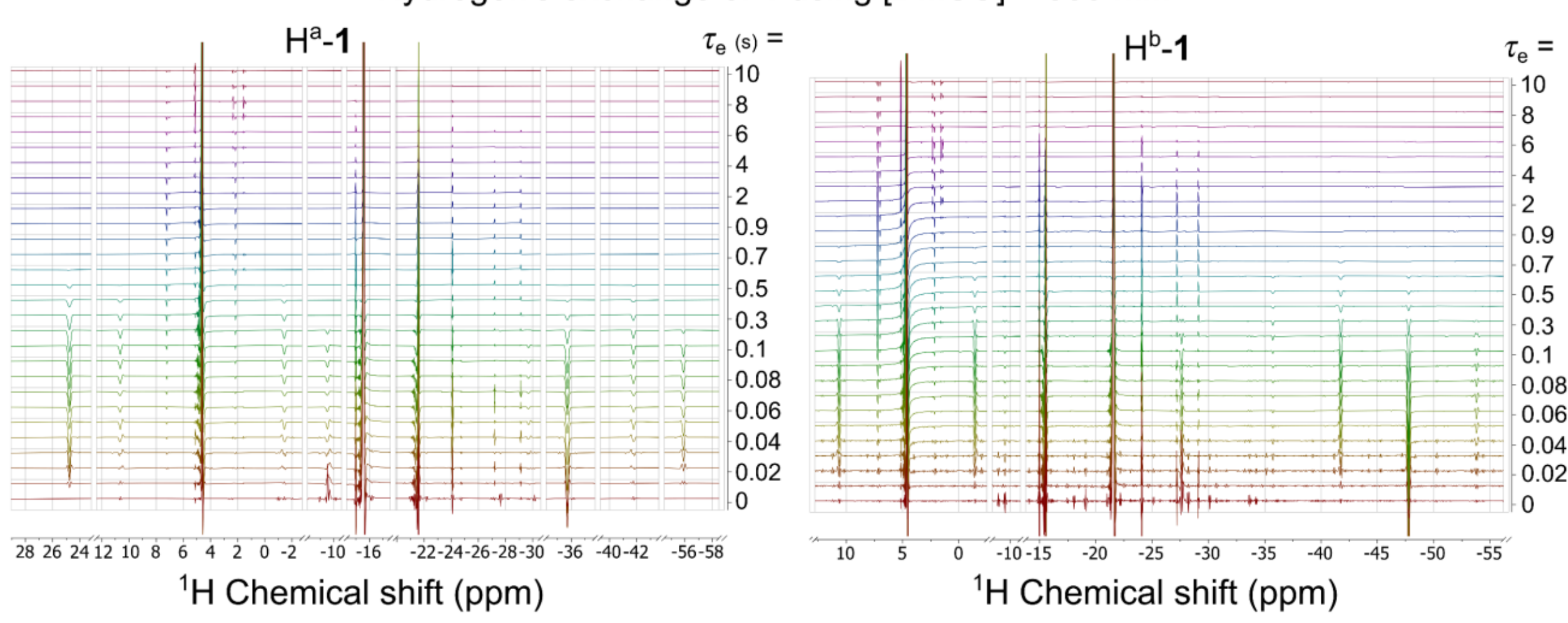

**Figure S23: Observation of hydride chemical exchange of H$^{a,b}$-1.** $^1$H NMR spectra at 265 K where selective pulse excites H$^a$-**1** (left) and H$^b$-**1** (right) using SEPP-SABRE pulse sequence. During the measurement of hydrogen exchange of H$^{a,b}$-**1** in the presence of 300 mM of DMSO, an unexpected polarization transfer from hydrogens of **1** to different complexes.

# 18. Electronic Structure Calculations

## 18.1. Computational details

If not mentioned otherwise, all calculations have been carried out at the B3LYP-D4/def2-TZVP CPCM (methanol) level of theory [54BeckeJCP-62CaldeweyherJCP]. Every geometry was fully optimized using tight convergence criteria and confirmed as a minimum structure via normal mode analysis. Note that while no extensive conformational searches have been carried out using refined search algorithms, several possible conformations of various structures have typically been screened before final optimization. For all calculations, the ORCA 6.1 program package has been used in [63NeeseWIRES2025,64NeeseWIRES2012]. If not stated otherwise, free energy corrections at 298.15K have been computed using the standard harmonic oscillator approximation as implemented in the thermochemistry module of ORCA [80GrimmeChemEurJ].

**Structure and Energetics of Compounds 2-4**

In a first series of calculations, the structures of compounds **2** to **4** were optimized, and relative Gibbs free energies were compared with implicit solvent inclusion using CPCM. Note that energy values are given in kJ/mol.

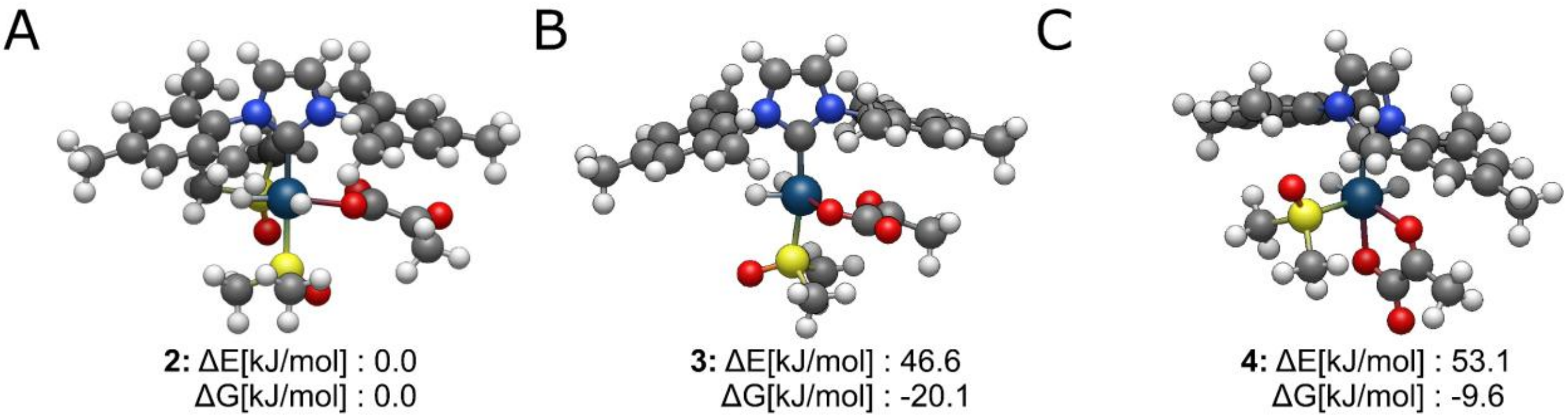

**Figure S24:** Optimized structures and relative free energies of complexes **2**-**4** obtained at the B3LYP-D4/def2-TZVP CPCM (methanol) level of theory. Note that for the relative energy of **2**, the free energy of DMSO has been computed at the same level of theory. For reference, free energies are shown in the table S21.

From the resulting free energies, the pyruvate complex **3** is the lowest energy structure - in contrast to what is found in experiment ($\Delta G^0_{2\leftarrow 3}$ = -5.3 kJ/mol in favor of **2**) - and the most stable compared to the other species, implying that **2** and **4** should not be observed at all (see ΔG in **Fig. S24**). For the electronic energies (**ΔE** in **Fig. S24**), however, **2,** which contains an additional DMSO, appears much more stable than **3** and **4**. This situation is often observed for Gibbs free energies of non-unimolecular reactions in solution. In the literature, it is well documented that in these cases the common entropy corrections contained in Gibbs free energies strongly overestimate the translational entropy [65FaliveneMolCat]. We will refer to this computational scheme as **CS1**.

Therefore, two additional computational schemes (**CS2** and **CS3**) have been applied in computing the relative free energies of **2**, **3,** and **4,** following the suggestions of Falivene et al. Ref [65FaliveneMolCat].

**CS2**: We computed the $\Delta G_{50\%}$ values, in which the entropy correction for all species was scaled by 0.5 (Ref. 66GarzaJCTC, 67YuJACS).

**CS3**: We computed the $\Delta G_{assoc}$ values, for which the standard entropy corrections were used, but the dissociating DMSO molecule was kept associated with the reacting complex in the second coordination sphere (see Fig. S25 for geometries). For the latter, translational entropy corrections mostly cancel.

The results with respect to compound **2** are shown in **Table S18**.

**Table S18:** Relative free energies in kJ/mol of compounds **3** and **4** with respect to compound **2**. Three computational schemes are compared. (**CS1**) ΔG denotes standard free energy corrections. For (**CS2**) $\Delta G_{50\%}$ the entropy correction has been scaled by 0.5. (**CS3**) $\Delta G_{assoc.}$ corresponds to the relative free energies of 3 and 4 encounter complexes with DMSO (see Fig. S25).

| Compound | **ΔG** | **$\Delta G_{50\%}$** | **$\Delta G_{assoc.}$** |
|---|---|---|---|
| **3** | -20.1 | 9.1 | 15.1 |
| **4** | -9.6 | 15.8 | 23.5 |

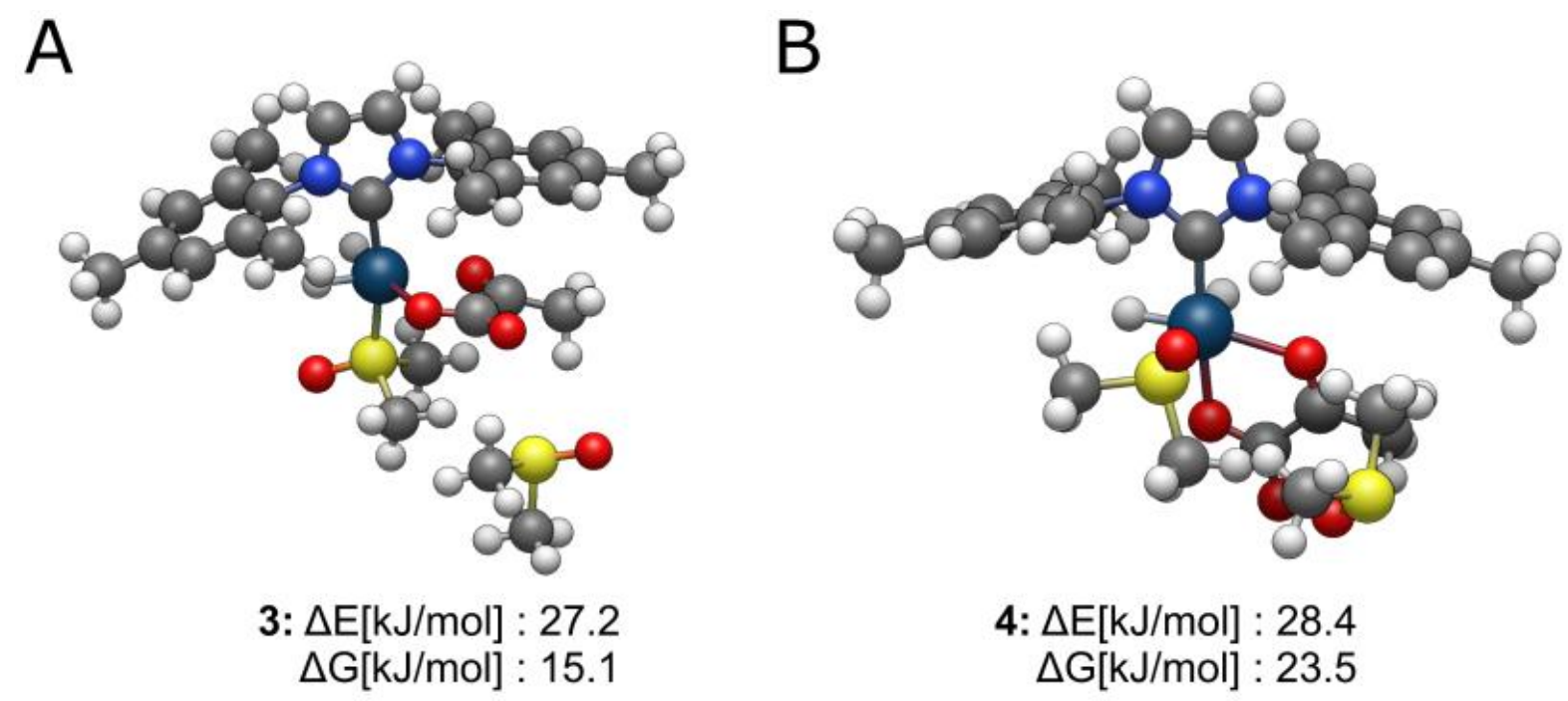

**Figure S25:** Geometries of compounds **3** and **4** optimized, including one weakly bound DMSO molecule in the second coordination sphere. Relative free energies and energies given in kJ/mol with respect to compound **2** (see **Fig. S24**).

As shown by the resulting Gibbs free energies (**Table S18**), the corrected values no longer favor the dissociation of DMSO from **2**. This clearly shows that the entropy corrections are a source of error in the computed results. For the discussion in the main manuscript, we have opted to report the $\Delta G_{50\%}$ values: $\Delta G_{50\%}$ and $\Delta G_{assoc.}$ clearly favor **2** over **3** by at least 9 kJ/mol, while **4** is unfavorable compared to **3** by at least 7 kJ/mol. These results are consistent with the fact that **2** is observed depending on the DMSO concentration and that it is unlikely that **4** is observed.

## 18.2. Additional influences on the relative stabilities of 2, 3, and 4

Experimental results indicate that, under certain conditions (for example, variation in [pyr], **Fig. 3E**), the relative concentrations of **2** and **3** can change, prompting additional computations of factors that may affect the balance between the two conformers. In this respect, it is noteworthy that sodium pyruvate is used in the experiment and has not been accounted for in calculations to date. An inspection of the electrostatic potentials (**Fig. 4**) already hints at another vital component, namely $Na^+$.

These electrostatic potentials suggest that the different compounds and conformers may exhibit strong stabilizing interactions with cations. The structures resemble a negatively charged pocket that facilitates explicit ion interactions. Consequently, all structures were reoptimized in the presence of $Na^+$, incorporating implicit solvation. As the measurements were carried out in methanol, we assessed a variety of adducts of **2**, **3,** and **4** with $Na^+$, DMSO, and methanol (**Fig. S26**; the list of structures is not exhaustive).

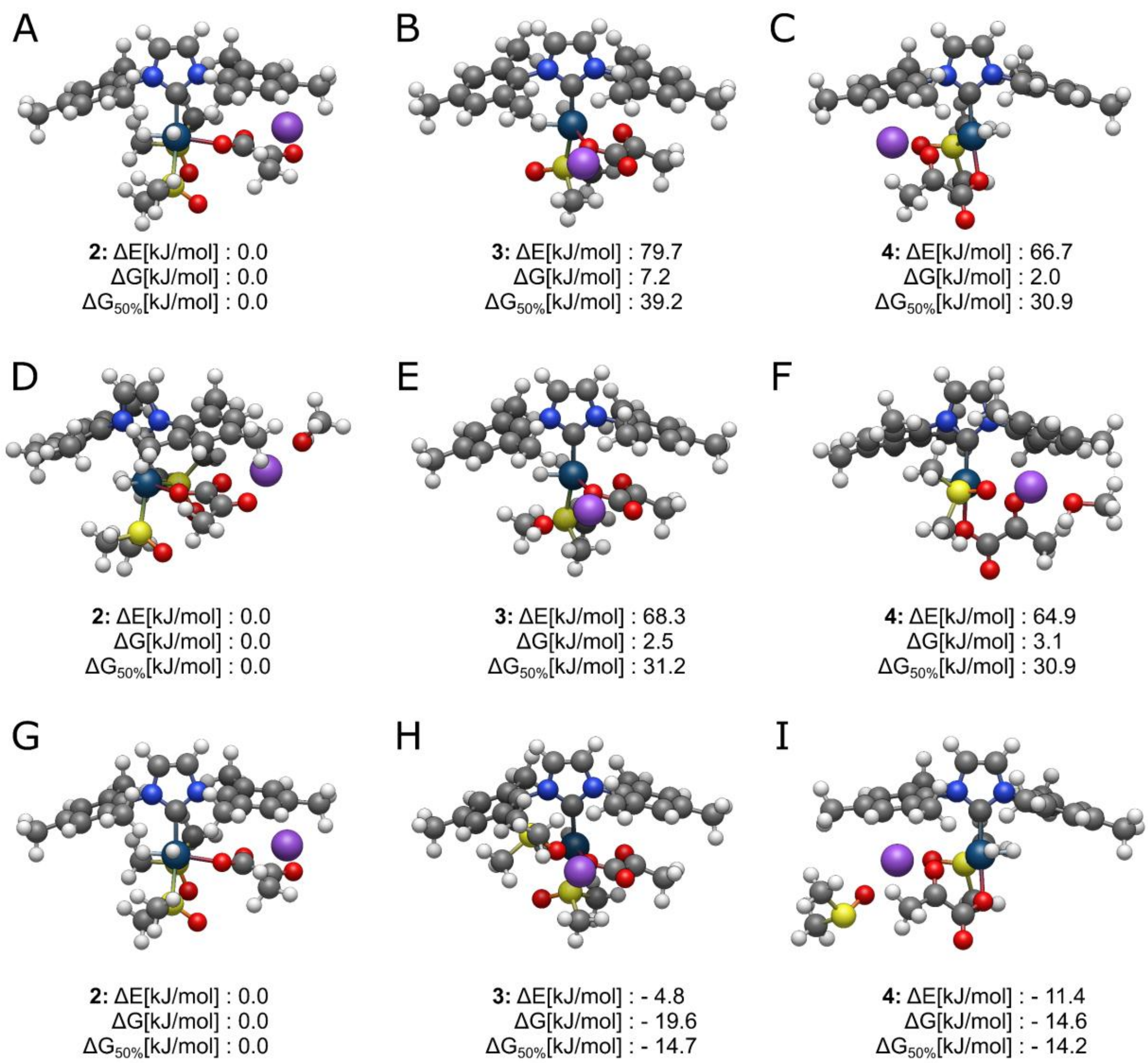

**Figure S26:** Optimized structures and relative free energies of complexes **2**-**4,** assuming sodium pyruvate yields the corresponding $Na^+$ adducts. Top row: (A, B, C) $Na^+$ adducts, middle row: (D, E, F) adducts with additional explicit solvent molecule (methanol), bottom row: (G, H, I) systems with DMSO in 2$^{nd}$ coordination sphere for **3** and **4**.

While these preliminary results on a narrow set of possible structures are far from quantitative, they indicate some important influences. While without the counter ion, **2** is favorable over **3** by about 16 kJ/mol (see middle column **Table S18**), with inclusion of the counter ion (see **Figure S26** top row), the difference increases to almost 40 kJ/mol (**Fig S26** top, using $\Delta G_{50\%}$ here). For the difference between **3** and **4**, compound **3** is favorable in energy by almost 7 kJ/mol without a counterion; with $Na^+$, this difference is reversed to more than -8 kJ/mol in favor of **4**. Note that the presence of $Na^+$ also slightly changes the minimum energy structures as it perturbs the balance of electrostatic repulsion and attraction in the ligands.

However, when including further species that are present in the solvent into the model (**Fig. S26**, middle and bottom), it becomes clear that the computed trends are very model-dependent. While **3** and **4** are always predicted to be close in energy, and for the majority of models, both are less stable than **2**, a more complete description of the effect of intermolecular interactions beyond implicit solvation is required for a more quantitative analysis. Nevertheless, the results still present a strong indication that the presence of counterions actually influences the stability of different species in the equilibrium in different ways, and might be applied to bias the system in a desired direction. Further studies are currently the focus of work in our group.

## 18.3. Formation energy of complex 2 from complex 1

From the experimental results, it can be assumed that **2** is formed from **1** after pyruvate dissociation and Cl addition. However, as their stoichiometries differ by the $Cl^-$ anion (and DMSO, which, however, is neutral), computed relative energies are much less reliable, as the common solvent models do not describe ions that undergo ion pair interactions very accurately. Note that, like in the previous section, this section is meant to assess possible influences and sources of error in the computational numbers, rather than derive accurate numbers.

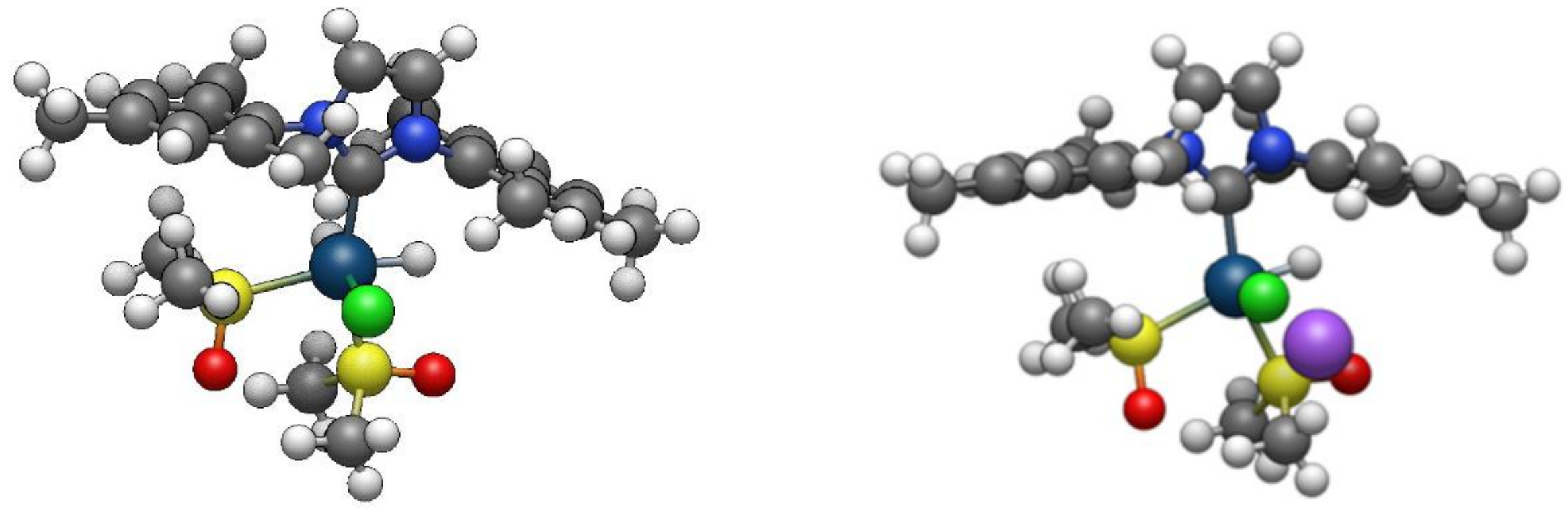

**Figure S27:** Optimized structures of complex **1** without (left), and with (right) $Na^+$ counterion.

To do so, we have computed the relative free energy differences using the computed energies (including CPCM for methanol) of sodium pyruvate, the chloride anion, the sodium cation, a sodium chloride ion pair, and the $Na^+$ educts (after preliminary screening of low energy conformers) of **1** and **3**, e.g. ΔG(**3**←**1**) = ΔG(**3**) - ΔG(**1**), at 298.15 K **(Table S19**), where **1** = [IrIMes(DMSO)$_2$(HH)Cl] and **3** = [IrIMesDMSO(pyr)(HH)]. From these numbers, it becomes obvious that a broad range of relative energies is obtained depending on how the reference energies are chosen.

In the extreme case of using $Na^+$ and $pyr^-$ for the educts and NaCl for the product side, **3** is more stable by 150 kJ/mol. If ions as reference systems are avoided, the relative energies change in favor of the educt side (like for [**3**+DMSO+NaCl ← [1+(Na pyr)]]), while the presence of an ion pair of **1** with $Na^+$ fully favors the educt side (as for [(**3** $Na^+$)+DMSO+NaCl] ← [(**1** $Na^+$)+(Na pyr)]]). Hence, reliable predictions for the relative free energies of **1** can only be made if more information about the final state of the ions can be obtained. This might include anything from precipitation of NaCl to strong interactions with other ions in solution.

**Table S19**: Relative free energy differences (ΔG, 298.15 K) for the interconversion between complexes **1** and **3** under different ionic environments.

| | $\Delta G^{298.15}$ (kJ/mol) |
|---|---|
| [**3**+DMSO+$Cl^-$] ← [**1**+$pyr^-$] | 21.8 |
| [**3**+DMSO+NaCl] ← [**1**+(Na pyr)] | -2.0 |

| [**3**+DMSO+NaCl] ← [**1**+$pyr^-$+$Na^+$] | -150.7 |
|---|---|
| [(**3** $Na^+$)+DMSO+NaCl] ← [(**1** $Na^+$)+(Na pyr)] | 51.0 |
| [(**3** $Na^+$)+DMSO+$Cl^-$] ← [(**1** $Na^+$)+$pyr^-$] | 74.9 |
| [(**3** $Na^+$)+DMSO+$Cl^-$] ← [**1**+(Na pyr)] | 53.6 |
| [(**3** $Na^+$)+DMSO+NaCl] ← [**1**+(Na pyr)+$Na^+$] | -118.9 |

## 18.4. Geometric parameters for 3

The structure of **3** calculated with B3LYP-D4/def2-TZVP CPCM(methanol) agrees with the structures and properties reported in Ref. 47MascittiPCCP obtained at the ω-B97XD(PCM) / LANL2TZ(Ir) 6-311G(d.p) level of theory (**Fig. S28**).

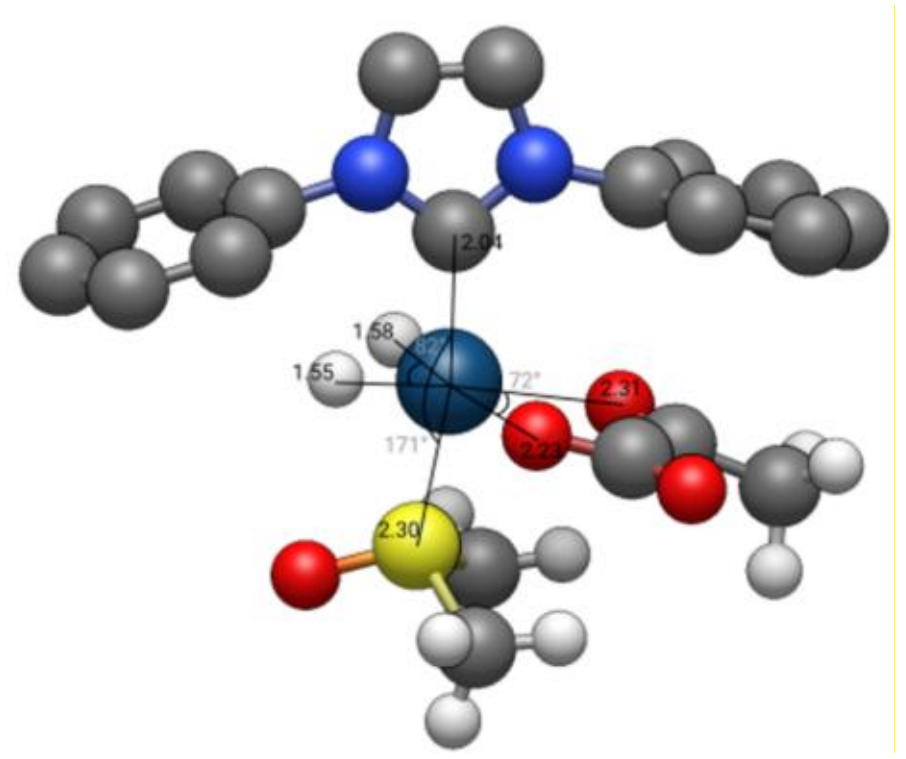

**Figure S28:** Geometrical parameters of **3** without the addition of $Na^+$ computed at the B3LYP-D4/def2-TZVP CPCM(methanol) level of theory.

## 18.5. Further DFT functional benchmarks

To assess the error introduced by DFT, the energies and energy difference between compounds **3** and **4** have been computed using single-point energy calculations using PBE, PBE0, TPSS, ωB97x, and B3LYP functionals (Refs. 81PerdewPRB-84ChaiPCCP) and B3LYP-D4/def2-SVP CPCM (methanol) optimized geometries with D4 dispersion correction and implicit solvation (**Table S20**, upper part).

The results indicate that, for these systems, typical DFT errors of 5 kJ/mol are observed, ensuring that the relative energies of different species are reproduced well, regardless of the functional chosen.

Comparing single-point calculations at the B3LYP-D4/def2-TZVP CPCM(methanol) geometry to the relative energies obtained from full optimizations using DFT-D4/def2-TZVP CPCM(methanol) (DFT = PBE, PBE0, TPSS, ωB97x, and B3LYP) shows that deviations are mostly due to the functional in the energy calculations and not so much due to differences in the structures (**Table S20**, lower part). Furthermore, B3LYP appears to accurately represent the range of results obtained. Also, the trend that **4** is always above **3** in energy is robust throughout the different results.

**Table S20**: Single-point energies (absolute electronic energies [a.u.]) and energy differences (ΔE[kJ/mol]) between compounds **3** and **4** calculated using different DFT functionals with B3LYP geometry (upper part, see text) and geometries optimized using the different functionals (lower part).

| | PBE | PBE0 | TPSS | WB97X | B3LYP |
|---|---|---|---|---|---|
| Geometry: B3LYP-D4/def2-TZVP CPCM<br>Single point energy: DFT-D4/def2-TZVP CPCM | | | | | |
| E(3) | -1923.701306930 | -1923.77022309 | -1925.879486509 | -1925.146623401 | -1924.707791733 |
| E(4) | -1923.699563319 | -1923.76765087 | -1925.877799114 | -1925.143430419 | -1924.705312465 |
| ΔE [kJ/mol] | 4.6 | 6.8 | 4.4 | 8.4 | 6.5 |
| Geometries optimized at the corresponding DFT-D4/def2-TZVP CPCM level of theory | | | | | |
| E(3) | -1923.703282727 | -1923.76936534 | -1925.880269513 | -1925.144541356 | -1924.7077917 |
| E(4) | -1923.704852988 | -1923.77189298 | -1925.881960145 | -1925.147787817 | -1924.707791733 |
| ΔE [kJ/mol] | 4.1 | 6.6 | 4.4 | 8.5 | 6.5 |

### Relaxed H-H distance potential energy curve scan

To assess the intramolecular dynamics of the hydrogens in the system, a relaxed surface scan for the H-H bond distance was carried out for **3** (see **Figure S29**). Note that the equilibrium distance between the hydrogen atoms is 2.05 Å. From this plot, it becomes evident that the potential of the H atoms moving in is, as expected, very shallow. Within 5-10 kJ/mol distance changes of almost 0.3 Å are possible, indicating a strong quantum nature in the dynamics of the hydrogen degrees of freedom.

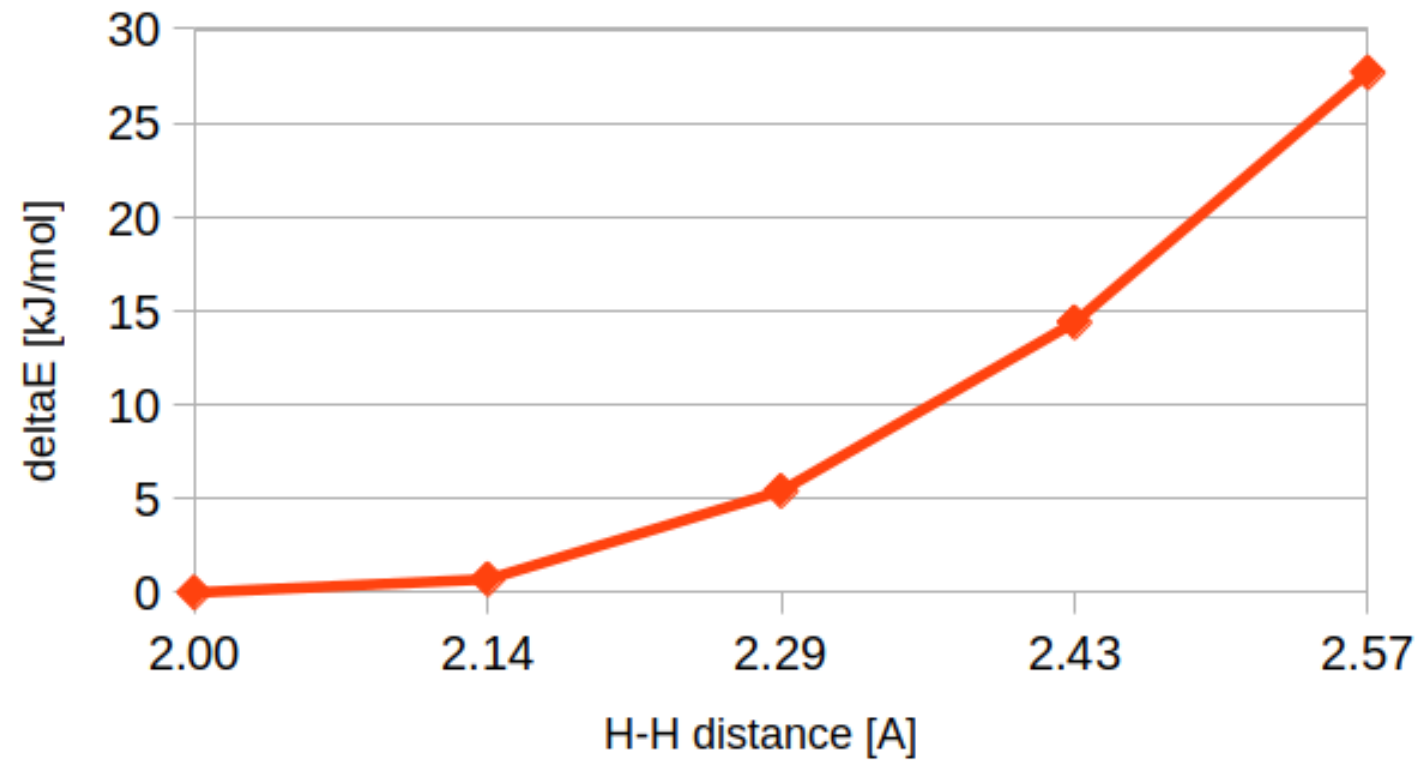

**Figure S29:** Relaxed H-H distance scan at the B3LYP-D4/def2-SVP CPCM(methanol) level of theory.

### Sample input files

Standard DFT geometry optimization and frequency calculation:

```
!B3LYP D4 RIJCOSX def2-TZVP def2/J CPCM(Methanol) TIGHTOPT FREQ
```

### Reference energies

For all relative Gibbs free energy calculations, the following reference systems have been used:

**Table S21**: Reference systems for all relative Gibbs free energy calculations.

| | G [a.u.] |
|---|---|
| $H_2$ | -1.17554706 |
| DMSO | -553.12635508 |
| MeOH | -115.69137577 |
| pyruvate anion | -341.91993197 |
| pyruvate protonated | -342.36424947 |
| chloride anion | -460.336856781981 |
| water | -76.43352387 |
| sodium chloride ($Na^+Cl^-$) model | -622.55046007 |
| sodium pyruvate | -504.1244614 |
| sodium cation | -162.147901128498 |

**Note:** XYZ coordinates of selected compounds are given as a separate zip file as part of the Zenodo DOI: https://doi.org/10.5281/zenodo.18449765.